\documentclass[a4paper,11pt]{article}

\pdfoutput=1
\usepackage[utf8]{inputenc}
\usepackage[T1]{fontenc}
\usepackage{fontawesome}
\usepackage{xspace}
\usepackage{jheppub}
\usepackage{comment}
\usepackage{graphicx,color,amsmath}
\usepackage{siunitx}
\usepackage[normalem]{ulem}
\usepackage{todonotes}
\usepackage{aas_macros}
\usepackage{fix-cm}    
\usepackage{tikz}
\usetikzlibrary{intersections, calc, quotes, angles, shapes.geometric}
\usepackage{tikz-3dplot}
\usepackage{wasysym}

\definecolor{planetcolor}{HTML}{fc8d62}
\definecolor{particlecolor}{HTML}{66c2a5}
\definecolor{eclipticcolor}{HTML}{66c2a5}
\definecolor{orbitcolor}{HTML}{fc8d62}

\makeatletter
\newcommand\HUGE{\@setfontsize\Huge{50}{60}}
\makeatother

\definecolor{deepblue}{rgb}{0.2,0.2,0.8}

\definecolor{deepred}{rgb}{0.8,0.2,0.2}

\newcommand{\vect}[1]{\boldsymbol{\mathbf{#1}}}

\newcommand{\dd}{{\rm d}}

\newcommand{\yr}{{\, {\rm yr}}}

\newcommand{\AU}{{\, {\rm AU}}}

\newcommand{\MeV}{{\, {\rm MeV}}}

\newcommand{\OO}{{\mathcal{O}}}

\definecolor{mypurple}{RGB}{164,64,214}
\definecolor{myindigo}{RGB}{55, 52, 235}
\definecolor{myblue}{RGB}{50,50,214}
\definecolor{myred}{RGB}{214,50,50}
\definecolor{mygreen}{RGB}{50,180,50}

\definecolor{linkcolor}{rgb}{0.7752941176470588, 0.22078431372549023, 0.2262745098039215}
\newcommand{\nbicon}{{\color{linkcolor}\faFileCodeO}\xspace}
\newcommand{\nblink}[1]{\href{#1}{\nbicon}}
\newcommand{\githubmaster}{\href{https://github.com/kenvantilburg/solar-basin-dynamics/}{\faGithub}\xspace}

\newcommand\githubicon[1]{\href{#1}{\faGithub}\xspace}

\begin{document}

\title{Orbital Dynamics of the Solar Basin}

\author[a]{Cara Giovanetti,}
\emailAdd{cg3566@nyu.edu}
\affiliation[a]{Center for Cosmology and Particle Physics, Department of Physics, New York University,
New York, NY 10003, USA}

\author[b]{Robert Lasenby,}
\emailAdd{rlasenby@stanford.edu}
\affiliation[b]{Stanford Institute for Theoretical Physics, Stanford University, Stanford, CA 94305, USA}

\author[a,c]{Ken Van Tilburg}
 \emailAdd{kenvt@nyu.edu}
 \affiliation[c]{Center for Computational Astrophysics, Flatiron Institute, New York, NY 10010, USA}

\date{\today}

\abstract{
We study the dynamics of the \emph{solar basin}---the accumulated population of weakly-interacting particles on bound orbits in the Solar System.  
We focus on particles starting off on Sun-crossing orbits, corresponding to initial conditions of production inside the Sun, and investigate their evolution over the age of the Solar System.
A combination of analytic methods, secular perturbation theory, and direct numerical integration of orbits sheds light on the long- and short-term evolution of a population of test particles orbiting the Sun and perturbed by the planets.
Our main results are that the effective lifetime of a solar basin at Earth's location is $\tau_{\rm eff} = 1.20\pm 0.09 \,\mathrm{Gyr}$, and that there is annual (semi-annual) modulation of the basin density with known phase and amplitude at the fractional level of 6.5\% (2.2\%). 
These results have important implications for direct detection searches of solar basin particles, and the strong temporal modulation signature yields a robust discovery channel.
Our simulations can also be interpreted in the context of gravitational capture of dark matter in the Solar System, with consequences for any dark-matter phenomenon that may occur below the local escape velocity.}
\maketitle


\section{Introduction}

Many theories of physics beyond the Standard Model (BSM) lead to astrophysical populations of new particles. Terrestrial laboratory experiments can search for a local flux of such particles at Earth. The best-known example is dark matter (DM), but there are other candidates, including new particles produced in the Sun and other stars~\cite{Sikivie_1983,Bibber_1989,PASCHOS_1994,Moriyama_1995,Arik_2011,Redondo_2013a, Armengaud_2014,Giannotti_2016,Giannotti_2017,Mastrototaro_2020,Di_Luzio_2022,Chang_2022}, supernovae~\cite{Engel_1990,lella2023,carenza2023}, or in cosmic-ray collisions and the early universe~\cite{Pospelov_2021,berlin_2023}.

Gravitational fields \emph{within} the Solar System typically have negligible effects on the total flux of such particles at Earth. For most studies of DM, the flux is assumed to be dominated by particles which are unbound to the Sun but are on bound trajectories through the galactic halo, as the escape velocity from the halo exceeds the local escape velocity of the Solar System by about a factor of 10. This is even truer for higher-velocity fluxes, such as those from supernovae, cosmic rays, or primordial dark radiation. Likewise, the vast majority of low-mass particles produced in the Solar core are generally produced at speeds well above the escape velocity of the Sun, so they exit the Solar System on nearly straight trajectories without losing an appreciable amount of kinetic energy.

However, some small fraction of particles produced in the Sun will be emitted onto bound orbits. This ``solar basin'' population~\cite{VanTilburg:2020jvl} may accumulate within the Solar System for billions of years, such that its energy density exceeds that of the unbound flux---its long lifetime compensating for the small volume of phase space corresponding to bound production. The same set of couplings responsible for their production may then also generate signals in direct detection experiments on Earth, or leave indirect signatures. 

Ref.~\cite{VanTilburg:2020jvl} identified these dynamics generally and presented a case study for axion-like particles coupled to electrons.  Ref.~\cite{Lasenby:2020goo} worked out in detail the case of kinematically-mixed dark photon production, where the solar-basin population provides \emph{leading} sensitivity over a large mass range. 
Resonant pair production of fermionic millicharged particles into the solar basin may be directly detected on Earth by proposed electromagnetic deflectors~\cite{Berlin:2021kcm}.
Indirect detection signatures from solar basin axions decaying to two X-ray photons very near the Solar limb were used in tandem with NuSTAR data to place strong constraints on axion couplings to electrons and photons~\cite{DeRocco:2022jyq}. Previously, similar indirect detection phenomenology of gravitationally-bound particles was also considered in the context of supernovae~\cite{Hannestad_2002} and solar coronal heating~\cite{DiLella_2003}, as well as direct detection of a solar basin of Kaluza-Klein towers~\cite{Morgan_2005,XMASS:2017sij}.

Prior to this work, there had been a large uncertainty regarding the present-day density of the solar basin near Earth due to the poorly known long-term evolution of test-particle orbits. While particles produced inside the Sun start out on highly elliptical, Sun-crossing orbits, they are perturbed by the gravitational influence of the planets. A combination of secular perturbations, sufficiently close encounters, and motional resonances may eventually eject the particle from the Solar System entirely. Refs.~\cite{VanTilburg:2020jvl,Lasenby:2020goo} adopted a variety of assumptions about this orbital evolution, with a plausible range from very short ejection dynamics on the order of the Lyapunov timescale of $ 10^7 \yr$~\cite{Laskar_1989,LASKAR_1990,Sussman_1992,Mogavero_2021,mogavero2023timescales}, to lifetimes of order the $4.6\,\mathrm{Gyr}$ age of the Solar System. While naive estimates and simulations based on toy models of phase-space diffusion of orbits~\cite{1991ApJ...368..610G,Anderson:2020rdk} seemed to point towards $\mathcal{O}(\mathrm{Gyr})$ lifetimes, it was not possible to be confident in such projections due to potential systematics arising from the omitted physics.

To remedy this situation, we have performed large-scale numerical simulations of the long-term evolution of test-particle orbits in a model Solar System consisting of the Sun, Venus, Earth, Jupiter, and Saturn. We used direct $N$-body numerical integration to evolve this system for $4.5 \, \mathrm{Gyr}$ of the lifetime of the Solar System, simulating hundreds of test particles in parallel. This paper describes in detail these simulations and their results. We supplement these results with semi-analytic methods to verify simplifying assumptions employed in our numerical simulations, and to address more detailed questions about temporal modulation. Our work may also have important implications for any DM process that is enhanced at low velocities, such as scattering via a light mediator  (see ref.~\cite{Essig:2022dfa} for a review).

We provide an executive summary of this work in section~\ref{sec:summary}, before providing the details of our analyses in subsequent sections.  In section~\ref{sec:numerical}, we describe the suites of numerical simulations performed to study the long-term evolution of the solar basin. Section~\ref{sec:secular} uses secular perturbation theory for a semi-analytic prediction of the short-term density variations, including annual and semi-annual modulation effects. We describe an interpretive analysis of basin evolution based on stochastic energy-changing processes in section~\ref{sec:stochastic}. In section~\ref{sec:analytical}, we enumerate a variety of analytical results and checks to validate the dynamics and assumptions of the prior sections. We conclude in section~\ref{sec:conclusions}. 

We also include three appendices containing supplemental information.  Appendix~\ref{app_elts} includes a refresher on orbital elements and the conventions for the action-angle variables in this paper, appendix~\ref{app:basin_density} reviews how the basin density is extracted from numerical data, and appendix~\ref{sec:scattering} details the calculation of the rate of gravitational scattering and includes analytic expressions omitted in the main body.

The code used to obtain the results of this study is available on GitHub, and a link (\githubmaster) below each figure provides the code with which it was generated. The large quantity of data generated by this analysis is difficult to host publicly, but is available upon request, and statistically reproducible with the simulation code provided.

Throughout our code base, we use units where $\SI{}{AU}=G_N M_{\odot}=1$, with $G_N$ being Newton's gravitational constant and $M_{\odot}$ the mass of the Sun, so that a circular orbit at $\SI{1}{AU}$ has a period of $1\,\mathrm{yr} = 2\pi$, though we will quote most results in their conventional units in the main text. We use dimensionless action variables when presenting results in action-angle variables, given by 
\begin{equation}
	\tilde J_1 = \sqrt{G_N M_{\odot} a}; 
	\qquad
	\tilde J_2 = \tilde L = \tilde J_1 \sqrt{1 - e^2};
	\qquad
	\tilde J_3 = \tilde L_z = \tilde J_2 \cos I, \label{eq:action-angle-tilde_intro}
\end{equation}
where $a$, $e$, and $I$ are the semi-major axis, the eccentricity, and the inclination of the orbit, respectively. We eschew the tildes in text.  More information about these conventions is included in appendix~\ref{app_elts}.

\section{Executive summary}
\label{sec:summary}

In this work, we determine the ``effective solar basin lifetime''~\cite{VanTilburg:2020jvl, Lasenby:2020goo} to be $ \tau_{\rm eff}(R = 1\, \mathrm{AU}) = 1.20\pm 0.09 \,\mathrm{Gyr}$. This universal constant can be used in tandem with a calculation of the (time-independent) solar basin energy density \emph{production} rate $\dot{\rho}_\mathrm{b}(R)$, which generically scales as $R^{-4}$, to determine the present-day solar basin energy density:
\begin{align}
    \rho_\mathrm{b}(R) = \dot{\rho}_\mathrm{b}(R) \tau_\mathrm{eff}(R). \label{eq:tau_eff}
\end{align}
Our work is thus crucial to connect direct detection experiments on Earth---whose signals scale as $\rho_\mathrm{b}(\SI{1}{AU})$---to the calculable production rate of axions~\cite{VanTilburg:2020jvl}, dark photons~\cite{Lasenby:2020goo}, millicharged particles~\cite{Berlin:2021kcm}, and other weakly coupled particles bound to the Sun. The effective lifetime depends on the distance $R$ from the Sun; in what follows, we will primarily concern ourselves with $\tau_\mathrm{eff}(R)$ at $R = 1\,\mathrm{AU}$ and omit the argument.

The value of $\tau_{\rm eff} = 1.20\pm 0.09 \,\mathrm{Gyr}$ extracted from the numerical simulations in section~\ref{sec:numerical} is not far from the ``optimistic'' scenario in refs.~\cite{VanTilburg:2020jvl, Lasenby:2020goo}, wherein the effective lifetime would equal the full lifetime of the Sun $\tau_\mathrm{eff} = t_\odot \approx 4.5\,\mathrm{Gyr}$~\cite{Bahcall:1995bt,2012Sci...338..651C}.\footnote{We conservatively take the fiducial age of the Sun to be slightly lower than these currently most precise estimates, because the overall structure of the Solar System is somewhat uncertain shortly after formation (see section~\ref{sec:phys_sys}).} This solidifies the solar basin direct detection limits on axions~\cite{VanTilburg:2020jvl} and dark photons~\cite{Lasenby:2020goo}, with the updated constraints from the latter providing the most stringent DM-independent bound on the kinetic mixing parameter over significant parts of the mass range $10 \,\mathrm{eV}$--$3\,\mathrm{keV}$. We expect this estimate of the basin lifetime applies to basins comprised of other BSM particles beyond axions and dark photons, as details about production in the Sun (cfr. figure~\ref{fig_pdf1}) or in-medium effects (cfr. figure~\ref{fig_lift1}) do not significantly impact the late-time basin density.

Our estimates of the present-day density, and therefore the effective lifetime, are obtained through $N$-body simulations, where test particle trajectories are calculated one at a time in the Solar System.  We calculate trajectories of particles emitted from the Sun, and observe whether they cross Earth, as well as trajectories that begin at Earth and evolve backward, and verify whether these cross the Sun at some point within the $4.5\,\mathrm{Gyr}$ history to confirm these orbits could correspond to particles emitted by the Sun. Density estimates obtained from these different integration strategies agree well with one another. 

We supplement our numerical simulations with analytic and semi-analytic approaches to verify our results, to confirm the approximations made in simulations are reasonable, and to identify experimental signatures.  Our secular perturbation theory analysis of section~\ref{sec:secular} reveals the annual and semi-annual modulation of the basin density at Earth, which are detectable ``smoking gun'' signals of a solar basin.  Our stochastic description of the solar basin in section~\ref{sec:stochastic}, including effects from close encounters with planets and diffusion of test particles throughout the Solar System, yields results in reasonable agreement with our numerical simulations.

Our results are relevant for constraining particle physics models (or finding evidence for them in the future) when combined with calculations of particle production rates in the Sun.  
We consider particles which can be singly produced inside the Sun, such as axions~\cite{VanTilburg:2020jvl} and dark photons~\cite{Lasenby:2020goo}. The results carry over straightforwardly to \emph{pair} production of e.g.~millicharges~\cite{Berlin:2021kcm}, since over the vast majority of phase space of basin production, the ``other'' particle escapes relativistically, effectively making it a single-production process for the solar basin.

\begin{figure}[t]
\begin{center}
\includegraphics[width=0.49\textwidth]{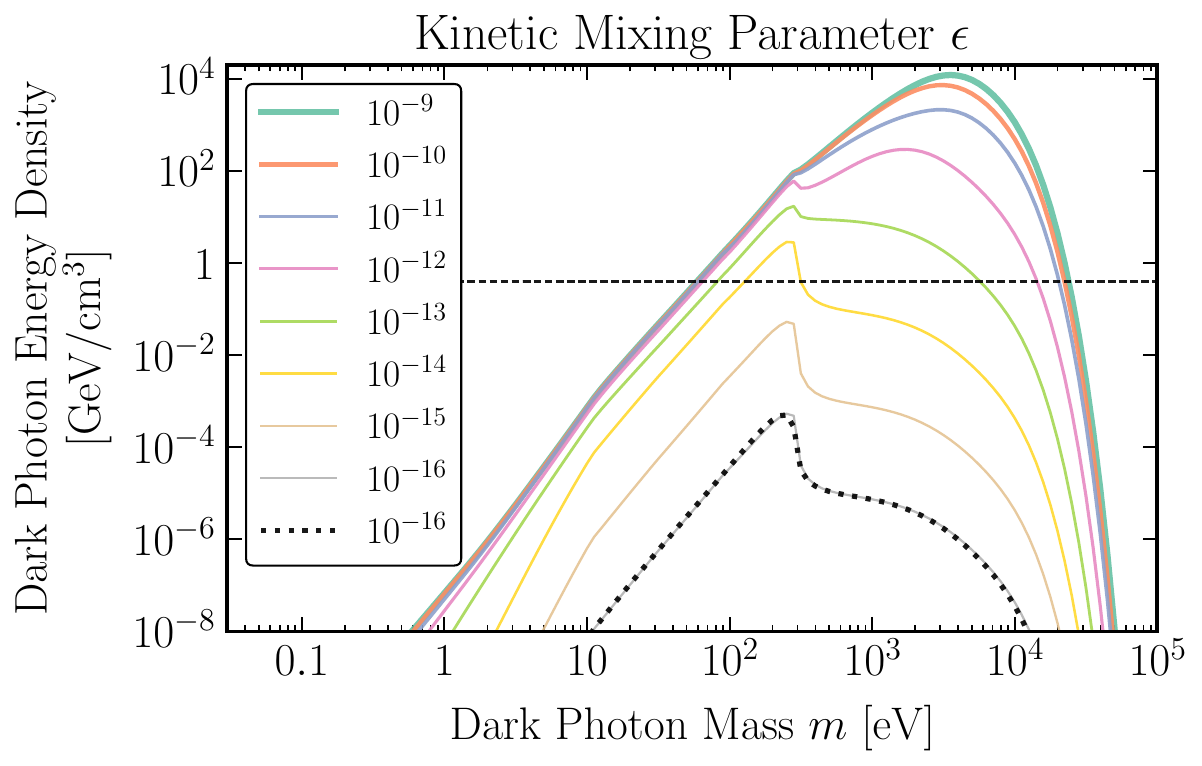}
\includegraphics[width=0.48\textwidth]{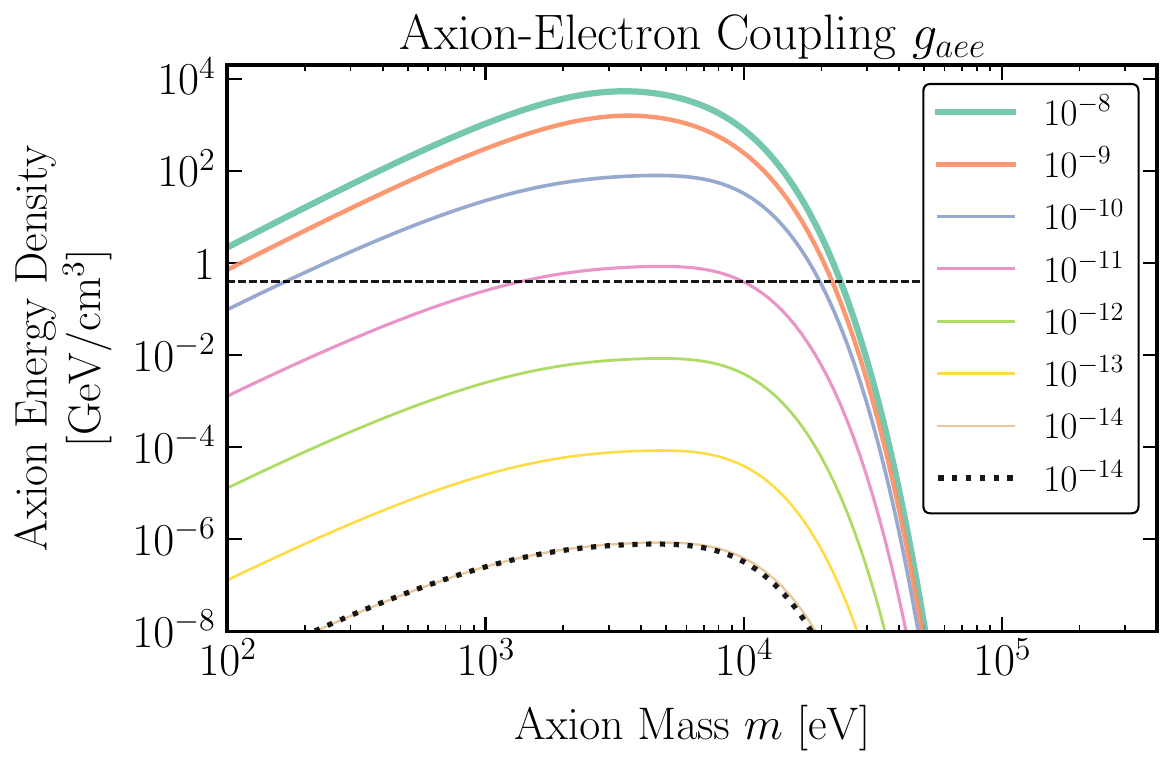}
\caption{\emph{Left panel:} Solar basin density at Earth from emission of dark photons in the Sun as a function of mass $m$, for different values of the kinetic mixing parameter $\epsilon$. The production rate are taken from ref.~\cite{Lasenby:2020goo}, while the orbital dynamics calculations are from this work.  
For $|\epsilon| \lesssim 10^{-14}$, saturation effects (see figure~\ref{fig_sat1}) are unimportant at all masses. In that regime, the density scales as $\epsilon^2$ and is simply the basin density production rate $\dot{\rho}_\mathrm{b}$ times the effective basin time $\tau_\mathrm{eff} = 1.2\,\mathrm{Gyr}$, as shown for $\epsilon = 10^{-16}$ by the black dotted curve. For values of $|\epsilon| \gtrsim 10^{-14}$, saturation effects become important for some masses, with the basin density asymptoting to a maximum value dictated by thermodynamic detailed balance.
\emph{Right panel:} same as in the left panel, but for an axion-like particle of mass $m$ with pseudoscalar coupling $g_{aee}$ to electrons with production calculations from ref.~\cite{VanTilburg:2020jvl}. Saturation effects are only important at coupling strengths already excluded by other observations ($|g_{aee}| \gtrsim 10^{-10}$). The black dotted curve shows that multiplication of the production rate with $\tau_\mathrm{eff}$ matches the full numerical result for $g_{aee} = 10^{-14}$. In both panels, the gray dashed line indicates the galactic DM density $\rho_\mathrm{DM} = 0.4\,\mathrm{GeV/cm^{3}}$ for reference. \githubicon{https://github.com/kenvantilburg/solar-basin-dynamics/blob/main/code/Summary/constraints.ipynb}}
\label{fig_rhoDP}
\end{center}
\end{figure}

In the left panel of figure~\ref{fig_rhoDP}, we show the solar basin density at Earth that would arise from dark photon production in the Sun, derived by combining the production rate calculations from \cite{Lasenby:2020goo} with the orbital dynamics calculations from this paper. When the coupling is small, we can ignore effects of dark photons being reabsorbed after emission, and the density at Earth scales $\propto \epsilon^2$ in the kinetic mixing parameter $\epsilon$. For larger couplings, the phase space around Earth can become nearly saturated for some range of dark photon masses. The density increases more slowly with $\epsilon$, eventually reaching the fully saturated value dictated by detailed balance for these couplings~\cite{Lasenby:2020goo}.

\begin{figure}[t]
	\begin{center}
\includegraphics[width=0.9\textwidth]{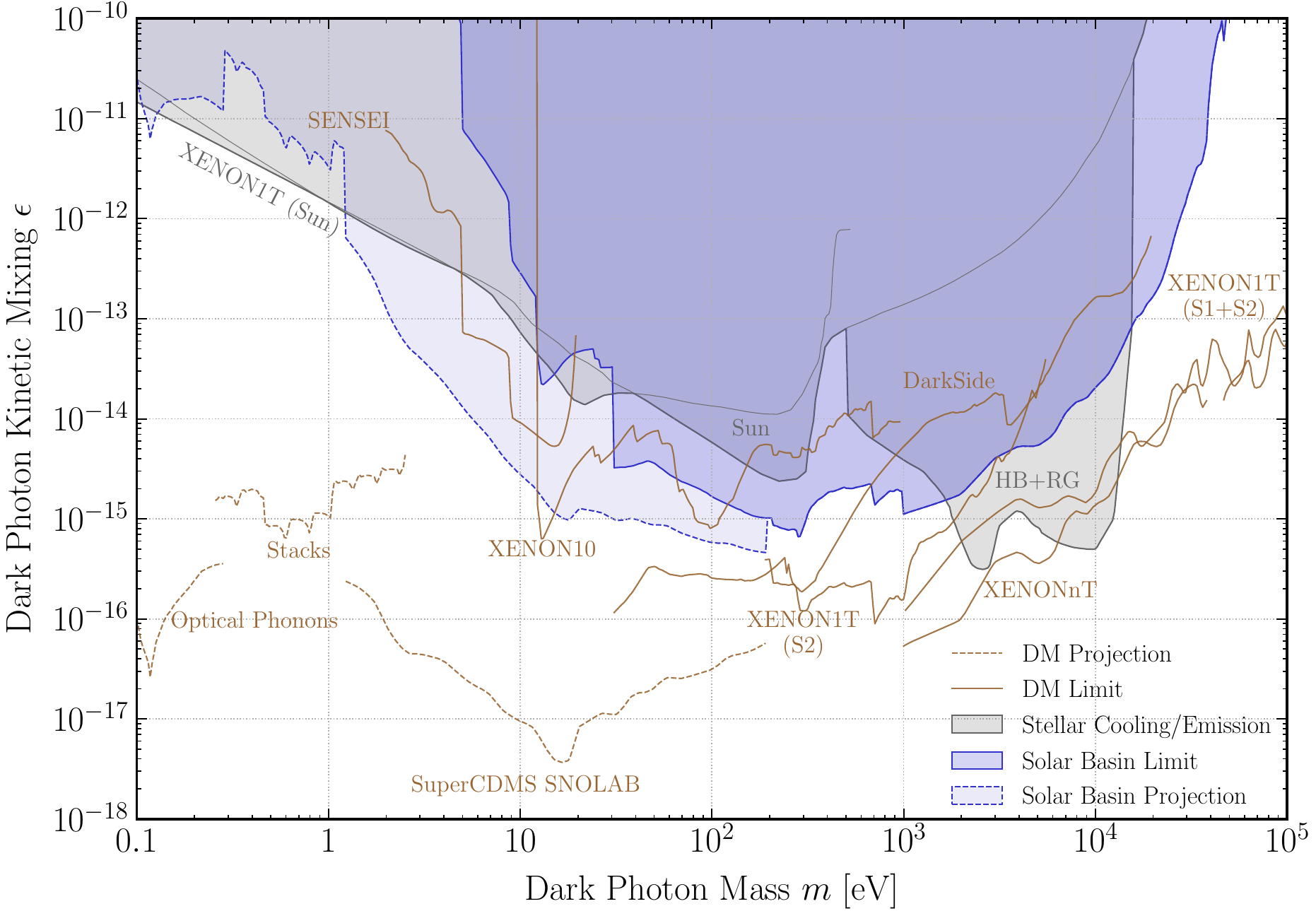}
\caption{Constraints and projected reach for the kinetic mixing $\epsilon$ of a dark photon with mass $m$. The brown solid lines indicate leading kinetic mixing limits from existing experiments~\cite{Bloch_2017,aprile2019light,barak2020sensei,Aprile_2020,Aprile_2022,Darkside50}, assuming that the dark photon makes up all of the DM\@.  The darker shaded blue region shows the constraints from non-detection of a solar basin population in Earth-based DM searches; it combines the production calculations from ref.~\cite{Lasenby:2020goo} with the orbital dynamics results from this paper.  Meanwhile, the brown dotted lines show the projected reach of future experiments~\cite{Knapen_2018,Baryakhtar_2018,supercdmscollaboration_2023}, again assuming the dark photon makes up all of the DM, while the lighter shaded blue region shows projected reach of these future DM detection experiments to the solar basin population. The gray shaded regions show constraints from non-detection of the relativistic solar axion flux~\cite{an2020new}, and stellar energy losses in the Sun~\cite{li2023} and asymptotic/red giant stars and horizontal branch stars~\cite{Dolan:2023cjs}.  These curves illustrate that, while the solar basin density is usually lower than the DM density, it can still be large enough to probe unconstrained parameter space. \githubicon{https://github.com/kenvantilburg/solar-basin-dynamics/blob/main/code/Summary/constraints.ipynb}}
\label{fig_epsDP}
\end{center}
\end{figure}

Future DM experiments based on liquid xenon as well as low-threshold targets should be sensitive to solar basin dark photons across a wide range of unexplored parameter space between $0.1 \,\mathrm{eV}$ and $10 \,\mathrm{keV}$, as illustrated in figure~\ref{fig_epsDP}. Our results further motivate novel experimental concepts exploiting the presence of a low-velocity population of new particles such as the ``direct deflector'' of ref.~\cite{Berlin:2021kcm}. We also identify smoking-gun temporal modulation signatures that can discriminate against prosaic backgrounds and, potentially, a DM origin for a tentative detection, and thus feasibly lead to a \emph{discovery} of a solar basin.  This annual modulation is derived and illustrated in figure~\ref{fig:modulation_real} in section~\ref{sec:secular}.

\begin{figure}[t]
\begin{center}
\includegraphics[width=0.9\textwidth]{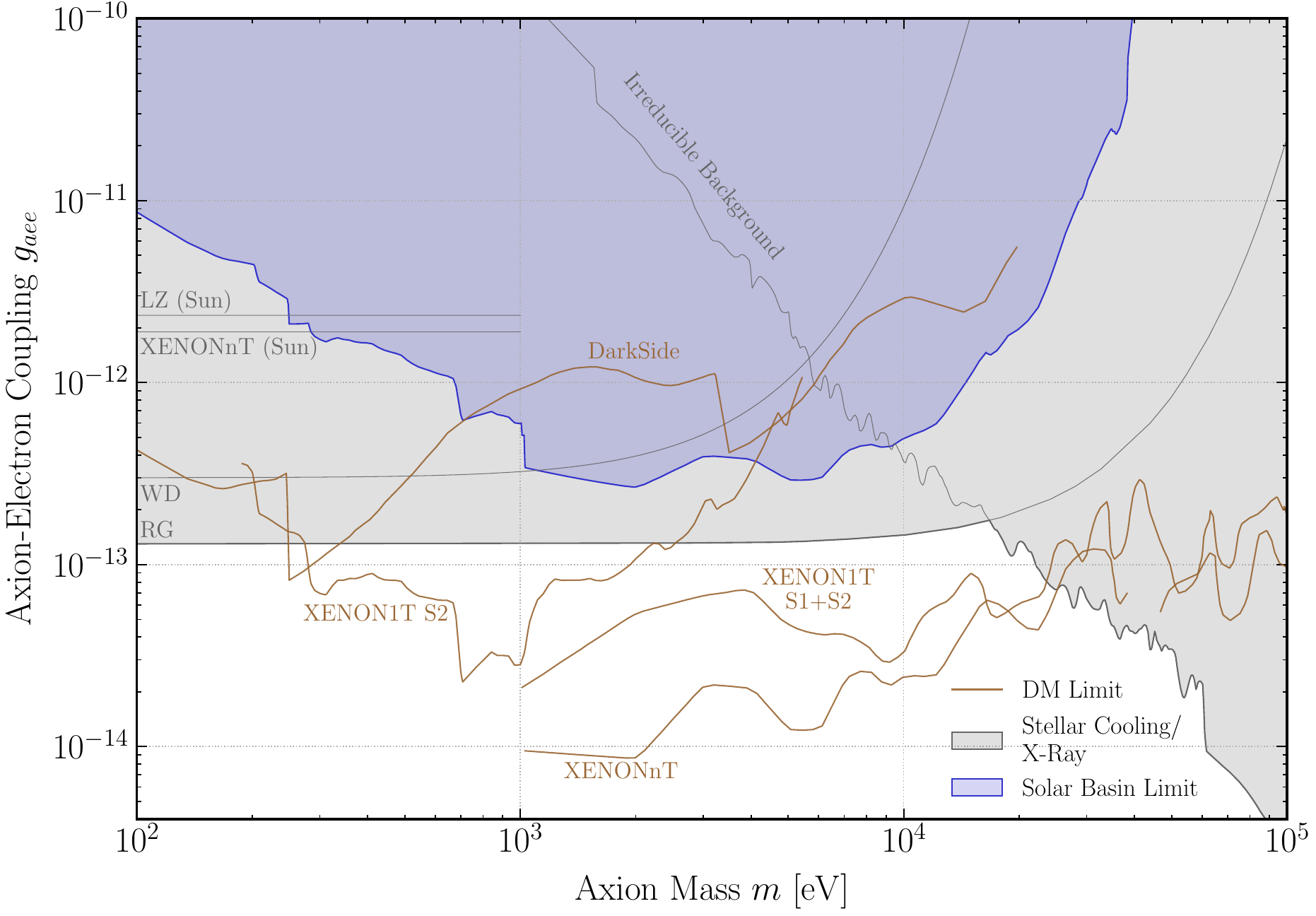}
\caption{Constraints and projected reach for the axion-electron coupling $g_{aee}$ of an axion-like particle with mass $m$. The blue shaded region shows the constraints from non-detection of a solar basin population in Earth-based experiments; it combines the production calculations from ref.~\cite{VanTilburg:2020jvl} with the orbital dynamics results from this paper. The gray shaded region shows constraints from stellar energy loss in red giants (RG)~\cite{2020PhRvD.102h3007C} and white dwarfs (WD)~\cite{MillerBertolami:2014rka}, and the X-ray flux from decays of the irreducible (photophobic) axion background~\cite{Langhoff:2022bij}, as well as solar axion constraints from XENONnT~\cite{Aprile_2022} and LZ~\cite{Aalbers_2023}. The solid brown lines show the axion coupling limits from various experiments assuming that the axion makes up all of the DM~\cite{aprile2019light, Aprile_2020, Aprile_2022, Darkside50}, though in this scenario the X-ray flux constraints would tighten further~\cite{Takahashi:2020bpq}. \githubicon{https://github.com/kenvantilburg/solar-basin-dynamics/blob/main/code/Summary/constraints.ipynb}
}
\label{fig_gaee}
\end{center}
\end{figure}

Axion-like particles with a pseudoscalar coupling to electrons are another quintessential class of BSM particles which could be produced in the Sun and then detected in terrestrial experiments, either through its relativistic flux~\cite{Redondo:2013wwa} or its solar basin~\cite{VanTilburg:2020jvl}.  
The right panel of figure~\ref{fig_rhoDP} shows the solar basin density at Earth for different axion masses and couplings, showing similar saturation effects to the dark photon case.  Resulting constraints on the axion-electron coupling $g_{aee}$ are shown in figure~\ref{fig_gaee}.

In previous estimates from refs.~\cite{VanTilburg:2020jvl, Lasenby:2020goo}, the large uncertainty in the effective solar basin lifetime $\tau_{\rm eff}$ (for which values as low as $10^7 \yr$ were considered) meant that these constraints only exceeded those from unbound stellar fluxes over a small range in dark photon masses. The results from this paper, which allow us to be confident that $\tau_{\rm eff} \gtrsim \mathrm{Gyr}$,  extend this to a wider mass range, and open the possibility of improved discovery potential over a wide range of smaller dark photon masses, from future experiments such as SuperCDMS~\cite{supercdmscollaboration_2023}.
The long solar basin lifetime also implies that any contribution from \emph{Earth}'s basin will be subdominant at all masses, even below those of Earth's core temperature; see refs.~\cite[Fig.~2]{VanTilburg:2020jvl} and~\cite[Fig.~3] {Lasenby:2020goo}.
Finally, the high value of $\tau_\mathrm{eff}$ implies that experimental searches for axion absorption are close to exceeding astrophysical constraints on stellar energy loss, so that future experiments will likely have leading sensitivity to keV-mass axions, independent of cosmological production.

\section{Numerical simulations}
\label{sec:numerical}

In this section, we describe our numerical simulations for the long-term evolution of a solar basin. 
In section~\ref{sec:setup}, we outline the setup of our numerical integration, including details of the physical system (section~\ref{sec:phys_sys}), the numerical integration algorithm (section~\ref{sec:algorithm}), and the initial conditions and integration strategies (section~\ref{sec:int_strat}).
We report the low-level results of our simulation runs in section~\ref{sec:sim_res}, including the density at Earth (section~\ref{sec:dens_est}), saturation effects (section~\ref{sec:saturation}), and an annual modulation analysis (section~\ref{sec:ann_mod}, more in section~\ref{sec:secular}).  The results from these sections were used for the direct detection constraints and discovery prospects of a solar basin in the executive summary of section~\ref{sec:summary}.
Finally, in section~\ref{sec:dens_capture}, we briefly describe how our results could be employed for the time-reversed problem, that of gravitational capture into bound Solar System orbits. 

\subsection{Setup} 
\label{sec:setup}

\subsubsection{Physical system}
\label{sec:phys_sys}

We perform $N$-body Solar System simulations in which five bodies---the Sun, Venus, Earth, Jupiter, and Saturn---are treated as massive gravitating bodies, and the basin particles are treated as (nongravitating) test particles. 

We treat the Sun as unchanged in mass, mass profile, and composition over its $4.5\,\mathrm{Gyr}$ history, so that it assumes its present-day state throughout our simulations.
Because of a considerable fraction of the Sun's nuclear fuel has been spent already, the Sun had a different density and temperature profile earlier in its life~\cite{Feulner_2012}, affecting both the particle production rate, and the orbits of Sun-crossing particles. As estimated in ref.~\cite{Lasenby:2020goo}, effects from the changing particle production rate on the present-day Earth-crossing density are likely to be at the $\lesssim \OO(10\%)$ level. However, changes to the density profile from, e.g., a lesser helium fraction in the young Sun would considerably change only the \emph{individual} particle orbits, and not their statistical properties~(cfr.~section~\ref{sec:solarphi}). 

The planetary arrangement was significantly different in the early Solar System, and evolved to its current configuration through dissipative, collisional processes~\cite{2005Natur.435..466G,Tsiganis_2005,Morbidelli_2005}, and possibly external gravitational influences. We do not attempt to include such effects in our simulations.  Plausible models suggest the Solar System attained its current configuration close to its present-day set of orbits $3.8\,\mathrm{Gyr}$ ago at the latest~\cite{2001Icar..152..205C,2009Icar..203..644R,2005Natur.435..466G,2005Sci...309.1847S,2017OLEB...47..261Z}. Even a very different early Solar System would thus only impact particles emitted during the first $10\%$ of the Solar System's history. As we will see below, the vast majority of these particles would have been ejected by now, and so this effect is unlikely to have large effects on Earth-crossing density of the solar basin today.

A full simulation would ideally include all of the planets in the Solar System, and even smaller bodies such as asteroids and moons. Such a simulation would be prohibitively expensive. Even a simulation incorporating all of the inner planets would be significantly slower---the fast orbit of Mercury in particular requires smaller time steps to integrate accurately.  As we are most interested in the basin density at Earth, it is only necessary to include the Sun, Venus, Earth, Jupiter, and Saturn.  The basin density at Earth is mostly set by basin particles with barely Earth crossing semi-major axes ($0.5~\AU \lesssim a \lesssim 1~\AU$), as we will explore in~\ref{sec:sim_res}. The inner Solar System orbits of most interest to us are little affected by Neptune, Uranus, and the rest of the outer Solar System. We furthermore omit Mercury, Mars, and smaller objects, as the main effect of the planets in the inner Solar System is to cause diffusion in phase space.  This diffusion is by far most efficiently driven by Venus and Earth due to their higher mass and smaller semi-major axes (than that of Mars), cfr.~section~\ref{sec:stochastic}.

We treat the planets as point masses, since encounters with impact parameters smaller than their radii are exceptionally rare.  Most of the phase space redistribution occurs through many ``weak'' encounters (\textit{i.e.}, via small-angle gravitational scattering) rather than the few strongest encounters~(section~\ref{sec:close_encounters}). The finite size of the Sun, however, is important: the non-$1/r$ potential leads to significant orbital precession for particles with perihelia inside the Solar interior, which in turn averages out secular perturbations from the planets (section~\ref{sec:solarphi}).  This slows the evolution of the particle orbits' energy and angular momentum. To capture these effects, we model the Sun's gravitational potential as that coming from an enclosed mass function
\begin{equation}
M(R) =  c_1 M_\odot \left(2 - \left[2 + \frac{c_2R}{R_{\odot}} \left(2 + \frac{c_2R}{R_{\odot}}\right)\right] e^{-\frac{c_2R}{R_{\odot}}}\right),\label{eq:sun_mass_function}
\end{equation}
where $c_1\approx 0.501$, $c_2 \approx 10.285$, and $M_{\odot}$ the total mass of the Sun, for $R < R_\odot$ (in heliocentric coordinates), and $M(R) = M_\odot$ for $R \geq R_\odot$.  This expression is in agreement with recent models in the literature \cite{vinyoles2017new}.  

We will discuss in section~\ref{sec:solarphi} why we expect our statistical results to be unaffected by unmodeled effects such as gravitational corrections from general relativity and the non-sphericity of the Sun.

\subsubsection{Algorithm}
\label{sec:algorithm}

Since the Solar System is chaotic, with typical Lyapunov times of order $10^7 \,\mathrm{yr}$ for the test particles of interest~\cite{Laskar_1989, LASKAR_1990,Sussman_1992, Mogavero_2021, mogavero2023timescales}, one cannot aim to compute trajectories precisely over the full history of the Solar System. However, certain nearly-conserved quantities such as an orbit's energy can be stable on much longer timescales, so it is important to use an integration method that respects these conservation laws.

Particles emitted with very small aphelia are never perturbed onto Earth-crossing orbits (see sections~\ref{sec:int_strat} and~\ref{sec:sim_res}); the test particles of interest are thus emitted on (initially) very eccentric orbits.  An adaptive step size method, which can take smaller time steps during the very fast motion close to or through the Sun, is thus vital.\footnote{An alternative possibility would be to analytically transport particles through the region close to the Sun, but it would be difficult to avoid introducing boundary effects.} While symplectic integration methods are often used for orbit integration, implementing an adaptive time step mechanism that does not violate symplecticity is not trivial. Though such methods exist \cite{Preto_1999,Hairer_2005,Richardson_2011,duruisseaux_2021}, they still incur errors from machine-precision arithmetic. These shortcomings mean that non-symplectic higher-order integrators with adaptive stepsize can have competitive performance.

In fact, even in situations where one might have expected symplectic integrators to be well suited, such as integrating well-behaved, energy-conserving systems such as the outer Solar System, the very low errors achievable through adapative-stepsize high-order integrators can make them the method of choice. For example, IAS15, a 15th-order adaptive step size Gauss-Radau integrator, can achieve machine-precision-limited accuracy on such problems~\cite{2015MNRAS.446.1424R}.

Machine-precision-limited accuracy is obviously a desirable goal, if achievable, and we investigated the IAS15 integrator for our simulations. However, for the orbits of interest, the typical  speed achieved on our computer systems was $\sim 100 ({\rm sim\,yr}) / ({\rm cpu\,sec})$ for each particle, with some particles evolving significantly more slowly. It would have taken multiple years to simulate the $4.5\,\mathrm{Gyr}$ history of the Solar System without selection bias. We found the DPRKN12 integrator~\cite{dprkn} from the \texttt{DifferentialEquations.jl} package~\cite{diffeqjl}, a 12th-order explicit adaptive Runge-Kutta-Nystr\"om method, to be a good compromise between accuracy and speed.

\subsubsection{Integration strategies}
\label{sec:int_strat}

We adopt two independent, qualitatively different strategies to integrate the test particle trajectories. Hereafter, we refer to these separate strategies as ``forward'' and ``backward'' runs; they are described below.

\paragraph{Forward runs}
The most obvious way to simulate the orbital evolution after solar particle production is to initialize particles
inside the Sun---with positions, velocities, and injection times randomly chosen from the appropriate distributions---and then evolve them forward in time within the Solar System. Some particles will be ejected from the Solar System between the time of production and the present, but others will survive until today. For the latter, we keep track of their close encounters with Earth and thus estimate the Earth-crossing density.

Forward-simulated particles are injected from a random locus at radius $R = R_\odot$, \textit{i.e.}~from a uniform distribution on the surface of the Sun. At each injection point, the phase space is also taken to be uniform in 3D velocity space, corresponding to a production process whose matrix element is constant in the ``soft limit'' of vanishing velocity $\vect{v} \to 0$~\cite{VanTilburg:2020jvl}. For computational efficiency, we restrict the \emph{magnitude} of the injection velocity to be such that the initial semi-major axes of the produced test particles fall in the range of $0.4~\mathrm{AU} < a < 4~\mathrm{AU}$, while the \emph{angle} of the injection velocity is isotropic. Particles with lower injection speeds do not cross Venus' orbit and therefore are unlikely to ever reach Earth-crossing orbits (see below and section~\ref{sec:sim_res}).  Further, these orbits would require much smaller time steps to integrate faithfully.  Particles with injection speeds higher than this range are either unbound from the start or have such large semi-major axes that they are quickly ejected from the Solar System by Jupiter. 

The injection times are also uniformly distributed between the present day and $4.5\,\mathrm{Gyr}$ in the past.  The starting configuration of the four planets at each injection time is obtained by simulating the planetary orbits backward in time from the present day to the injection time (with the same DPRKN12 algorithm).

The forward runs were performed on 256 CPUs.  Each CPU simulated the mock Solar System with 5 active particles (the Sun and four planets) and \emph{one} test particle in the solar basin. Test particles were simulated until ejection from the Solar System (a status assigned if they attain a distance of 30~AU or more from the Sun) or until they reached the present day. The state of the simulation, namely the positions and velocities of the planets and test particles, was saved in time steps of $10^3$ years. 

Once a particle's simulation was finished, another particle was started using the same CPU\@. In total, 2540 simulations were started in this way, of which 2284 finished. All of the 256 initial simulations ran to completion. Since there is a significant bias in how long simulations with different initial conditions take to complete---for example, particles with smaller initial semi-major axes $a$ generally take longer to be ejected---this bias can be propagated to our results if we use all of the completed runs in our analysis.  We therefore make a conservative cut on our results and mostly analyze the ``original'' 256 forward runs throughout this paper to avoid any completion bias. 

\paragraph{Backward runs}
For roughly the other half of the simulations, we employed the ``opposite'' procedure. Particles were initialized close to Earth's present-day position, with random velocities (restricted to bound trajectories), and then evolved \emph{backward} in time. This protocol samples directly from particle trajectories that eventually hit Earth, as opposed to the forward runs, which sample from trajectories at particle production. At each Sun crossing, a particle in the backward runs \emph{could have been produced}, and it is automatically conditioned on crossing Earth, where it could have been detected.

For the backward simulations, particles were injected at random locations on a shell $0.002\,\mathrm{AU} \approx 47 \, R_\oplus$ from the center of Earth (about the Earth-Moon separation), with velocities chosen at random within the velocity-space ball of bound trajectories. The initial displacement from Earth's surface at $R_\oplus$ was chosen to avoid potential numerical errors from initialization near a point particle (Earth in the simulation), but is sufficiently small that significant basin density differences are not expected. The injection time was chosen uniformly at random within the past 100 years to cover a range of Earth positions relative to the other planets. A particle's simulation ran until it was either ejected from the Solar System, or $4.5 \,\mathrm{Gyr}$ elapsed.

As in the forward runs, once a particle's simulation was finished, another backward simulation for a different particle was started on the same CPU\@. In total, 5609 simulations were started, of which 5382 finished. All of the 256 initial simulations ran to completion.  Arranged in order of starting time, the first 3879 of the 5609 total simulations completed, with the remainder of finished simulations distributed across later start times.  We will primarily use the 256 initial simulations as an unbiased sample. Where a larger sample is desirable, we will use the first 2048 simulations started---the largest power-of-two sample size available. The latter procedure still technically involves some selection bias.  However, since initial conditions with longer completion times generally correspond to longer ejection times, this bias is conservative in that it will \emph{underestimate} the basin energy density. Furthermore, the empirical probability for simulation number 2049 through 3879 to \emph{not} have run to completion must be less than $10^{-3}$ (otherwise at least one would not have finished), so the selection bias in our sample of the first 2048 runs is exceedingly small.

\paragraph{Relative merits of forward and backward integration}
Aside from our two integration strategies serving as a useful cross-check, the forward and backward runs complement each other with regards to several observables of interest. 

Overall, the backward runs are more versatile and provide multiple benefits. 
Firstly, since Earth's geometric cross-section is far smaller than that of the Sun, the backward runs are naively more efficient at sampling Sun-crossing trajectories than the forward runs are at sampling Earth-crossing trajectories (per unit simulation time), thus yielding sharper estimates for the effective basin lifetime (cfr.~section~\ref{sec:sim_res}).\footnote{For the forward runs, we actually consider intersections with a larger shell around Earth, rather than strictly Earth-crossing trajectories, to mitigate precisely this issue.}
Secondly, while definite choices about particle production (production location, energy) must be made for the forward runs, the backward runs can all be analyzed differently for varying production assumptions. 
Finally, following the previous point, the backward runs also enable us to analyze the Earth-crossing flux from other sources, such as gravitationally captured DM particles, or a primordial abundance of DM in the Solar System.

On the other hand, the forward runs naturally provide more data points for particle encounters with Earth since each particle may have many encounters over the period under consideration.  This potentially provides insight into more fine-grained quantities, such as short-term temporal modulation of the basin density and late-time distribution in phase space. 

\paragraph{Particle Injections}
As discussed above, we restrict the magnitude of the injection velocity of test particles such that the initial semi-major axes of test particles falls between $0.4 \, \mathrm{AU} < a < 4 \, \mathrm{AU}$.  We confirmed that the lowest $a$ value attained is $0.35 \AU$ in 256 forward and 2048 backward simulations. This distance is set by close encounters with Venus: the smallest distance between Venus and the Sun is $0.72 \AU$, so if a particle is knocked onto a high-eccentricity orbit via a close encounter with Venus, it has $a \gtrsim 0.36 \AU$, and secular evolution does not change $a$ until at least third order in perturbation theory.  If a particle is injected with $a \le 0.35 \AU$, we would not expect it to become Earth crossing, so there is no need to inject particles at smaller $a$ for our purposes.
We have checked that estimates of the effective basin lifetime are not significantly affected by this decision, by dropping all of the injected particles with $a \in [0.4,0.45]\, \mathrm{AU}$.  We similarly do not expect that our conclusions would differ significantly if has we injected down to $a = 0.35 \AU$.

\subsection{Results}
\label{sec:sim_res}

\begin{figure}[t]
\begin{center}
\includegraphics[width=\textwidth]{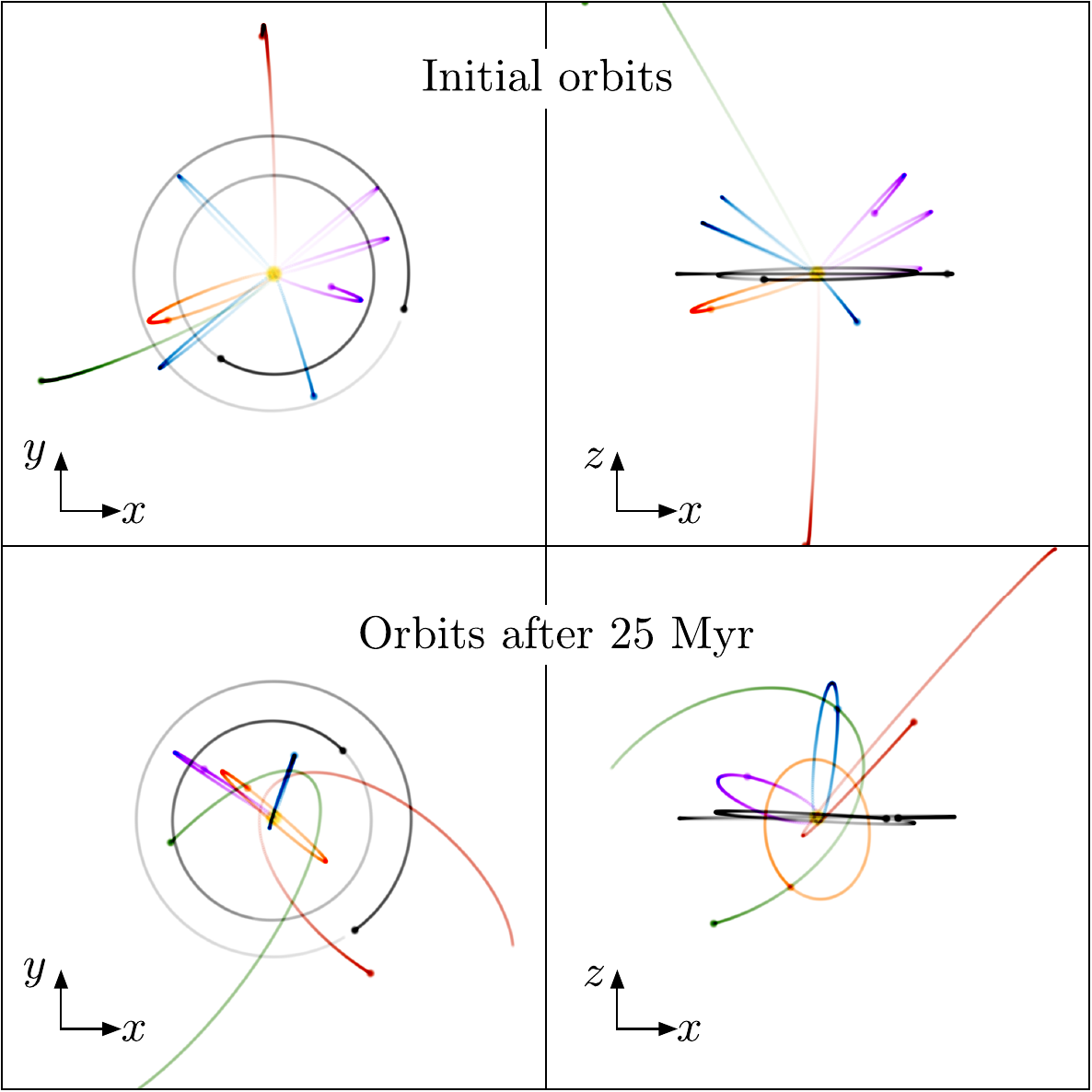}
\caption{Diagrams illustrating the orbits of solar basin particles shortly after production inside the Sun (upper row), and 25 million years later (bottom row). The yellow dot corresponds to the Sun, the colored trails correspond to one year of solar basin particle orbits, and the black trails correspond to the orbits of Venus and Earth. While the initial orbits of the solar basin particles are all Sun-crossing, gravitational perturbations from the planets drive them onto different orbits, as illustrated in the bottom row. In the left (right) column, the line of sight is oriented along the $z$-axis ($y$-axis) so that the ecliptic is face-on (edge-on).}
\label{fig_orbits1}
\end{center}
\end{figure}

Figure~\ref{fig_orbits1} shows examples of the evolution of particle orbits during this process. At first, particles are emitted on Sun-crossing orbits. As illustrated in the top row, their orbits will initially precess rapidly due to the non-$1/r$ potential inside the Sun (section~\ref{sec:solarphi}). Over time, gravitational perturbations from the planets modify the orbits; the bottom row of figure~\ref{fig_orbits1} shows the particle orbits after 25 Myr of evolution, showing that they are no longer Sun-crossing, and have significantly changed most of their orbital elements. However, the particle energy, corresponding to the orbit's semi-major axis $a$, is mostly constant over these timescales (the largest change in semi-major axis in figure~\ref{fig_orbits1} is $6\%$).

\begin{figure}[t]
\begin{center}
\includegraphics[width=\textwidth]{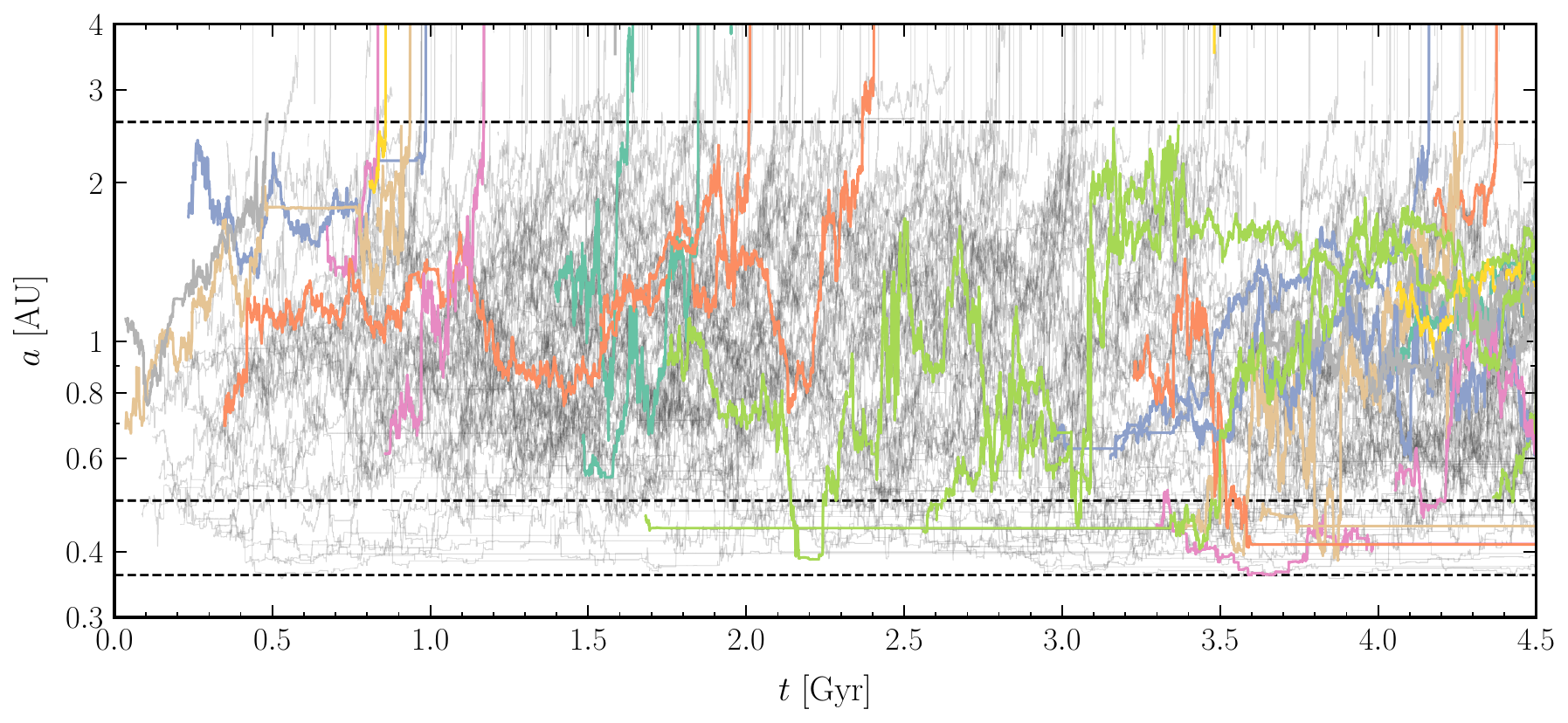}
\caption{Evolution of the semi-major axes $a$ for 256 test particles under the gravitational influence of the Sun, 
Venus, Earth, Jupiter and Saturn. The particle trajectories are evolved forward in time, where $t=\SI{0}{Gyr}$ corresponds to the birth of the Solar System and $t=\SI{4.5}{Gyr}$ is the present day.  Particles are initialized at uniformly random times, and at starting locations on the Sun's surface, with initial velocities corresponding to semi-major axes between $0.4\,\mathrm{AU}$ and $4\,\mathrm{AU}$. 188 of these particles are ejected from the Solar System during this evolution, but 68 survive until the present day. A representative set of 32 particle trajectories is shown in color so that their evolution is easier to see. Close encounters with Jupiter, Earth, and Venus are possible for particles with semimajor axes above half the planetary orbital radii, which are shown as black dashed lines at $0.34$, $0.5$, and $\SI{2.6}{AU}$. For $a > \SI{2.6}{AU}$, Jupiter ejects particles on timescales of a few $\SI{}{Myr}$. \githubicon{https://github.com/kenvantilburg/solar-basin-dynamics/blob/main/code/Numerical_integration/trajectories.ipynb}}
\label{fig_asun}
\end{center}
\end{figure}

As mentioned above, we find that Jupiter-crossing particles are usually ejected from the Solar System on relatively short timescales, and that this is by far the most efficient means of ejection. In addition, once $a$ is large enough, the other orbital elements generally change on fast enough timescales that the particle becomes Jupiter-crossing and is ejected relatively quickly. The lifetime of test particles within the Solar System is therefore set by the time taken for their semi-major axes to become Jupiter-crossing $(a > R_J / 2)$. This behavior is evident in our forward simulations and is illustrated in figure~\ref{fig_asun}, which show the secular evolution of the test particles' semi-major axes over time, for the 256 initial runs. We can see that $a$ evolves mostly in a quasi-random-walk fashion over timescales of hundreds of Myr, with the evolution rate significantly increasing for larger $a$. This energy-changing behavior will motivate the stochastic description of section~\ref{sec:stochastic}.

\begin{figure}[t]
\begin{center}
\includegraphics[width=0.49\textwidth]{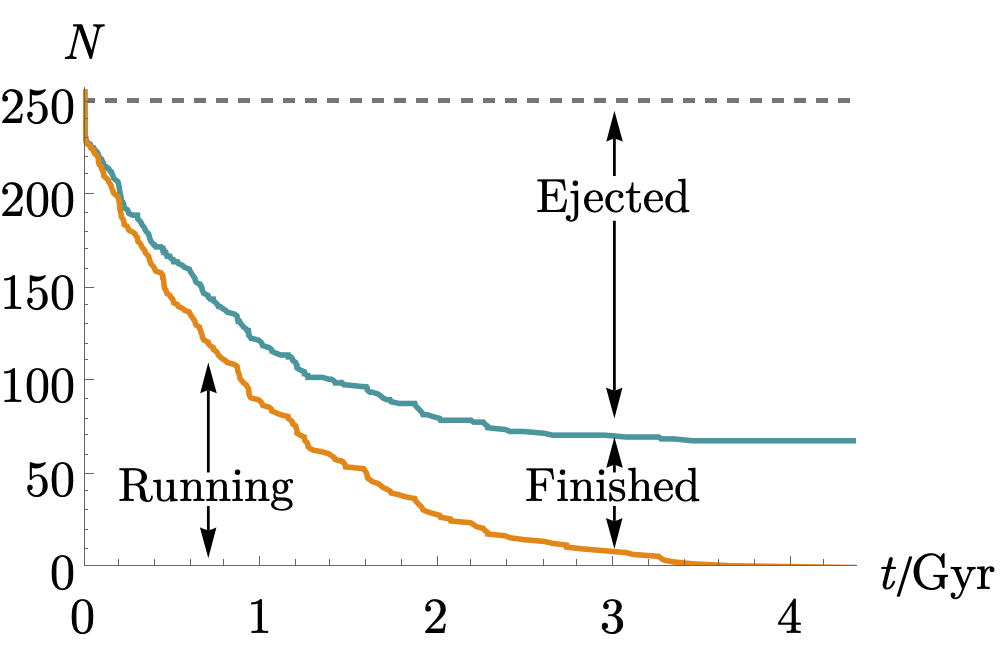}
\includegraphics[width=0.49\textwidth]{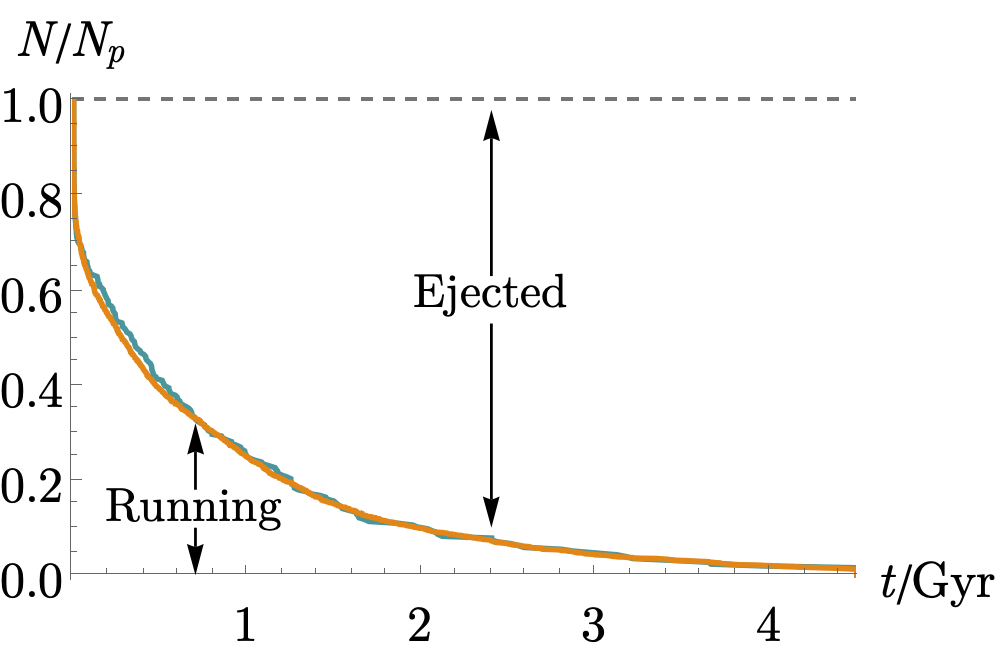}
\caption{\emph{Left panel:} plot of particle lifetimes for the forward simulations described in section~\ref{sec:sim_res}. The horizontal axis shows to the time elapsed since the start of a given particle's simulation. The teal curve corresponds to the number of particles that have not been ejected prior to that point, while the orange curve depicts the number of simulations still running up to the given time. The gap between the 
curves corresponds to the number of simulations which finished (\textit{i.e.}~ran up to the present) without undergoing an ejection event up to the given time.
\emph{Right panel:} plot of particle lifetimes for the backward simulations described in section~\ref{sec:sim_res}. The teal curve corresponds to the first $N_p = 256$ simulations, and the orange curve to the first $N_p = 2048$ simulations. Both curves initially fall very steeply, due to the fast ejection of Jupiter-crossing particles, then fall with a decay rate of order ${\rm Gyr}^{-1}$. \githubicon{https://github.com/kenvantilburg/solar-basin-dynamics/blob/main/code/PaperPlots.nb}}
\label{fig_eej1}
\end{center}
\end{figure}

Most (188 out of 256) of the particles in the forward simulations are ejected during their evolution.  This is depicted in the left panel of figure~\ref{fig_eej1}.  The teal curve illustrates the number of particles that have not been ejected by a given simulation time; the gap between this curve and $N_p=256$ shows the number of particles that have been ejected and the distribution of ejection times.  68 particles survived until the present day and could contribute to the Earth-crossing density. As we discuss in section~\ref{sec:dens_est}, we can use the trajectories of these particles to estimate the Earth-crossing density today.

\begin{figure}[t]
\begin{center}
\includegraphics[width=\textwidth]{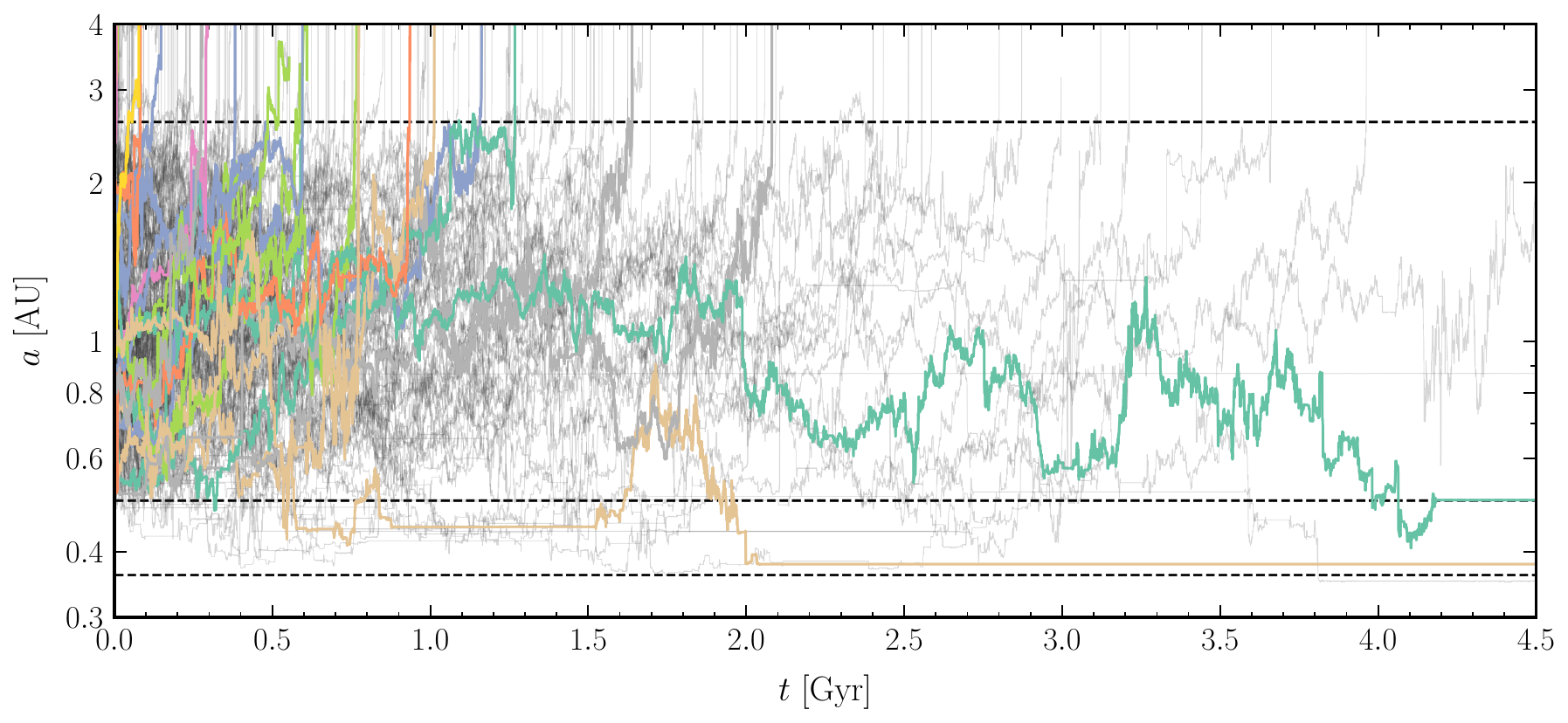}
\caption{Evolution of the semi-major axes $a$ for 256 test particles under the gravitational influence of the Sun, Venus, Earth, Jupiter and Saturn. The particles are evolved backward in time (with $t$ the look-back time and $t=0\,\mathrm{Gyr}$ the present day), starting from locations close to Earth's position, with random sub-escape velocities. All but 5 of these particles are ejected from the Solar System within $4.5\,\mathrm{Gyr}$ during this backward evolution, corresponding (in forward evolution) to having been captured from unbound halo particles. 209 of them pass through the Sun during the integration time, corresponding to trajectories that particles could have taken to reach Earth after emission from the Sun (in forward evolution).  A representative set of 32 particle trajectories is shown in color so that their evolution is easier to see.  Close encounters with Jupiter, Earth, and Venus are possible for particles with semimajor axes above half the planetary orbital radii, which are shown as black dashed lines at $0.34$, $0.5$, and $\SI{2.6}{AU}$. \githubicon{https://github.com/kenvantilburg/solar-basin-dynamics/blob/main/code/Numerical_integration/trajectories.ipynb}}
    \label{fig_aearth}
	\end{center}
\end{figure}

Figure~\ref{fig_aearth} shows the secular evolution of the test particles' semi-major axes over time for the 256 initial runs of the \emph{backward} simulations. All but 5 of these particles are ejected from the Solar System during the $4.5\,\mathrm{Gyr}$ simulation time.  The distribution of ejection times is plotted as the teal curve in the right panel of figure~\ref{fig_eej1}. Initially Jupiter-crossing orbits are ejected on relatively short timescales (in agreement with the analytical estimates of section~\ref{sec:ejections}), while essentially all later ejections are due to the semi-major axis increasing due to secular and inner-Solar-System perturbations, a much slower process, until the particle is Jupiter-crossing. For the first 2048 runs, all but 26 of the particles were ejected, with the distribution of ejection times plotted as the orange curve in the right-hand panel of figure~\ref{fig_eej1}.

\subsubsection{Earth-crossing density estimates}
\label{sec:dens_est}

\paragraph{Forward runs} 
For the forward simulations, we simulated few enough particles that it is very unlikely for any particles to be Earth-crossing precisely at the present day (Earth's interior takes up only a $10^{-13}$ fraction of the volume within $1\,\mathrm{AU}$ of the Sun).
However, if we assume that the statistics of Earth crossings are similar over some timescale, we can track all of the Earth crossings that occur during the final $t_f$ period of the simulation.  Similarly, if we assume the precise size of Earth is not very important, as long as it is small enough compared to Solar System scales, then we can look at particles which cross some larger sphere centered on Earth's position.  

Making both of these approximations enables us to obtain sufficient statistics to estimate the present-day Earth-crossing density precisely. We track the motion of a particle $p$ during the final $t_f$ period of the simulation, and find the time $\hat{t}_{\rm{ball},p}$ during which $p$ is within a distance $r_{\rm{ball}}$ of Earth. By relating the number of simulated particles to the number that would have been emitted in a particular particles physics model, we can estimate the Earth-crossing density for that model.

A useful, model-independent way to express our density estimates is in terms of the ``effective basin time'' $\tau_{\rm eff}$ defined in eq.~\eqref{eq:tau_eff}~\cite{VanTilburg:2020jvl,Lasenby:2020goo}. This is the constant of proportionality between the present-day density at Earth, and the rate of change of this density assuming that particles remain on their initial trajectories (without re-absorption). 
If each particle emitted from the Sun remained on its initial trajectory, then the average time spent per particle in a ball of radius $r_{\rm{ball}}$, at distance $R$ from the Sun, is 
\begin{equation}
\bar t_{\rm ball}(R) \simeq \frac{1}{2} \frac{r_\mathrm{ball}^3}{R^4}
\frac{t_f}{1/a_{\rm min} - 1/a_{\rm max}},
\label{eq_tball}
\end{equation}
where $a_{\rm min}$ and $a_{\rm max}$ are the minimum and maximum semi-major axes of the emitted particles that we consider, assuming $a_{\rm min} < R/2$ and $a_{\rm max} \gg R/2$. This expression is derived in appendix~\ref{app:density_estimators}.  We can estimate $\tau_{\rm eff}$ via
\begin{equation}
\hat \tau_{\rm eff}(R) \simeq t_\odot \frac{\sum_p \hat{t}_{\rm{ball},p}}{N_p \bar t_{\rm ball}}. \label{eq_tau_eff_forward}
\end{equation}
where $t_\odot = 4.5 \,\mathrm{Gyr}$ is the lifetime of the Solar System, $N_p$ is the number of simulated particles, and $\sum_p$ represents the sum over the simulated particles. To perform this estimate with the simulations, we saved the state of the solver at regular intervals of $10^3\,\mathrm{yr}$. We then re-ran the simulations at higher resolution starting from these checkpoints, noting each time the test particle passed within $r_{\rm{ball}}$ of Earth.  This gives an estimate of $\tau_{\rm eff} = 1.20 \,\mathrm{Gyr}$ at $R = 1\,\mathrm{AU}$, based on this averaging procedure over the final $t_f = 10^7 \yr$ for the first 256 forward runs.

To estimate the statistical uncertainty in our estimate, we use a bootstrap method. Given our $N_p = 256$ particles, each individual orbit $p$ gives us an (imprecise) estimate $\hat{\tau}_{\mathrm{eff},p} = t_\odot (\hat{t}_{\mathrm{ball},p} / \bar{t}_\mathrm{ball})$.
We then create a single bootstrap sample $s$ by choosing $N_p$ random numbers $s_j$ uniformly and independently from $p = 1,\dots,N_p$ (sampling with replacement), so that $\hat{\tau}_{\mathrm{eff},s} = \frac{1}{N_p} \sum_{j=1}^{N_p} \hat{\tau}_{\mathrm{eff},s_j}$.
Drawing many such samples gives us an empirical estimate for the probability distribution function corresponding to choosing $N_p$ random particles, from the uniform probability distribution on the actually-computed particles. The quantity we are actually interested in, of course, is the probability distribution corresponding to choosing $N_p$ random particles from the full initial distribution, rather than our $N_p$ samples from it. However, the theory of bootstrap estimators \cite{Efron_1979} shows that, in the limit of large $N_p$, the bootstrap-estimated PDF becomes a good approximation to the true one.

\begin{figure}[t]
\begin{center}
\includegraphics[width=0.5\textwidth]{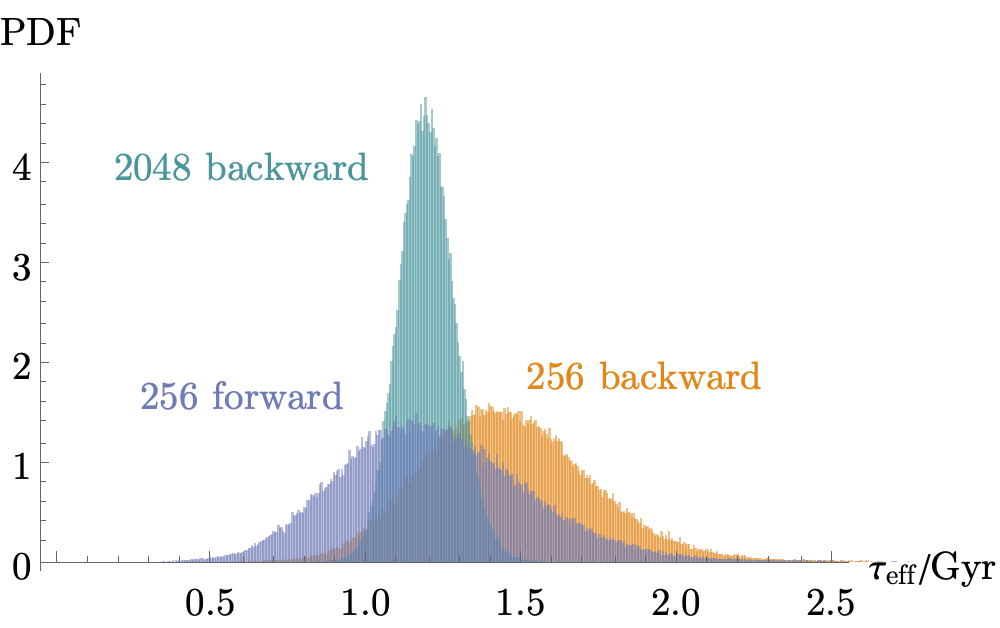}
\includegraphics[width=0.23\textwidth]{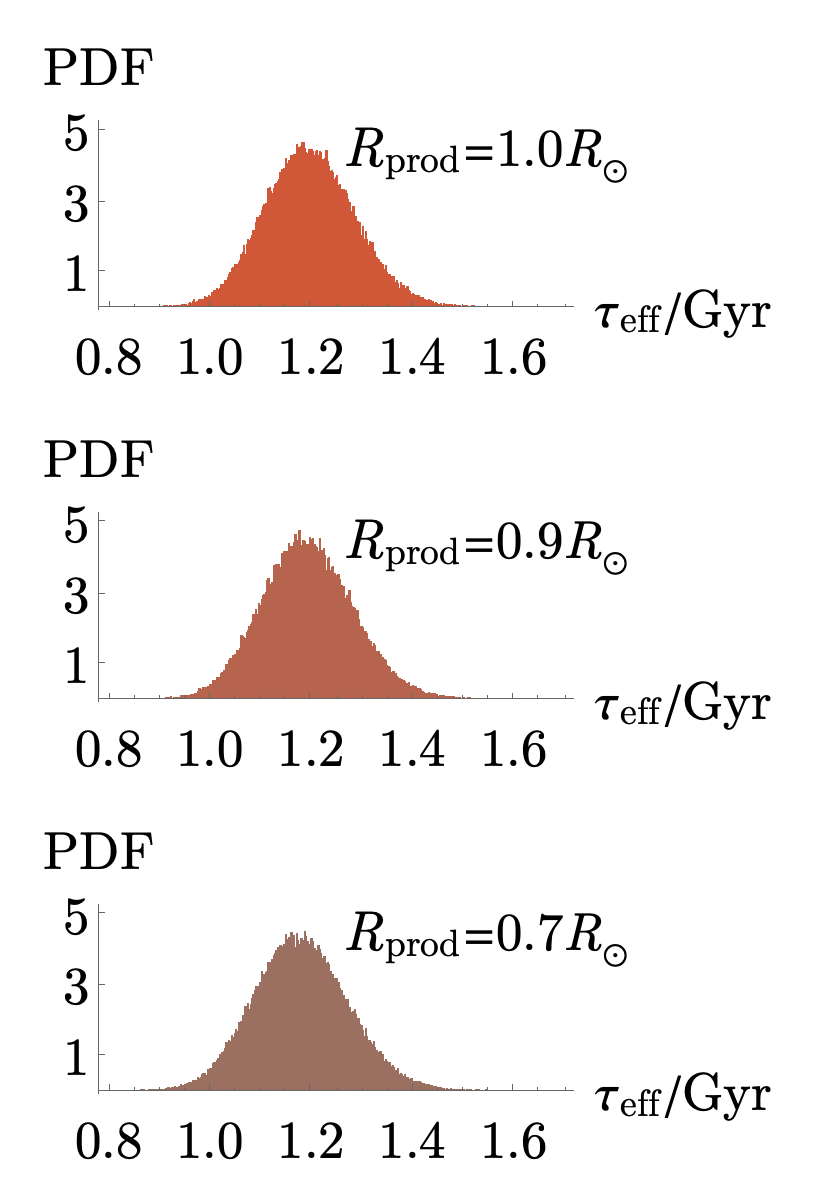}
\includegraphics[width=0.23\textwidth]{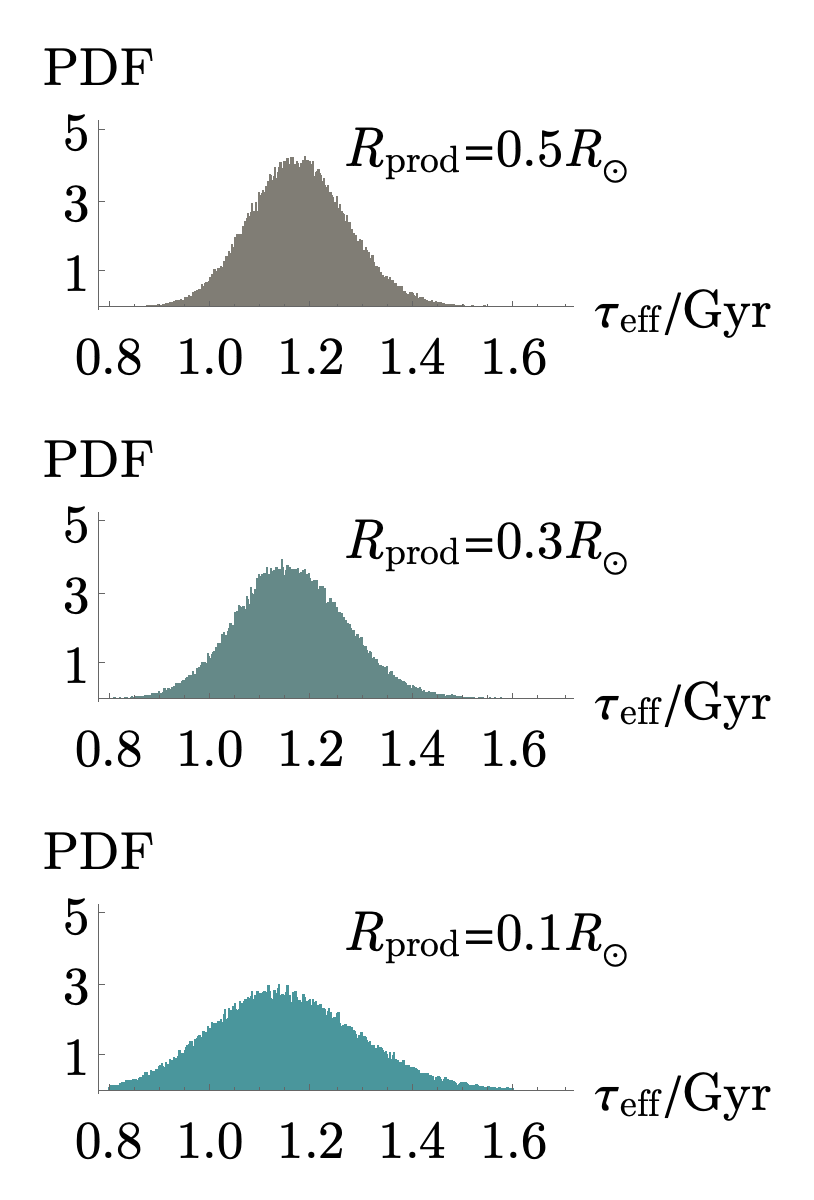}
\caption{Bootstrap-estimated probability density functions for the effective solar basin lifetime $\tau_{\rm eff}$, for particles emitted from close to the Solar surface. The purple histogram corresponds to the estimate from the last $10^7\,\mathrm{yr}$ of the 256 forward simulations. The orange histogram corresponds to the estimate from the first 256 backward simulations, while the teal histogram corresponds to the first 2048 backward simulations. The latter gives the most precise estimate, with $\tau_{\rm eff} = 1.20\pm 0.09 \,\mathrm{Gyr}$. The right-hand side illustrates the PDF for the 2048 backward simulations for different choices of emission radius, illustrating that the result does not depend strongly on this choice. \githubicon{https://github.com/kenvantilburg/solar-basin-dynamics/blob/main/code/PaperPlots.nb}
}
\label{fig_pdf1}
\end{center}
\end{figure}

Figure~\ref{fig_pdf1} shows the bootstrap-estimated probability distribution for the Earth-crossing density, where we look at intersections with a sphere of radius $r_\mathrm{ball} = 0.01 \, \mathrm{AU}$ around Earth during the final $t_f = 10^7 \, \mathrm{yr}$ of the simulation.  We can use this data to update our result for the effective lifetime of the basin with an uncertainty, finding $\tau_{\rm{eff}}=1.20^{+0.31}_{-0.27} \SI{}{Gyr}$ at $68\%$ CL.
As we discuss below, these estimates are compatible with those from the backward runs.

\paragraph{Backward runs}
For the backward simulations, a large fraction---209/256 and 1681/2048 for our fiducial samples---of the injected particles passed through the Sun at some point during their evolution. If we reverse the arrow of time again (to the ``correct'' direction), these correspond to trajectories that particles could have taken to reach Earth subsequent to emission from the Sun. The number of Sun crossings for a given backward trajectory allows an estimate of the present-day Earth-crossing phase space density for that trajectory.

The phase space density $f$ in a transiting wavepacket in a given location of the Sun evolves as~\cite{Lasenby:2020goo}
\begin{align}
\dot f = \Gamma_{\rm prod} \left[1 + \left(1 - e^{m/T}\right) f\right],
\label{eq:fdot}
\end{align}
where $m$ is the particle mass, $T$ is the Solar temperature, and $\Gamma_{\rm prod}$ is the in-medium particle production rate.
For a given theory, the proper procedure would be to integrate over the particle's trajectory within the Sun, evolving $f$ given the properties of the medium that it passes through. Here, we make the approximation that particle emission and absorption only occur at discrete radii inside the Sun. For some models with resonant production (e.g.~dark photons, millicharged particles), this can be good approximation by itself---for others, we can sum over the appropriate weighted combination of different radii.

The change in a wavepacket's phase space density during a transit of a thin shell at radius $R_\mathrm{prod} < R_\odot$, ignoring re-absorption (see section~\ref{sec:saturation}), is 
\begin{equation}
\delta f \simeq \int \dd t \, \Gamma_{\rm prod}[R(t)]
= \int \frac{\dd R}{v_R} \, \Gamma_{\rm prod}(R)
\simeq \frac{1}{v_{R_\mathrm{prod}}} \int_\mathrm{shell} \dd R \, \Gamma_\mathrm{prod}(R)
\equiv \frac{\Gamma_\mathrm{prod} \Delta R_\mathrm{prod}}{v_{R_\mathrm{prod}}}
\label{eq:delta_f}
\end{equation}
where $v_{R_\mathrm{prod}}$ is the radial velocity of the particle at that radius from the Sun's center, which depends on $e$ and (weakly) on $a$. The shell's thickness is assumed so small that  $v_R \simeq v_{R_\mathrm{prod}}$ can be treated as constant across the shell. We derive in appendix~\ref{app:density_estimators} that for production dominated by a \emph{single} shell, the effective basin time can be estimated as
\begin{align}
	\hat\tau_{\rm eff}(R) = \frac{8}{9} R \left(\frac{R}{R_\mathrm{prod}}\right)^2 
	\frac{v_{\mathrm{esc},R}}{v_{\mathrm{esc},R_\mathrm{prod}}}
	\left(\frac{1}{N_p}\sum_p \sum^{(p)}_i \frac{1}{v_{R_\mathrm{prod}}}\right), \label{eq_tau_eff_backward}
\end{align}
where we are mostly interested in $R = 1\,\mathrm{AU}$. As before, $\sum_p$ represents the sum over the $N_p$ simulated particles, and $\sum^{(p)}_i$ is over passages of the particle $p$ through the shell of radius $R_{\rm{prod}}$, occurring with radial velocity $v_{R_{\rm{prod}}}$.  The prefactor includes $v_{\mathrm{esc},R}$ and $v_{\mathrm{esc},R_\mathrm{prod}}$, the escape velocities at $R$ and $R_\mathrm{prod}$, respectively.  

As illustrated in the right-hand panel of figure~\ref{fig_pdf1}, the estimates of the probability density functions for $\tau_{\rm{eff}}$ for shells of different radii are similar. We conclude that $\tau_\mathrm{eff}$ is essentially the same for different particles physics models: for example, axion-like particles produced in the solar core and low-mass dark photons resonantly produced in a shell near the solar surface will both have the same effective basin lifetime, within our statistical uncertainties.

As above, we can estimate the uncertainty in our estimates via a bootstrap procedure. The bootstrap-estimated probability distributions from the first 256 particles and the first 2048 particles are shown in the left panel of  figure~\ref{fig_pdf1}. These are in good agreement with the estimate from the forward runs; at $68\%$ CL, the first 256 particles give $ \tau_{\rm eff} = 1.44^{+0.27}_{-0.25}\,\mathrm{Gyr}$, while the first 2048 particles give $\tau_{\rm eff} = 1.20\pm 0.09 \,\mathrm{Gyr}$.  The different forward and backward methods of calculating the present-day density at Earth act as somewhat independent checks on each other.

\begin{figure}[t]
\begin{center}
\includegraphics[width=0.4\textwidth]{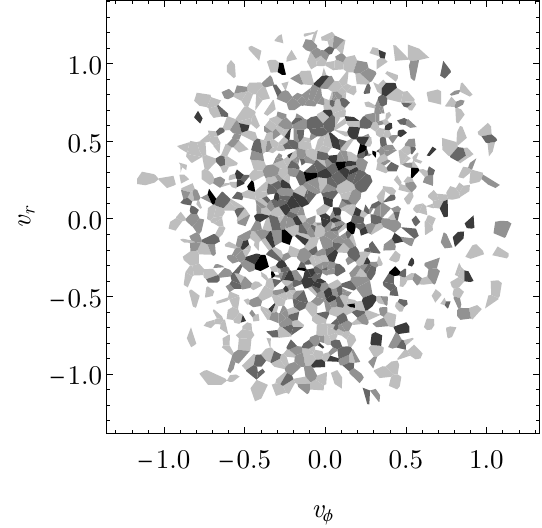}
\includegraphics[width=0.4\textwidth]{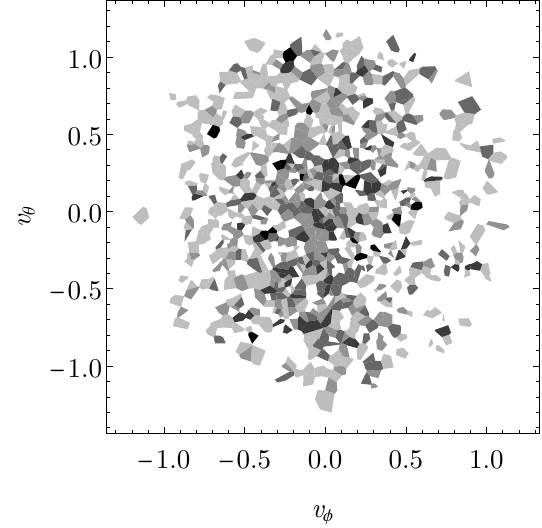}
\includegraphics[width=0.4\textwidth]{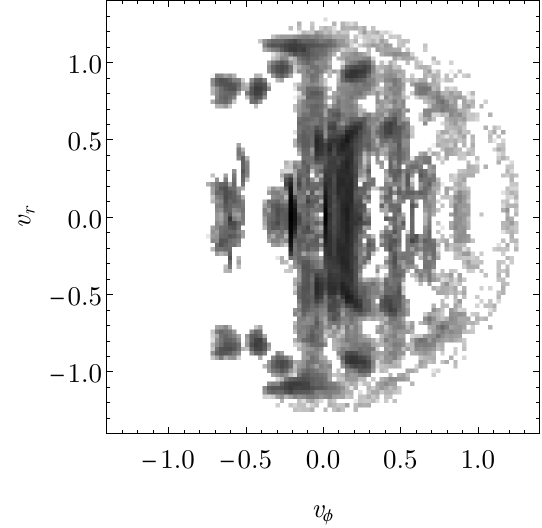}
\includegraphics[width=0.4\textwidth]{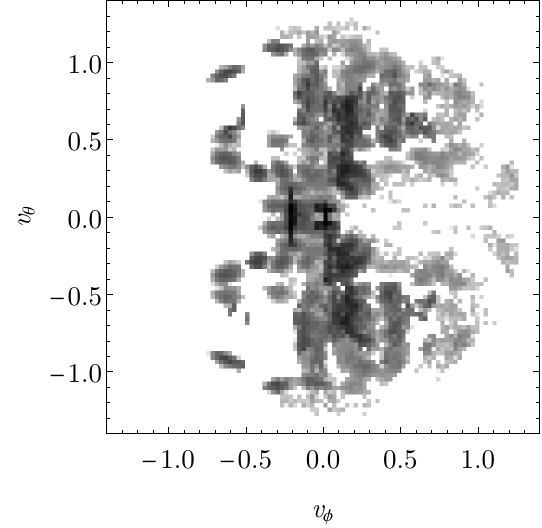}
\caption{\emph{Top row:} Voronoi plots of the estimated velocity phase at Earth from particles emitted at the Solar surface, using the first 2048 backward simulations described in section~\ref{sec:int_strat}. These are plotted in spherical coordinates with the ecliptic as the equatorial plane, normalized to $v_0 = 2\pi \AU / \yr$, so that Earth's velocity is approximately $(v_r,v_\theta,v_\phi) = (0,0,1)$. The gray level of each cell is set by the (logarithm of the) estimated density at Earth for that particle trajectory, with darker cells corresponding to higher densities. 
\emph{Bottom row:} As above, but showing the estimated phase space density from the first 256 forward simulations, obtained by accumulating phase space density whenever a particle passes within $0.05 \AU$ of Earth. The different velocities obtained from the same particle at different times are clearly highly correlated (leading to apparent artificial clustering), but the basic features of a higher density at small $v_\phi$, and very little density at large and negative $v_\phi$, are common between the forward and backward results. \githubicon{https://github.com/kenvantilburg/solar-basin-dynamics/blob/main/code/PaperPlots.nb}
}
\label{fig_veldist1}
\end{center}
\end{figure}

\subsubsection{Saturation density}
\label{sec:saturation}

If the production rate in the Sun is large enough, then re-absorption of basin particles can also become important, cfr.~the terms proportional to $f$ in equation~\ref{eq:fdot}. Assuming that particles are emitted from Solar material at temperature $T$, the phase space density in the Sun, and consequently elsewhere in the Solar System, is at most the bosonic thermal occupation number $f_T = (e^{m/T}-1)^{-1}$. Equilibrium is achieved at $\dot{f} = 0$, indicating a detailed balance between the Sun and its solar basin. If particles emitted from the Sun eventually access all of the Earth-crossing velocity phase space, this would lead to a maximum ``saturation'' density at Earth of $\rho_{\rm sat,full} = \frac{4}{3} \pi m^4 v_{\mathrm{esc},R}^3 f_T$.

However, since the velocity distribution of Earth-crossing particles emitted from the Sun is not uniform, significant re-absorption starts occurring at densities well below $\rho_{\rm sat,full}$. If all particles remained on their initial, highly eccentric orbits, then depending on the emission radius inside the Sun, only $\OO(10^{-3})$ to $\OO(10^{-2})$ of the velocity phase space volume at Earth would be occupied, reducing the saturation density by the corresponding amount~\cite{Lasenby:2020goo}.

\begin{figure}[t]
\begin{center}
\includegraphics[width=0.7\textwidth]{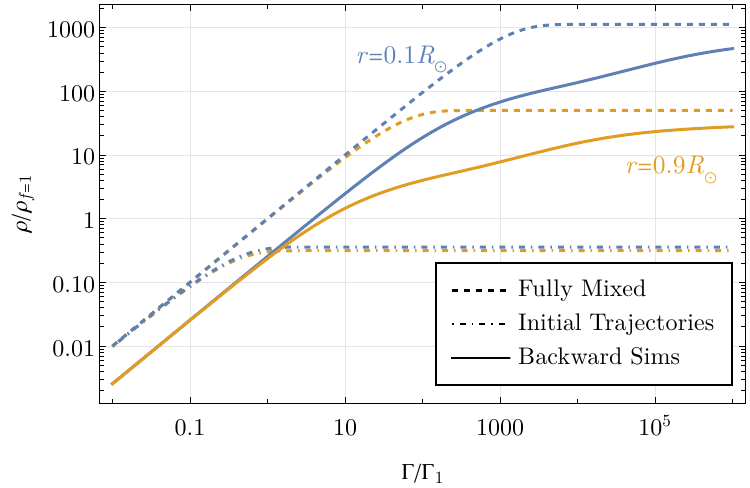}
\caption{Basin energy density $\rho$ at Earth for production rates $\Gamma$ high enough that Solar re-absorption is important. The density is normalized by $\rho_{f=1} = \frac{4}{3} \pi m^4 v_{\mathrm{esc},R}^3$ (occupation number $f$ equal to unity over the full phase space at Earth), and the production rate by $\Gamma_1$ need to achieve that density.
Orange (teal) curves correspond to emission from a shell of radius $R_\mathrm{prod} = 0.9 \, R_\odot$ ($R_\mathrm{prod} = 0.1 \, R_\odot$), where the solar temperature is $T \approx 50 \,\mathrm{eV}$ ($T \approx 1.1 \,\mathrm{keV}$), for a solar basin particle of mass $m = 1\,\mathrm{eV}$.
While the saturation occupation number---and thus the saturation density for a fully mixed phase space---is higher by a factor $T(0.1 \, R_\odot)/T(0.9 \, R_\odot) \approx 22$ for the emission from the higher-temperature shell, the smaller size of this shell means that the initial orbits occupy a smaller phase space volume. These competing effects mean that their saturation densities \emph{without} phase space mixing are comparable. The solid curves correspond to density estimates for the phase space distribution arising from particle evolution in the Solar System, derived from the first 2048 backward simulations and applying the correction factor in eq.~\ref{eq:f_abs}. Dashed (dot-dashed) curves show the density for maximally (minimally) efficient phase space mixing, assuming the orbital energy is conserved and no ejections occur. \githubicon{https://github.com/kenvantilburg/solar-basin-dynamics/blob/main/code/PaperPlots.nb}
}
\label{fig_sat1}
\end{center}
\end{figure}

Our simulations show that the velocity distribution at Earth sits between these two extremes after $t_\odot$, as we depict in figure~\ref{fig_veldist1}. The perturbed solar basin explores much more of the phase space than the initial trajectories immediately after production, but is far from fully mixed.
If a backward-simulated particle accumulates a phase space density $f_0$ over its evolution ignoring re-absorption, then the re-absorption-corrected phase space density is
\begin{equation}
f = f_T \left(1 - e^{-f_0/f_T}\right). \label{eq:f_abs}
\end{equation}
Figure~\ref{fig_sat1} illustrates these effects for emission from shells at different production radii inside the Sun. The dashed curves correspond to the density at Earth that would arise from a fully-mixed phase space (in orbital elements other than $a$) with no ejection, while the dot-dashed curves correspond to particles remaining on their initial trajectories. At small enough emission rates, our data-driven estimates correspond to smaller densities, due to semi-major axis $a$ evolution (including ejections) which reduces the effective particle lifetime to $\tau_{\rm eff} \approx 1.2 \,\mathrm{Gyr} < 4.5 \,\mathrm{Gyr}$.
For higher emission rates, the phase space evolution means that the density can exceed the saturation density of the phase space volume of the initial trajectories; eventually tending to the fully-mixed saturation density for extremely high emission rates. 

In appendix~\ref{app:sat_density}, we outline precisely our algorithm for determining the basin density as a function of mass and production rate, based on the time spent at different solar radii by backward-simulated particles. The method presented there was used to calculate the dark photon and axion densities of figure~\ref{fig_rhoDP}, including re-absorption/saturation effects.

\subsubsection{Annual modulation}
\label{sec:ann_mod}

Even if planetary perturbations were unimportant, the eccentricity of Earth's orbit means that  the solar basin density at Earth would change over the course of the year. For example, the initial population of solar basin orbits gives rise to a $\propto 1/R^4$ density distribution, where $R$ is the distance from the Sun~\cite{VanTilburg:2020jvl,Lasenby:2020goo}. Since the current orbital eccentricity of Earth is $e_\oplus \simeq 0.017$, this would result in about a $14\%$ density variation over an orbit, \textit{i.e.}~fractional amplitude of $6.8\%$ with a one-year period and largest at perihelion ($M=0$).

\begin{figure}[t]
\begin{center}
\includegraphics[width=0.47\textwidth]{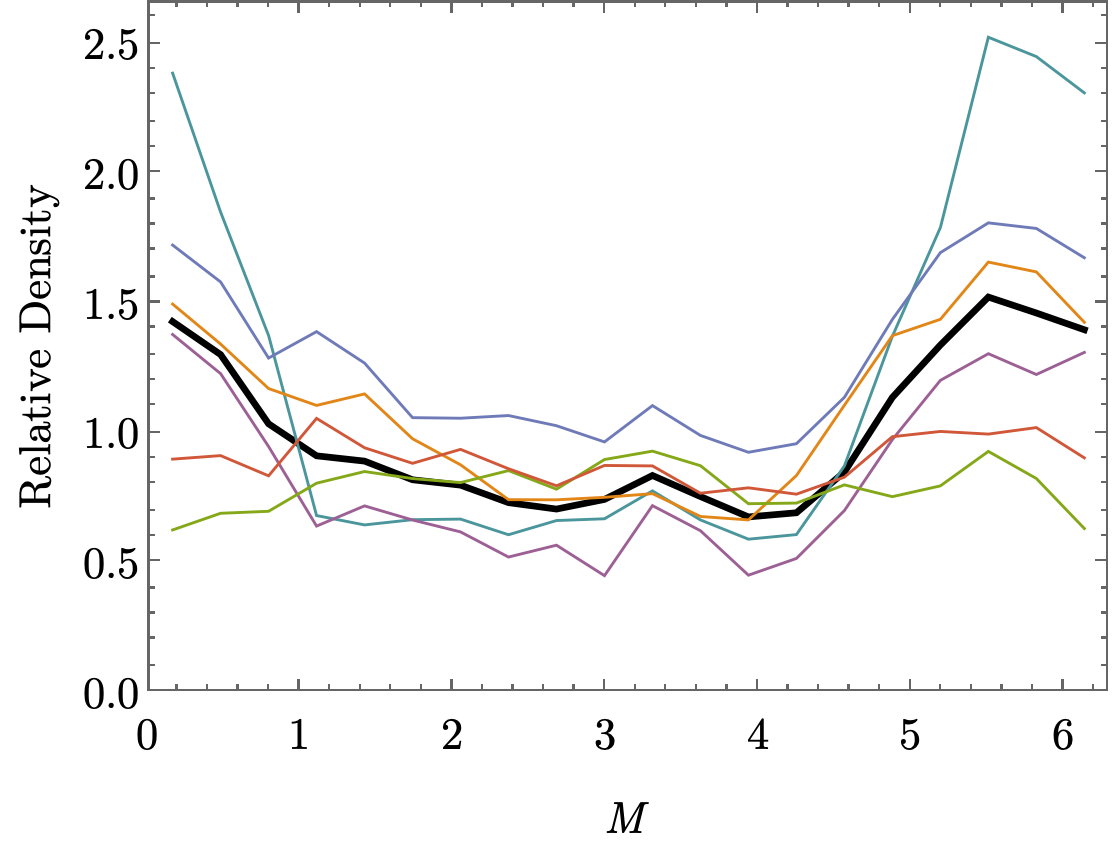}
\includegraphics[width=0.47\textwidth]{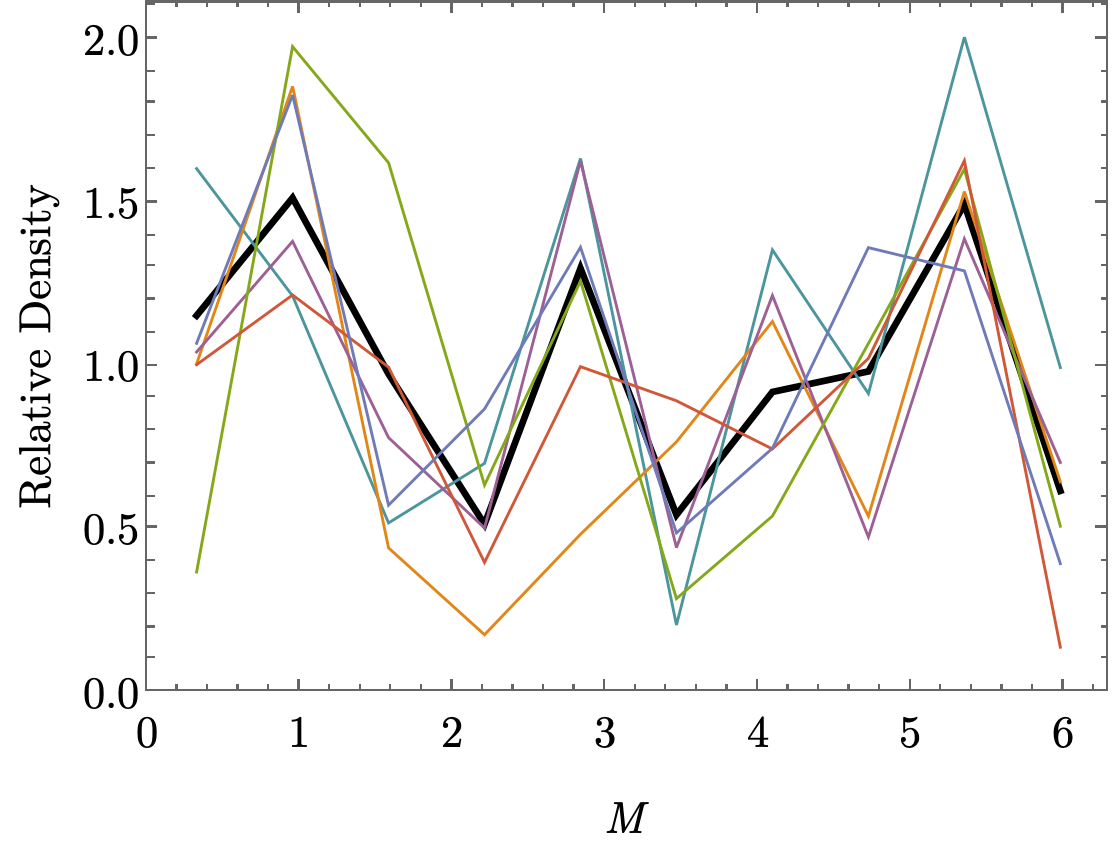}
\caption{Estimates of the solar basin density at Earth at different points along Earth's orbit, as measured by Earth's mean anomaly $M$. The left panel shows estimates obtained from the first 256 forward runs, and the right panel from the first 2048 backward runs (with the same methodology and from the same $t_f = 10^7\,\mathrm{yr}$ sample as in figure~\ref{fig_pdf1}). The thick black curves correspond to the accumulated density from the simulated particles, while the coloured curves correspond to bootstrap samples selecting a subset (with replacement) of these particles. There are not enough samples in either of these direct numerical simulations (but see section~\ref{sec:secular} for an alternative method) to provide conclusive evidence of short-term temporal modulation. \githubicon{https://github.com/kenvantilburg/solar-basin-dynamics/blob/main/code/PaperPlots.nb}}
\label{fig_modulation}
\end{center}
\end{figure}

We investigated the temporal modulation of the basin density over Earth's orbit from the same samples as those used in section~\ref{sec:dens_est} to estimate the Earth-crossing density. We recorded the mean anomaly $M$ of every ``Earth crossing'' in the final $t_f = 10^7\,\mathrm{yr}$ of the first 256 forward simulations. Similarly, we can estimate the density as a function of the ``starting'' mean anomaly $M$ of the backward simulation, since the start time was chosen uniformly randomly over the last $100\,\mathrm{yr}$. The results normalized to the mean density are shown as the thick black curve in figure~\ref{fig_modulation}. The thin colored lines depict six other bootstrap samples of the same simulation data, using the same method as in section~\ref{sec:dens_est}. We do not obtain significant evidence for temporal modulation using this method, likely due to insufficient statistics. This is not surprising, since we were able to determine the \emph{time-averaged} density to $\OO(10\%)$, while temporal modulation is a more fine-grained question at the few-percent level. The forward runs yield slightly lower noise despite the lower number of simulated orbits because each trajectory may have many Earth crossings over $t_f = 10^7\,\mathrm{yr}$. However, since secular timescales are shorter than $t_f$ (sections~\ref{sec:secular} and~\ref{sec:kozai}), there is a concern that temporal variation would be washed out even with increased statistics.

\subsection{Density from gravitational capture}
\label{sec:dens_capture}

In addition to modeling particle emission from the Sun, our simulations can also be applied to other scenarios, such as halo DM particles captured gravitationally (or by other interactions)~\cite{Peter:2009mi,gould1988direct}, or a primordial abundance of DM in the Solar System~\cite{Anderson:2020rdk}. While \emph{present-day} capture from the galactic halo cannot significantly enhance the density of such particles at Earth due to Liouville's theorem, it can lead to a low-relative-velocity population that may be of interest for certain types of interactions~\cite{Peter:2009mi, gould1988direct, Essig:2022dfa, Berlin:2019uco, VanTilburg:2024xib,Iles_2024}.

\begin{figure}[t]
\begin{center}
\includegraphics[width=0.4\textwidth]{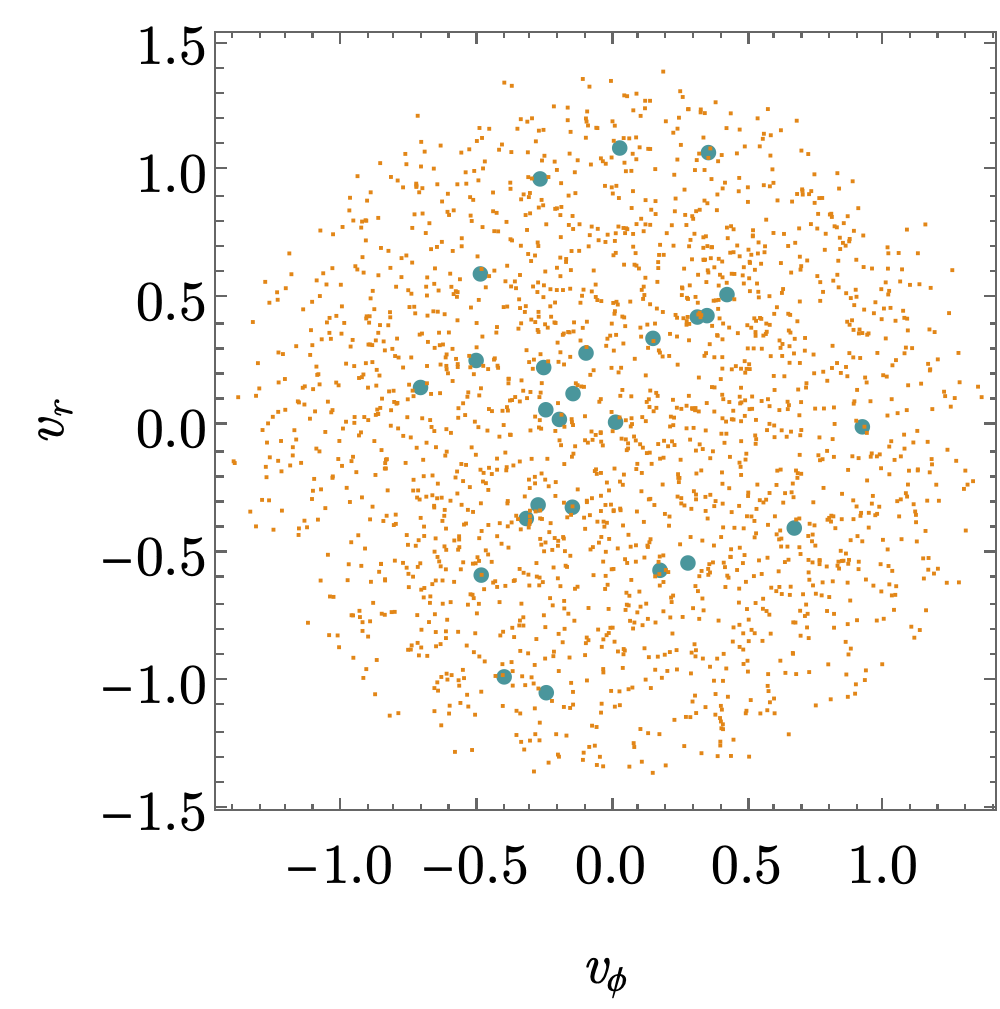}
\qquad
\includegraphics[width=0.4\textwidth]{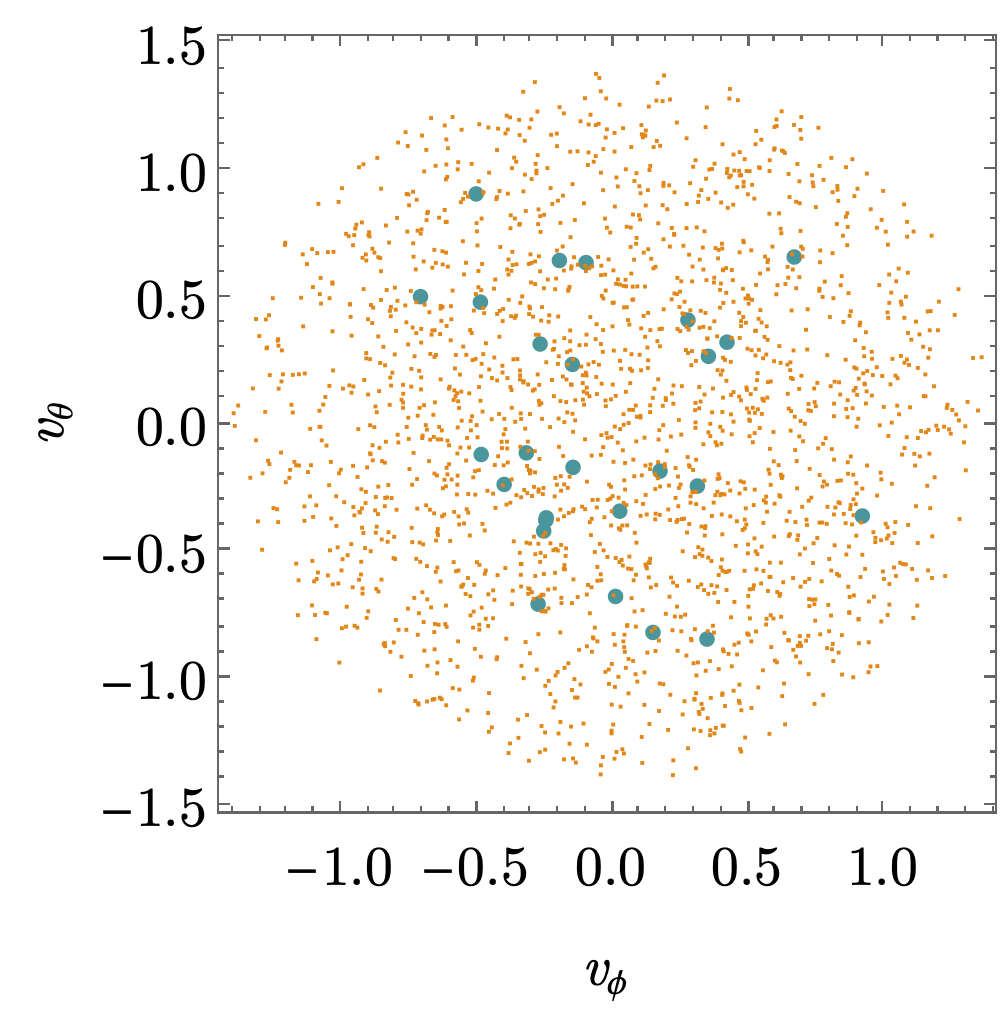}
\caption{Plots of initial particle velocities relative to Earth for the first 2048 backward simulations. These are plotted in spherical coordinates with the ecliptic as the equatorial plane, normalized to $v_0 = 2\pi \AU / \mathrm{yr}$, so that Earth's velocity is approximately $(v_r,v_\theta,v_\phi) = (0,0,1)$. The large blue dots correspond to particles which are not ejected from the Solar System during the backward evolution, while the small orange dots correspond to particles which are ejected. If we consider evolving forward in time, then the orange dots correspond to trajectories which are gravitationally captured from unbound halo trajectories.  We do not see evidence for the ``hole'' structure proposed in ref.~\cite{1991ApJ...368..610G}. \githubicon{https://github.com/kenvantilburg/solar-basin-dynamics/blob/main/code/PaperPlots.nb}}
\label{fig_ejv}
\end{center}
\end{figure}

For very weakly-interacting particles, the backward simulations model this scenario directly---each backward run that is ejected from the Solar System corresponds to a particle gravitationally captured from an unbound orbit. The fact that the vast majority (251/256 and 2022/2048 for our fiducial samples) of the simulated particles are ejected during the backward runs shows that gravitationally captured particles diffuse to fill almost all of the Earth-crossing phase space, as illustrated in figure~\ref{fig_ejv}. In particular, we do not see any evidence for the ``hole'' structure proposed in ref.~\cite{1991ApJ...368..610G}, where it was claimed that a large volume of Earth-crossing phase space remains empty over the lifetime of the Solar System.

One reason for this difference, already noted in ref.~\cite{Anderson:2020rdk}, appears to be that ref.~\cite{1991ApJ...368..610G} did not take into account the effects of Kozai oscillations (section~\ref{sec:kozai}). For orbits which are not Jupiter-crossing, ref.~\cite{1991ApJ...368..610G} assumed that phase space evolution is driven by perturbations from Earth and Venus, with perturbations from a given planet preserving the particle's velocity relative to that planet. However, as we discuss in section~\ref{sec:kozai}, even for particles with semi-major axis $a < a_\mathrm{J}/2$, the gravitational effects of Jupiter can drive significant evolution of all of the particle's orbital elements, apart from its semi-major axis, on timescales short compared to the age of the Solar System.
The assumption that particles approximately maintain their velocity relative to Earth is not generally a good one. Our direct numerical simulations, both forward and backward, show that the inner planets can perturb particles onto Jupiter-crossing orbits in less (but not much so) than the lifetime of the Solar System in almost all of the Earth-crossing phase space, in contrast to the predictions from refs.~\cite{1991ApJ...368..610G, Anderson:2020rdk}. We can qualitatively understand this behavior (up to motional resonances) as diffusive evolution of orbital energies due to close-encounter gravitational scattering with Venus and Earth (section~\ref{sec:stochastic}).

If nongravitational particle interactions are significant, then scattering or absorption in the Sun may affect the Earth-crossing distribution.
For bosonic particles which can be absorbed inside the Sun, the effect of such absorption (and associated production processes) is to bring the phase space density $f$ closer to the saturation value $f_T$ associated with the Solar temperature. If $f_T$ is larger than the phase space density of the unbound DM halo---true for a dark photon in the mass range $50 \,\mathrm{eV} \lesssim m \lesssim 20 \,\mathrm{keV}$~\cite[Fig.~3]{Lasenby:2020goo}---absorption increases the Earth-crossing density; otherwise, absorption decreases it.

For particles which are \emph{scattered} inside the Sun, the effect of solar scattering events will depend on the mass of the particle. Unlike gravitational interactions, which (to a very good approximation) preserve the phase space volume of test particle trajectories, scattering processes can exchange entropy with SM particles, either heating or cooling the hidden-sector particles.
For sufficiently heavy particles captured into tightly bound orbits, this cooling can result in significantly enhanced densities inside the Sun, which could lead to e.g.~enhanced annihilation rates \cite{1985ApJ...296..679P, 2009PhRvD..79j3532P, 1987NuPhB.279..804S, Lundberg:2004dn}.

The density in Earth-crossing trajectories can generally not be enhanced in this way. For light DM ($m \lesssim 50 \MeV$), whose thermal velocity at the temperature of the solar core is significantly higher than the Solar System's escape velocity, this follows simply from phase space considerations. The phase space density of unbound halo DM is larger than the thermal phase space density corresponding to Solar temperatures, so scatterings in the Sun will tend to reduce the phase space density to bring it closer to the thermal value.

For heavier DM, whose thermal velocity at Solar core temperatures is much less than the Sun's escape velocity, we can consider the phase space volume $\mathcal{V}_\mathrm{b}$ of bound, Sun-crossing orbits which reach large radii (say, semi-major axes greater than $0.3\,\mathrm{AU}$). These particles, as well as Sun-crossing unbound halo particles, have speeds close to the escape velocity within the Sun. In dynamical equilibrium, the average phase space density $\bar{f}$ within some energy bin $a$ must satisfy:
\begin{equation}
	\sum_b \bar{f}_b \Gamma_{b\rightarrow a}
	= \bar{f}_a \sum_c \Gamma_{a \rightarrow c}.
\end{equation}
The LHS sum is dominated by higher-energy bins $b$ (with velocities up to just above the Solar escape velocity), and the RHS sum by lower-energy bins $c$. If the energy bin $a$ is within $\mathcal{V}_\mathrm{b}$ and the scattering process is not strongly dependent on velocity (particles in bins $a$ and $b$ have velocities almost equal to the Solar escape velocity), 
then we have $\Gamma_{b\rightarrow a} \simeq \Gamma_{a \rightarrow b^*}$, where the energies satisfy $E_b - E_a \simeq E_a - E_{b^*}$. 
Then,
\begin{equation}
	\bar{f}_a \simeq \frac{\sum_{b^*} \bar{f}_b \Gamma_{a \rightarrow b^*}}
	{\sum_c \Gamma_{a \rightarrow c}}
\end{equation}
Thus, $\bar{f}_a$ is at most as large as the $\bar{f}_b$ for larger energies. The average phase space density within $\mathcal{V}_\mathrm{b}$ will therefore not be appreciably larger than the DM halo density.\footnote{This is the same conclusion as reached in ref.~\cite{2012PhRvD..85l3514S}. However, their arguments based on ``indistinguishability of Solar scattering from scattering in a time-dependent potential'' and detailed balance do not obviously apply to the physical case, in which Solar scatterings can have irreversible
behaviour. Scattering events decrease the energy of a heavy particle with very high probability; detailed balance does not hold since there is particle flow from the unbound halo to the solar core. 
They also assume the existence of the phase space hole of ref.~\cite{1991ApJ...368..610G}, and so conclude the solar scattering can populate bound Earth-crossing orbits that would not otherwise be occupied, in disagreement with the results of our simulations.}
This argument breaks down for bound, Sun-crossing orbits with speeds well below the Solar escape velocity (corresponding to semi-major axes $\ll \AU$), for which density enhancements can occur.

Since gravitational capture will populate almost all of the bound Earth-crossing phase space at the halo DM phase space density, these arguments show that Solar scatterings can at best deplete this density. We leave a quantitative analysis of these effects---in particular, the parameters for which scatterings do significantly reduce the Earth-crossing density---to future work.


\section{Secular perturbation theory}\label{sec:secular}
In section~\ref{sec:numerical}, we tackled the \emph{full} problem of solar basin evolution using numerical integration of particle orbits, with almost all relevant effects ``turned on'' (in section~\ref{sec:analytical}, we discuss several negligible effects that were left out of the simulations). However, because direct integration of particle trajectories is computationally demanding, it is challenging to have sufficient statistics for a precise determination of temporal modulation of the basin density along Earth's trajectory.

In this section, we use secular perturbation theory to tackle this question of temporal modulation on annual time scales. Our main assumption is that motional resonances can be neglected over most of the relevant phase space, and that close-encounter scattering typically occurs on time scales much longer than secular time scales (see section~\ref{sec:stochastic}). In other words, most of the time, particles (planets included) follow simple Keplerian orbits which vary slowly in time.

Recall these orbits can be described by the standard orbital elements: $a$, $e$, $I$, $\bar{\omega}$, $\Omega$, and $M$, which are the semi-major axis, eccentricity, inclination, longitude of perihelion, longitude of the ascending node, and the mean anomaly, respectively.  A refresher on orbital elements is included in appendix~\ref{app_elts}.  The premise of secular perturbation theory is that those first five orbital elements evolve slowly (on secular time scales), and that the rapid oscillation of the mean anomaly (on orbital time scales) can be averaged over.

Following the treatment of ref.~\cite[Ch.~9]{fitzpatrick2012introduction}, the evolution of the alternative elements $h \equiv e \sin \bar{\omega}$, $k \equiv e \cos \bar{\omega}$, $p \equiv \sin{I} \sin{\Omega}$, and $q \equiv \sin{I} \cos{\Omega}$ of a test particle at semi-major axis $a$ is described by:
\begin{align}
\frac{\dd h}{\dd t} &=  C k  +\sum_i A_i k_i, \label{eq:sec:1}\\
\frac{\dd k}{\dd t} &= - C h -\sum_i A_i h_i, \label{eq:sec:2}\\
\frac{\dd p}{\dd t} &= - C q  +\sum_i B_i q_i, \label{eq:sec:3}\\
\frac{\dd q}{\dd t} &= C p -\sum_i B_i p_i;  \label{eq:sec:4}
\end{align}
where
$A_i \equiv -\frac{1}{4} n \alpha_i \bar{\alpha}_i \frac{m_i}{M_\odot} b_{3/2}^{(2)}(\alpha_i)$, $B_i \equiv +\frac{1}{4} n \alpha_i \bar{\alpha}_i \frac{m_i}{M_\odot} b_{3/2}^{(1)}(\alpha_i)$, and $C \equiv \sum_i B_i$, with $\alpha_i \equiv \min\{a/a_i,a_i/a\}$ and $\bar{\alpha}_i \equiv \min\{a/a_i,1\}$ (note that we do not use Einstein summation). 
We have also defined the test particle's mean orbital angular velocity $n \equiv \sqrt{G_N M_\odot/a^3}$ around the Sun, which does not change because $\dd a / \dd t = 0$ in this treatment. Finally, we need the following function:
\begin{align}
b_s^{(j)}(\alpha) = \int_0^{2\pi} \dd \psi \, \frac{\cos j \psi}{\left[1-2\alpha \cos \psi + \alpha^2 \right]^s}.
\end{align}
All quantities with subscript $i = 1, \dots, 8$ are the equivalent quantities for the eight planets in the Solar System, with masses $m_i$.
Their orbital elements obey similar evolution equations, with solution given by~\cite{fitzpatrick2012introduction}:
\begin{align}
h_i &= \sum_l e_{il} \sin\left(g_l t + \beta_l\right)  \label{eq:sec:5} \\
k_i &= \sum_l e_{il} \cos\left(g_l t + \beta_l\right)  \label{eq:sec:6} \\
p_i &= \sum_l I_{il} \sin\left(f_l t + \gamma_l\right)  \label{eq:sec:7} \\
q_i &= \sum_l I_{il} \cos\left(f_l t + \gamma_l\right). \label{eq:sec:8}
\end{align}
The eccentricity and inclination eigenvectors $e_{il}$ and $I_{il}$, secular eigenfrequencies $g_l$ and $f_l$, and phases $\beta_l$ and $\gamma_l$ are known and given in ref.~\cite[Ch.~9]{fitzpatrick2012introduction}.

The evolution eqs.~\ref{eq:sec:1}--\ref{eq:sec:4} are a coupled system of first-order, inhomogeneous differential equations, with forcing terms proportional to eqs.~\ref{eq:sec:5}--\ref{eq:sec:8}. Its solutions are therefore readily obtained: for a particle with initial conditions $a_0$, $e_0$, $I_0$, $\bar{\omega}_0$, and $\Omega_0$ at time $t_0$, the solution at time $t$ is
\begin{alignat}{2}
h(t) &= e_0 \sin\left[ +C (t-t_0)+ \bar{\omega}_0 \right]  - \sum_{i,l} \frac{A_i e_{il}}{C-g_l} \left\lbrace \sin\left[g_l t + \beta_l \right] - \sin\left[+C (t-t_0) + g_l t_0 + \beta_l\right] \right\rbrace\label{eq:sec:9}\\
k(t) &= e_0 \cos\left[ +C (t-t_0)+ \bar{\omega}_0 \right]  - \sum_{i,l} \frac{A_i e_{il}}{C-g_l} \left\lbrace \cos\left[g_l t + \beta_l \right] - \cos\left[+C (t-t_0) + g_l t_0 + \beta_l\right] \right\rbrace \label{eq:sec:10}\\
p(t) &= \sin I_0 \sin\left[ -C (t-t_0)+ \Omega_0 \right]  + \sum_{i,l} \frac{B_i I_{il}}{C+f_l} \left\lbrace \sin\left[f_l t + \gamma_l \right] - \sin\left[-C (t-t_0) + f_l t_0 + \gamma_l\right] \right\rbrace \label{eq:sec:11}\\
q(t) &= \sin I_0 \cos\left[ -C (t-t_0)+ \Omega_0 \right]  + \sum_{i,l} \frac{B_i I_{il}}{C+f_l} \left\lbrace \cos\left[f_l t + \gamma_l \right] - \cos\left[-C (t-t_0) + f_l t_0 + \gamma_l\right] \right\rbrace. \label{eq:sec:12}
\end{alignat}
By inspection of the solutions in eqs.~\ref{eq:sec:9}--\ref{eq:sec:12} for the test particle and eqs.~\ref{eq:sec:5}--\ref{eq:sec:8} for the planets, it is clear that there are correlations between the orbital elements of the two sets of bodies. For example, even if one averages over all possible emission times $t_0$, one still finds the correlation:
\begin{align}
\left \langle h(t) h_j(t) \right\rangle_{t_0} = - \sum_{i,l,l'}  \frac{A_i e_{il}e_{jl'}}{C-g_l} \sin\left(g_l t + \beta_l \right) \sin\left(g_{l'} t + \beta_{l'}\right),\label{eq:hh_corr}
\end{align}
\emph{independent} of the initial longitude of perihelion $\bar{\omega}_0$.
Similar such correlations exist between all orbital elements, e.g.~$\langle k(t) p_j(t)\rangle$, $\langle q(t) q_j(t)\rangle$, etc.

Note that all correlations of the type in eq.~\ref{eq:hh_corr} ``lose memory'' of the initial conditions $e_0$, $I_0$, $\bar{\omega}_0$, and $\Omega_0$, after averaging (with uniform weights) over possible emission times $t_0$, as is appropriate for near-constant production in the Sun. The correlations only depend on the distribution of semi-major axis $a_0$, through the test particle's secular frequencies $A_i$, $B_i$, and $C$. They are small in an absolute sense only because the inclinations and eccentricities in the Solar System are small, but they are not small in a relative sense: the size of the cross-correlations between the orbital elements of the test particle and any planet $j$, e.g.~$\langle h(t) h_j(t) \rangle$, can be of order the square of the planets' orbital elements, e.g.~$h_j(t)^2$. Lastly, these cross-correlations themselves change over time, with a rate given by (sums and differences) of secular angular eigenfrequencies, as is clear from eq.~\ref{eq:hh_corr} for example.

Within the above framework of secular perturbation theory, we can numerically estimate the temporal modulation of the basin density precisely. Our numerical experiments consist of $10^4$ ``runs''. For each individual run, we injected $10^3$ ``particles'' (really, secular orbits) with semi-major axes randomly drawn from the distribution $f(a) \propto 1/a^2$ with $a < 10\,\mathrm{AU}$, initial eccentricities from $f(e_0) = 2 e_0$, inclinations from $f(I_0) \propto \sin I_0$, and $\omega_0$ and $\Omega_0$ from uniform distributions. For each such set of initial orbital elements, the particle was injected at $10^3$ different start times $t_0$, randomly drawn from a uniform distribution between $t_0 = -4.5\,\mathrm{Gyr}$ and $t_0 = 0$. In total, we thus consider $10^4 \times 10^3 \times 10^3 = 10^{10}$ distinct particle injections, statistics which are infeasible for direct numerical integration using the methods from section~\ref{sec:numerical}.

We used eqs.~\ref{eq:sec:9}--\ref{eq:sec:12} to compute the present-day $t=0$ orientation of all secular orbits. For all $10^6$ orbits ($10^3$ different orbital elements, $10^3$ values of $t_0$) in each of the $10^4$ runs, we sampled $10^3$ points per orbit equidistant in time (\textit{i.e.}~mean anomaly) and filled a 3D histogram in $xyz$-coordinates with bin size $\Delta x = \Delta y = \Delta z = 0.024 \, \mathrm{AU}$ to create a solar basin density field. The ecliptic plane is taken to be $z=0$ in this histogram.  The 2D slice of $|z| < 0.012 \, \mathrm{AU}$ around the ecliptic was then interpolated to obtain the solar basin density along Earth's orbit, for each of the $10^4$ runs. We checked that the systematic error due to the finite size of the $z$ bins was subdominant to the statistical error below by comparing the density field in adjoining $z$ bins.

\begin{figure}
    \centering
    \includegraphics[width=0.8\textwidth]{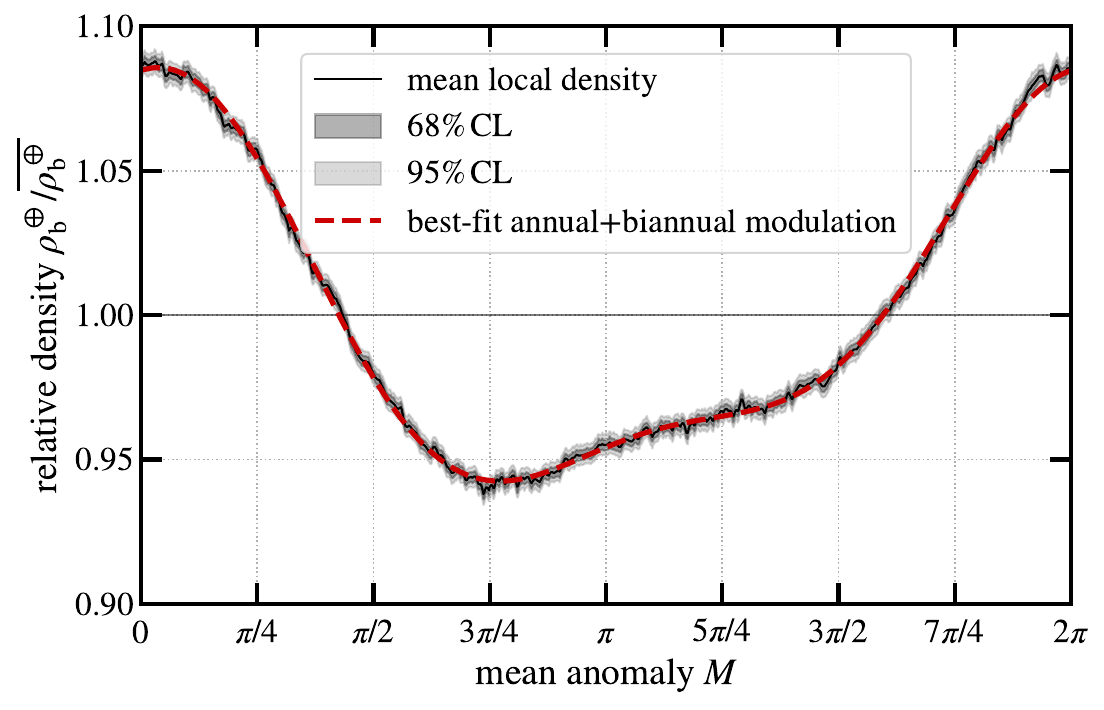}
    \caption{Solar basin density variation as a function of mean anomaly $M = 2 \pi t / \mathrm{yr}$ along Earth's orbit at $t \approx 0$ (epoch J2000). The black line is the mean across all simulated secular orbits, and the dark (light) gray bands depict the statistical 68\% (95\%) CL intervals. The red dashed line visualizes eq.~\ref{eq:modulation_real_2}, isolating the only two statistically significant Fourier components with periods of $1\,\mathrm{yr}$ and $0.5\,\mathrm{yr}$. \githubicon{https://github.com/kenvantilburg/solar-basin-dynamics/blob/main/code/Secular_PT/secular_pt.ipynb}}
    \label{fig:modulation_real}
\end{figure}

The resulting solar basin density field $\rho_\mathrm{b}^\oplus$ along Earth's orbit at epoch J2000 is shown as a function of mean anomaly $M$ in figure~\ref{fig:modulation_real}, relative to the mean across one year. By repeating the procedure separately for $10^4$ runs, we could calculate the mean fractional density variation (black) as well as the 68\% and 95\% CL intervals (dark and light gray bands), with a typical (bootstrapped) statistical error of about 0.2\%. The observed peak-to-through variation is 14.8\%, with the maximum achieved 14 days after perihelion (January 17), and the minimum 146 days after perihelion (May 29). The amplitude of the \emph{annual} modulation is in line with the expectation of a basin density field that falls off approximately as $\rho \propto R^{-4}$ combined with Earth's eccentric orbit with eccentricity $e$, which would yield the naive prediction of a $6.8\%$ fractional annual modulation. However, distortion(s) at higher frequency---from orbital correlations of the type in eq.~\ref{eq:hh_corr}---are visually obvious in figure~\ref{fig:modulation_real}.

\begin{figure}
    \centering
    \includegraphics[width=0.8\textwidth]{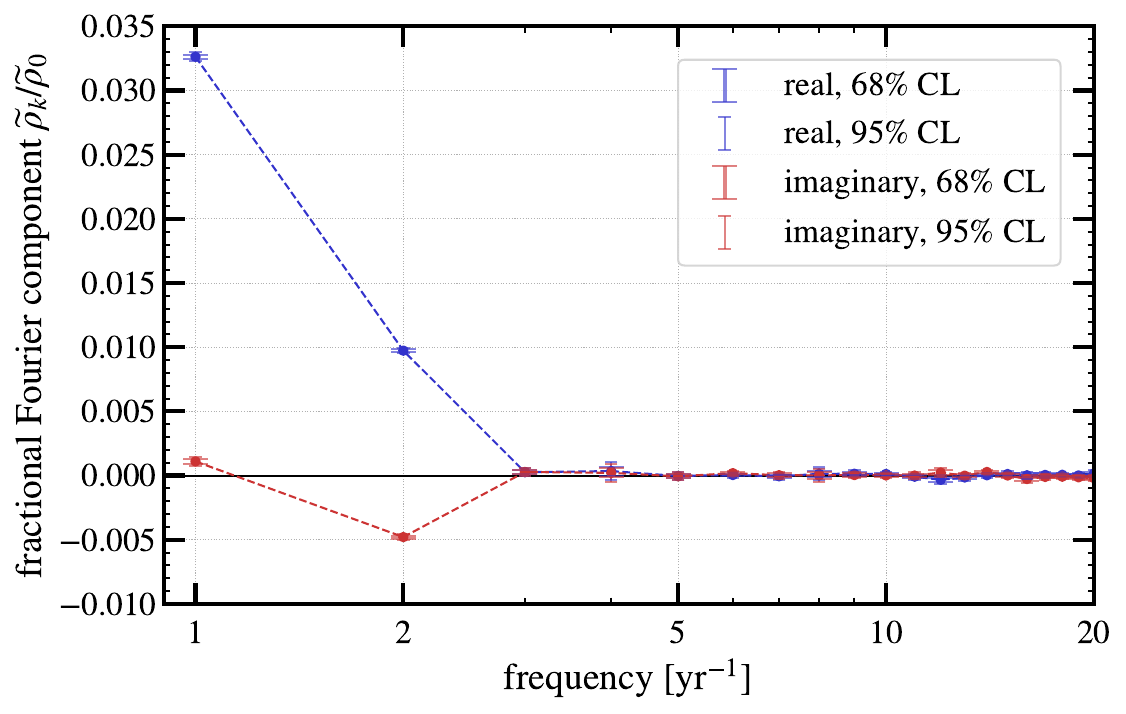}
    \caption{Discrete Fourier transform coefficients $\widetilde{\rho}_k$ of the basin density data as in eq.~\ref{eq:DFT_coeff}. The Fourier frequencies are integer multiples $k$ of an inverse year. Both real (blue) and imaginary (red) parts of the amplitudes are shown, along with their 68\% and 95\% CL intervals, and are normalized relative to the zero-frequency coefficient $\widetilde{\rho}_0$. The only modes with nonzero amplitude are those with $k \leq 2$, which are observed at very high signal-to-noise ratio. \githubicon{https://github.com/kenvantilburg/solar-basin-dynamics/blob/main/code/Secular_PT/secular_pt.ipynb}}
    \label{fig:modulation_DFT}
\end{figure}

We quantify the temporal modulation in figure~\ref{fig:modulation_real} using the discrete Fourier transform (DFT) in figure~\ref{fig:modulation_DFT}. Specifically, we discretely sample $\left(\rho_\mathrm{b}^\oplus\right)_m \equiv \rho_\mathrm{b}^\oplus(t_m)$ at $n = 1{,}024$ times $t_m = (m/n) \, \mathrm{yr}$ for $m = 0, \dots, n-1$. The (complex) DFT coefficients are defined as:
\begin{align}
    \widetilde{\rho}_k = \sum_{m=0}^{n-1} \left(\rho_\mathrm{b}^\oplus\right)_m \exp \left\lbrace - 2\pi i \frac{m k}{n} \right\rbrace; \quad k = 0, \dots, n-1. \label{eq:DFT_coeff}
\end{align}
The real and imaginary parts of these coefficients for $k \geq 1$ are plotted in blue and red, respectively, in figure~\ref{fig:modulation_DFT}. 
The components are normalized relative to the (real) zero-frequency coefficient 
$\widetilde{\rho}_0 = \sum_m \left(\rho_\mathrm{b}^\oplus\right)_m = n {\overline{\rho}_{\mathrm{b}}^\oplus} $. As in figure~\ref{fig:modulation_real}, the procedure was repeated for all $10^4$ runs separately; the dots indicate the mean of the components over all runs, and the error bars the bootstrapped 68\% and 95\% CL intervals.

Using this procedure, we find both annual and semi-annual modulation, with Fourier components
\begin{align}
    \frac{\widetilde{\rho}_k}{\widetilde{\rho}_0}
    = 10^{-2} \begin{cases} 
    3.26^{+0.02}_{-0.02} + i \, 0.11^{+0.02}_{-0.02} & (k = 1) \\
    0.97^{+0.01}_{-0.01} - i \, 0.48^{+0.01}_{-0.01} & (k=2),
    \end{cases}
\end{align}
but no significant evidence for power at higher frequencies $k \geq 3$, apart from the complex conjugates 
$\widetilde{\rho}_{n-1} = \widetilde{\rho}_1^*$ and $\widetilde{\rho}_{n-2} = \widetilde{\rho}_2^*$. 
The basin density variation at Earth's location as a function of time of year, relative to the mean, is thus:
\begin{align}
    \frac{\rho_\mathrm{b}^\oplus (t)}{\overline{\rho}_{\mathrm{b}}^\oplus} 
    &= 1 + \sum_{k=1}^\infty 2 \left| \widetilde{\rho}_k \right| \cos\left[2 \pi k \frac{t}{\mathrm{yr}} + \arctan \left(\frac{\mathrm{Im} \, \widetilde{\rho}_k}{\mathrm{Re} \, \widetilde{\rho}_k}\right) \right] \label{eq:modulation_real_1} \\
    &\approx 1 + 0.065 \cos\left( 2 \pi \frac{t + 2.0 \, \mathrm{day}}{\mathrm{yr}} \right) + 0.022 \cos\left( 4 \pi \frac{t - 26.6 \, \mathrm{day}}{\mathrm{yr}} \right). \label{eq:modulation_real_2}
\end{align}
Equation~\ref{eq:modulation_real_2} is overlaid as the red dashed line in figure~\ref{fig:modulation_real}, and can be seen to account for essentially all of the statistically significant temporal modulation. Because of the small but statistically significant imaginary part 
$\mathrm{Im} \, \widetilde{\rho}_1 \neq 0$, there is a small offset of the annual modulation maximum from perihelion (January 3, where $M=0$ by construction), appearing instead about $2$ days later. Due to secular correlations of test particles and planets, the basin density has a preferred axis, leading to \emph{semi-annual} modulation with an offset from perihelion of about $27$ days. Numerically, this semi-annual modulation is only about 3 times smaller in amplitude than the annual modulation. The temporal variation in eq.~\ref{eq:modulation_real_2} is a smoking-gun prediction of a solar basin signal that is qualitatively different from that of a DM signal, which lacks semi-annual modulation and has weaker annual modulation with a different phase offset~\cite{lee2014effect}.

\section{Stochastic description}
\label{sec:stochastic}

In this section, we attempt to describe the long-term evolution of solar basin particles' semi-major axes using a stochastic description. Test particles are subject to secular perturbations (section~\ref{sec:secular}), motional resonances, and close encounters with planets. Here, we will model the resulting diffusion through phase space \emph{only from the latter process}, \textit{i.e.}~gravitational scattering. Our modeling is reminiscent of the phase-space diffusion studies performed in refs.~\cite{1991ApJ...368..610G, Anderson:2020rdk} in the context of gravitational capture of DM in the Solar System. 

We find that the long-term changes in orbital energy, or equivalently semi-major axis, \emph{are roughly captured} by repeated quasi-random close encounters, with a fiducial estimate for the effective basin lifetime of order $\si{Gyr}$, similar to the one found in section~\ref{sec:numerical}. Our findings here do not constitute a proof or even a robust validation of that section because of strong assumptions and omissions, but it does qualitatively point to the primary mechanism that predicts an effective basin lifetime shorter---but not much so---than the age of the Solar System. Our results below indicate that motional resonances likely do play an important role, and for this reason alone direct numerical integration is necessary for a precise estimate of the solar basin density at late times. In this section, we also make the strong assumption that the phase space is fully equilibrated in the dimensions orthogonal to the semi-major axis at all times. This is certainly an oversimplification, as the filling of phase space happens gradually and never reaches completion.

In section~\ref{sec:close_encounters}, we give an analytic formula of the differential gravitational scattering probability for a fully equilibrated phase space (with some of the derivation relegated to appendix~\ref{sec:scattering}), and demonstrate that it is a qualitatively good descriptor of energy-changing processes in our numerical simulations. Since the cumulative effects of \emph{many weak encounters} dominate the overall dynamics, this scattering probability is then converted to a diffusion function and an ejection rate in a Fokker-Planck equation for the long-term semi-major axis evolution in subsection~\ref{sec:diffusion}.

\subsection{Close encounters with planets} \label{sec:close_encounters}
Most of the energy changes (away from motional resonances) are driven by close-encounter gravitational scattering between a basin particle and the planets.  In this subsection, we calculate these dynamics analytically, first for bound-bound scattering in section~\ref{sec:bound-bound} which will lead to diffusion, then for bound-unbound scattering, \textit{i.e.}~ejections, in section~\ref{sec:ejections}.

\subsubsection{Bound-bound scattering}\label{sec:bound-bound}
In appendix~\ref{sec:scattering}, we derive the differential gravitational scattering cross-section
\begin{align}
\frac{\dd\sigma}{\dd \cos\phi_{\textrm{out}}} = 2\pi \frac{(G_N M_{\textrm{P}})^2}{w^4}\frac{1-\cos\phi_{\textrm{in}}\cos\phi_{\textrm{out}}}{\left(\cos\phi_{\textrm{in}} - \cos\phi_{\textrm{out}}\right)^3}
\end{align}
for a planet of mass $M_\mathrm{P}$ with circular velocity $v_\mathrm{P}$ at radius $R$, to scatter a particle from some incoming velocity $\vect{v}_\mathrm{in}$ to an outgoing velocity $\vect{v}_\mathrm{out}$. The $\phi_\mathrm{in,out}$ angles are related to the incoming/outgoing velocities as $\cos \phi = (v^2-v_\mathrm{P}^2-w^2)/(2 v_\mathrm{P} w)$, and $w = |\vect{v}_\mathrm{in} - \vect{v}_\mathrm{P}| = |\vect{v}_\mathrm{out} - \vect{v}_\mathrm{P}|$ is the relative speed in the encounter. From this differential cross-section, we then derive in appendix~\ref{sec:scattering} the differential scattering rate between incoming and outgoing semi-major axes $a_\mathrm{in,out}$:
\begin{align}
\frac{\dd \Gamma}{\dd a_{\textrm{in}}\dd a_\mathrm{out}} = \int \dd \Omega_\mathrm{in} \, f(\vect{v}_{\textrm{in}},R) \frac{\dd\sigma}{\dd \cos\phi_{\textrm{out}}} \frac{(G_N M_\odot)^2}{4 a_\mathrm{in}^2 a_\mathrm{out}^2} \frac{v_\mathrm{in} }{v_\mathrm{P}}, \label{eq:dGammadaout}
\end{align}
where $f(\vect{v}_\mathrm{in},R)$ is the velocity phase space density at radius $R$, normalized such that $n(R) = \int \dd^3 \vect{v}_\mathrm{in} \, f(\vect{v}_\mathrm{in},R)$ is the number density. 
If the angular phase space is fully mixed (across the angles of $\vect{v}_\mathrm{in}$ in the $\int \dd \Omega_\mathrm{in}$ integral), then the phase space density at radius $R$ for a single particle with semi-major axis $a_\mathrm{in}$ is
\begin{align}
f(\vect{v},R) = f(v,R) \simeq \frac{1}{4\pi^3}\frac{1}{(G_N M_\odot)^{3/2} a^{1/2}} \delta(a - a_\mathrm{in}) \Theta\left(2 a - R\right), \label{eq:f_single}
\end{align}
where we used
\begin{align}
    v_\mathrm{P} = \sqrt{\frac{G_N M_\odot}{R}}, \qquad v = \sqrt{2 G_N M_\odot \left( \frac{1}{R} - \frac{1}{2a}\right)}.
\end{align}
Plugging in eq.~\ref{eq:f_single} into eq.~\ref{eq:dGammadaout} finally yields:
\begin{align}
    \frac{\dd \Gamma}{\dd a_\mathrm{out}}(a_\mathrm{in}) 
    &= \frac{1}{16 \pi^3} \sqrt{\frac{G_N M_\odot}{a_\mathrm{in}}} \frac{1}{a_\mathrm{in}^2 a_\mathrm{out}^2} \sigma_\mathrm{out}  \label{eq:dGammadaout_single},
    \qquad
    \sigma_\mathrm{out} \equiv \int \dd \Omega_\mathrm{in} \, \frac{\dd\sigma}{\dd \cos\phi_{\textrm{out}}}.
\end{align}
The scattering cross-section $\sigma_\mathrm{out}$ can be expressed in (piecewise) closed form, though we do not include this expression as it is not illuminating.

Let us now compute the differential scattering probability $\dd P_\mathrm{scatter}$ for a single particle to change its semi-major axis by an amount $\Delta a \equiv a_\mathrm{out} - a_\mathrm{in}$ in a small range $\dd (\Delta a)$  
over some small time step $\dd t$:
\begin{align}
\dd P_\mathrm{scatter}\left(a_\mathrm{in},\Delta a, \dd t\right) 
&\simeq \dd (\Delta a) \, \dd t \, \frac{\dd \Gamma}{\dd a_\mathrm{out}} \label{eq:dPscatter} \\
&\simeq \dd t \, \dd (\Delta a)\, \frac{F(a_\mathrm{in})}{|\Delta a|^3} \Theta\left(|\Delta a| - (\Delta a)_\mathrm{min}\right) \Theta\left(2a - R\right) + \mathcal{O}\left(\frac{1}{|\Delta a|^{2}}\right). \nonumber
\end{align}
In the second line, we have expanded eq.~\ref{eq:dGammadaout} in terms of small $|\Delta a|$, \textit{i.e.}~minor, ``soft'' interactions, which will turn out to dominate over major, ``hard'' encounters. We have also introduced the cutoff $(\Delta a)_\mathrm{min} \equiv \sqrt{F(a_\mathrm{in}) \dd t}$ fixed by imposing a normalized scattering probability: $\int_{-\infty}^{+\infty} \dd(\Delta a) \, \dd P_\mathrm{scatter}/{\dd (\Delta a)} = 1$.
The numerator function $F(a_\mathrm{in})$ is expressed in closed form in eq.~\ref{eq:F} and is continuous, but its first and second derivative has a discontinuity at $a_\mathrm{in} = R$, and its second derivative furthermore diverges at $a_\mathrm{in} = R/2$. 

We observe two key behaviors from the the analytic result of eq.~\ref{eq:dPscatter}. Firstly, the planetary scattering probability scales roughly as the combination $P_\mathrm{scatter} \propto M_\mathrm{P}^2/R$ for $ a_\mathrm{in}\gg R/2$ , which is why Venus will contribute comparably to Earth in terms of energy changes of basin particles (even slightly more than Earth for $a \lesssim \mathrm{AU}$), despite its smaller mass. For the same reason, Mercury and Mars have negligible effects. 

Secondly, the scattering probability also scales as $P_\mathrm{scatter} \propto \dd t /  \Delta a^{3}$, so \emph{small-angle} scattering will turn out to dominate the overall dynamics. As we will see in section~\ref{sec:diffusion} below, the diffusion function is naively logarithmically divergent, signalling that every $e$-fold in $\Delta a$ contributes equally to the diffusive dynamics. This implies that the accumulated effects of minor scattering events with small $\Delta a$ dominate over major scattering events with large $\Delta a$ and over ejections, both by a large logarithmic factor. At least within a stochastic framework where motional resonances are neglected, this suggests a diffusive evolution, described in section~\ref{sec:diffusion}.

\subsubsection{Ejections}\label{sec:ejections}
The cross-section for a particle to be ejected from the solar system due to gravitational scattering with a planet of mass $M_{\rm{P}}$ has already been calculated in ref.~\cite{Levin_2006}:
\begin{alignat}{2}
\sigma_\mathrm{ej} = \pi \frac{\left(G_N M_\mathrm{P}\right)^2}{w^4} \frac{\sin^2 \phi_\mathrm{ej}}{\left(\cos \phi_\mathrm{in} - \cos \phi_\mathrm{ej} \right)^2},
\end{alignat}
with $\cos \phi_\mathrm{ej} = (v_\mathrm{ej}^2 - v_\mathrm{P}^2 - w^2)/(2v_\mathrm{P} w)$. (Indeed, the notation here and in appendix~\ref{sec:scattering} for bound-bound scattering is heavily borrowed from ref.~\cite{Levin_2006}.) The resulting ejection rate is then
\begin{alignat}{2}
\Gamma_\mathrm{ej}
&= \int \dd^3 \vect{v}_\mathrm{in} \, f(\vect{v}_\mathrm{in}) w \sigma_\mathrm{ej} 
\simeq \frac{1}{4\sqrt{2}\pi^3} \frac{1}{R^{1/2}a_\mathrm{in}^3} \sqrt{2 a_\mathrm{in} - R} \int \dd \Omega_\mathrm{in} \, w \sigma_\mathrm{ej}, \label{eq:Gamma_ej}
\end{alignat}
with the second equality for a fully mixed phase space conditional on the basin particle having a semi-major axis $a_\mathrm{in}$ as in eq.~\ref{eq:f_single}. The latter integral can be evaluated analytically and is given explicitly in eqs.~\ref{eq:Gamma_ej_1}--\ref{eq:Gamma_ej_3}. Parametrically, one has $\Gamma_\mathrm{ej} \sim (M_\mathrm{P}/M_\odot)^2 v_\mathrm{P} / R$ for $a_\mathrm{in} \sim R$. Therefore, the ejection time is $(M_\odot/M_\mathrm{P})^2$ orbital times, which is longer than the age of the Sun in the inner Solar System, but much shorter beyond Jupiter's orbit.

\subsection{Diffusive evolution}\label{sec:diffusion} 
Because the scattering dynamics for changes in semi-major axis are governed by \emph{quasi-random, minor} encounters, we model them via the Fokker-Planck equation:
\begin{equation}
    \frac{\partial p(a,t)}{\partial t} = \frac{\partial^2}{\partial a^2} \left[D(a)p(a,t)\right] - \Gamma_\mathrm{ej}(a)p(a,t) + \Gamma_\mathrm{prod}(a), \label{eq:fokker_planck}
\end{equation}
where $p(a,t) \equiv {\dd N}/{\dd a}$ is the distribution of basin particle number $N$ over semi-major axis $a$, as function of time $t$. The second term on the RHS is the ejection rate of eq.~\ref{eq:Gamma_ej}.
The third term incorporates production inside the Sun, which follows a distribution $\Gamma_\mathrm{prod} = c/a^2$ for most particles, including axions~\cite{VanTilburg:2020jvl} and dark photons~\cite{Lasenby:2020goo}; the constant $c$ is proportional to the basin production rate but will cancel out in what follows. For specificity, we take the production function to be:
\begin{align}
    \Gamma_\mathrm{prod}(a) = \frac{\mathrm{AU}^2/_\odot}{a^2} S_2(a,a_\mathrm{min},a_\mathrm{min}+\sigma_a) S_2(-a,-a_\mathrm{max},-a_\mathrm{max} + \sigma_a) \label{eq:Gamma_prod}
\end{align}
so that $p$ is dimensionless and the solution to eq.~\ref{eq:fokker_planck} would be $p(\mathrm{AU},t_\odot) = 1$ if there were no diffusion nor ejections. We also multiplied with the sigmoid-like clamping functions (``smoother steps''):
\begin{align}
S_2(x,x_\mathrm{x_\mathrm{min}},x_\mathrm{max}) = 
\begin{cases}
0 & 0 < \tilde{x} \\
6 x^5 - 15 x^4 + 10 x^3 & 0 \leq \tilde{x} \leq 1 \\
1 & 1 < \tilde{x}; 
\end{cases}
\qquad \tilde{x} \equiv \frac{x - x_\mathrm{min}}{x_\mathrm{max} - x_\mathrm{min}},
\end{align}
so that we can study the problem on a finite interval $a \in [a_\mathrm{min}, a_\mathrm{max}]$ and consistently enforce boundary conditions $p(a_\mathrm{min},t) = p(a_\mathrm{max},t) = 0$.

The diffusion function $D(a)$ can be extracted from $\dd P_\mathrm{scatter}$ as:
\begin{align}
    D(a) \simeq \frac{\dd \, \mathrm{Var}\left[(\Delta a)^2 \right]}{2 \dd t} = \int_{(\Delta a)_\mathrm{min}}^a \dd(\Delta a) \,(\Delta a)^2  \, \frac{\dd P_\mathrm{scatter}}{\dd t \, \dd(\Delta a)} = F(a) \ln \left( \frac{a}{(\Delta a)_\mathrm{min}} \right). \label{eq:D_diffusion}
\end{align}
As discussed in section~\ref{sec:bound-bound}, the contributions to the variance in $\Delta a$ are nearly scale independent, because $\dd P_\mathrm{scatter} \propto 1/|\Delta a|^3$. Because of the logarithmic divergence, \emph{many} scales in $\Delta a$ contribute, and we can expect the leading energy-changing behavior to be diffusive. The dependence on UV and IR cutoffs is very weak; as a sensible guess, we take them to be $|\Delta a| < a$ (beyond which our approximations break down) and $|\Delta a| > (\Delta a)_\mathrm{min}$ from below eq.~\ref{eq:dPscatter} with $\dd t = 1\,\mathrm{yr}$ to have a normalized, approximately independent scattering probability over a typical orbit.

\begin{figure}
    \includegraphics[width=0.48\textwidth]{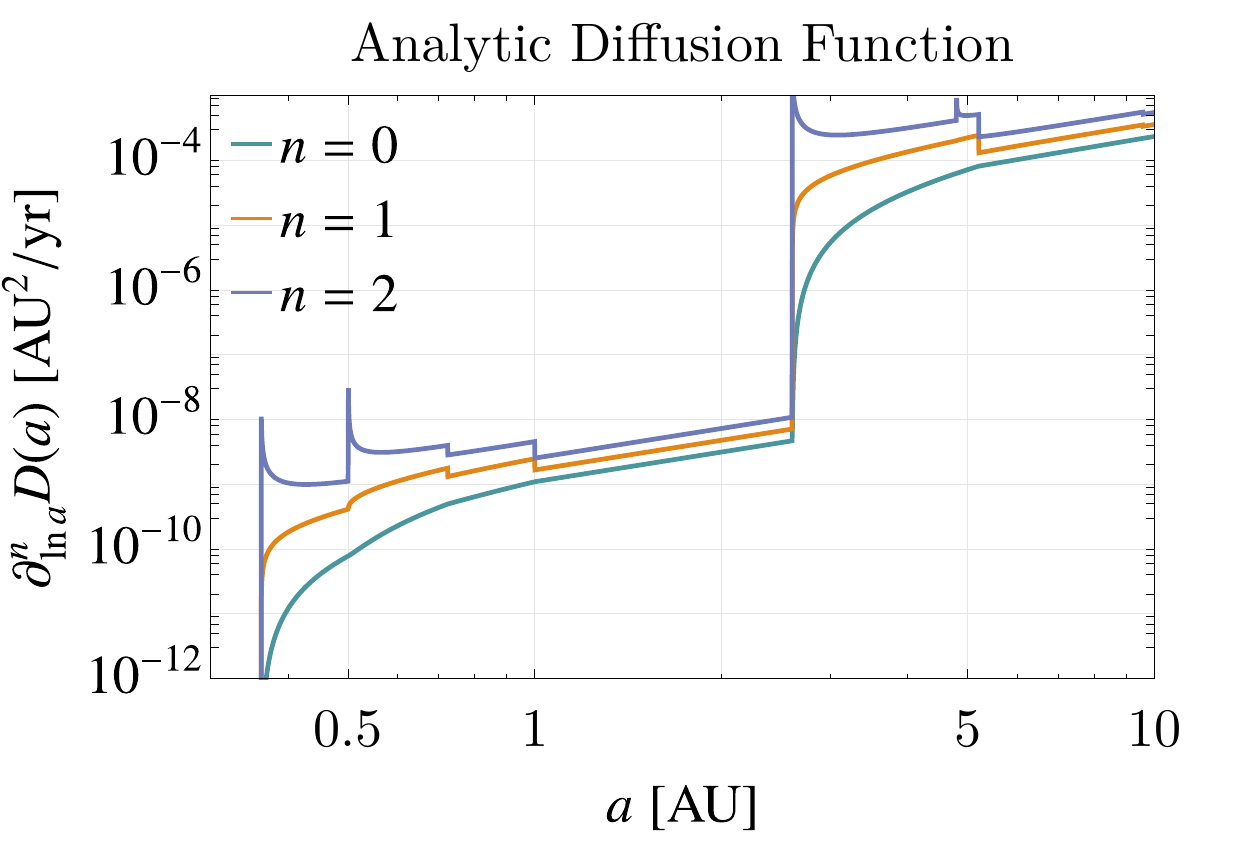}
    \includegraphics[width=0.48\textwidth]{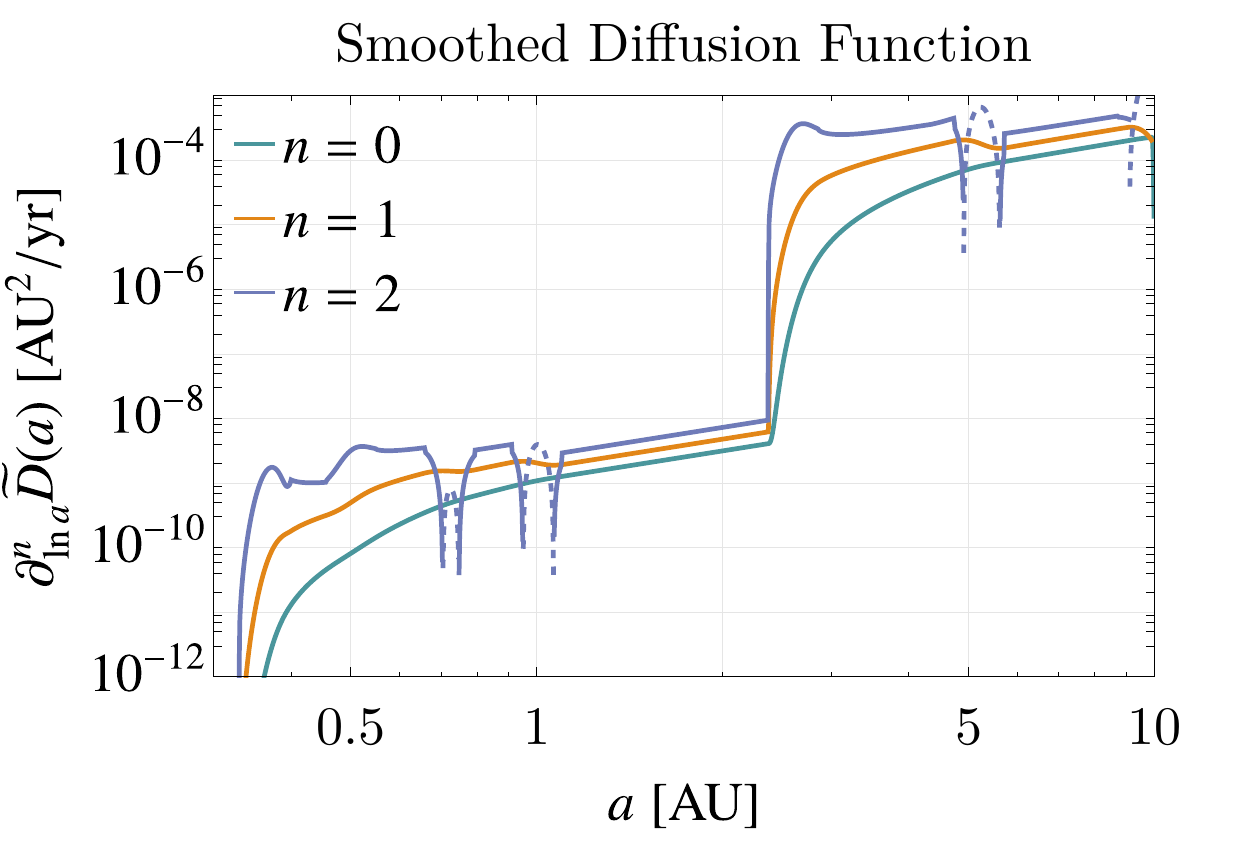}
    \caption{\textit{Left panel:} Diffusion function $D(a)$ from eq.~\ref{eq:D_diffusion} (blue-green) as a function of semi-major axis $a$, as well as its first two logarithmic derivatives (orange and purple). For $a \gtrsim 2.6\,\mathrm{AU}$, energy-changing diffusion through phase space is dominated by Jupiter. In the inner Solar System, it is driven roughly equally by Venus and Earth. \textit{Right panel:} Smoothed diffusion function $\widetilde{D}(a)$ from eq.~\ref{eq:D_tilde} employed to cure the discontinuities and divergences in the derivatives of $D(a)$. \githubicon{https://github.com/kenvantilburg/solar-basin-dynamics/blob/main/code/Stochastic_description/diffusion.nb}}
    \label{fig:diffusion}
\end{figure}

The resulting diffusion function $D(a)$ and, since they enter into eq.~\ref{eq:fokker_planck}, its first two (logarithmic) derivatives, are plotted in the left panel of figure~\ref{fig:diffusion}. The discontinuities in the derivatives at $a = R$, and the divergence of the second derivative at $a = R/2$ are not physical, as those would be regulated by motional resonances at those locations. To avoid pathological behavior in eq.~\ref{eq:fokker_planck}, we smooth the diffusion function with a Gaussian kernel in $\ln a$ space:
\begin{align}
\widetilde{D}(a) = 
\int \dd (\ln a') \, \frac{\exp \left\lbrace - \frac{(\ln a - \ln a')^2}{2\sigma_{\ln a}^2} \right\rbrace}{\sqrt{2\pi} \sigma_{\ln a}} D(a'), \label{eq:D_tilde}
\end{align}
where we use a relatively narrow smoothing of $\sigma_{\ln a} = 0.02$. 
We plot this smoothed diffusion function $\widetilde{D}(a)$ as the blue-green curve in the right panel of figure~\ref{fig:diffusion}. By eye, it is essentially indistinguishable from the unsmoothed, analytic version in the left panel, except at the sharp transition at $a \approx 2.6\,\mathrm{AU}$, near half of Jupiter's semi-major axis. The effect on the derivatives is much stronger: the first derivative (orange) is much smoother near $a = R/2$ and $a=R$ for each planet, while the second derivative's divergence at $a = R/2$ is regulated, and discontinuity at $a = R$ is replaced by a negative fluctuation (purple curve, dashed for negative values). 

\begin{figure}
\centering
        \includegraphics[width=0.7\textwidth]{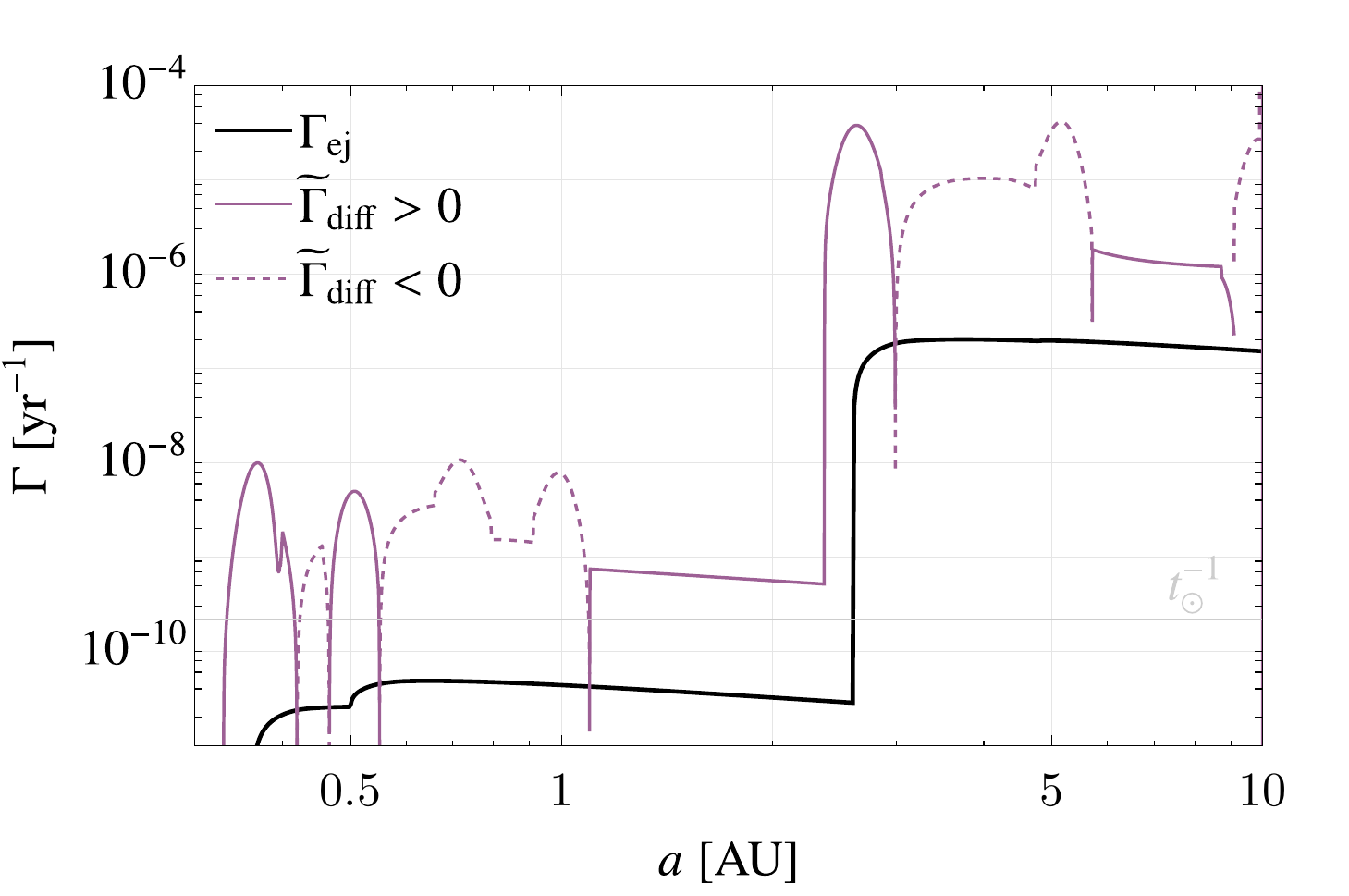}
    \caption{Ejection rate $\Gamma_\mathrm{ej}$ (black) from eq.~\ref{eq:Gamma_ej} a function of semi-major axis $a$, as well as the \emph{initial} diffusion rate $\widetilde{\Gamma}_\mathrm{diff}$ from eq.~\ref{eq:Gamma_diff} (purple). Positive values (solid curve) for the latter indicate density increases, while negative values signify decreases. Even in the inner Solar System, the diffusion rate exceeds the inverse age $t_\odot^{-1}$ of the Sun (light gray). \githubicon{https://github.com/kenvantilburg/solar-basin-dynamics/blob/main/code/Stochastic_description/diffusion.nb}}
    \label{fig:Gamma}
\end{figure}

While diffusion does not change the particle number, it does change its distribution from the injected $\Gamma_\mathrm{prod} \propto 1/a^2$. The time scale on which it does so, at least initially (before backreaction), is
\begin{align}
    \widetilde{\Gamma}_\mathrm{diff} = \frac{\partial_a^2 \left[ \widetilde{D}(a) a^{-2} \right]}{a^{-2}}. \label{eq:Gamma_diff}
\end{align}
This function is plotted in purple in figure~\ref{fig:Gamma}. Positive values indicate an \emph{increase} in the phase space density due to net diffusion into that part of phase space (typically from upscattering from more deeply bound orbits), while negative values signal a \emph{decrease} in phase space density from diffusion to other semi-major axes.

Note that in the region most relevant for the basin energy density at Earth, $0.5\,\mathrm{AU} < a < 1 \, \mathrm{AU}$, $\widetilde{\Gamma}_\mathrm{diff}$ is mostly negative and larger in magnitude than the inverse age of the Sun, so we can expect a significant reduction in the basin energy density.  We also plot eq.~\eqref{eq:Gamma_ej} as the black curve in figure~\ref{fig:Gamma} to demonstrate that the ``soft'' diffusive encounters described by $\widetilde{\Gamma}_\mathrm{diff}$ dominate over ejections when determining the basin lifetime.

\begin{figure}
        \includegraphics[width=0.9\textwidth]{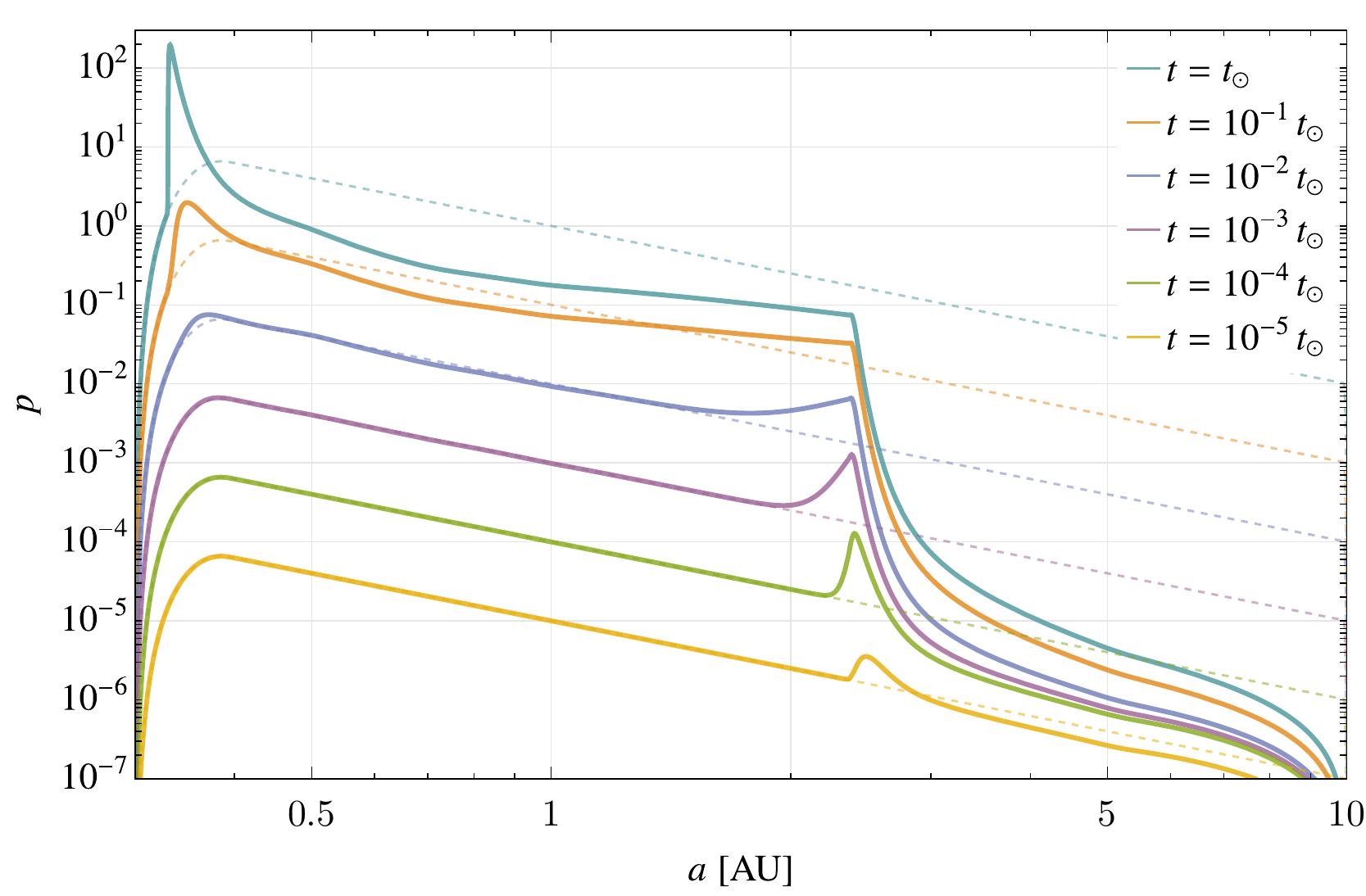}
    \caption{
    Distribution $p(a,t)$ of basin particles as a function of semi-major axis $a$, for various times $t$ equal to the $t_\odot = 4.5\,\mathrm{Gyr}$ age of the Solar System (top curve teal) and small fractions thereof (lower curves). The solid lines depict the results obtained by integrating the Fokker-Planck equation from eq.~\eqref{eq:fokker_planck}, with  the production function $\Gamma_\mathrm{prod}$ of eq.~\eqref{eq:Gamma_prod}, the ejection rate from eq.~\eqref{eq:Gamma_ej}, and the smoothed diffusion function of eq.~\eqref{eq:D_tilde}. The dashed lines neglect both ejections and diffusion. Ejections are responsible for the suppression of the basin phase space distribution for $a \gtrsim 2.6\,\mathrm{AU}$. In this simple model, diffusion acts to roughly push $a \lesssim 1\,\mathrm{AU}$ orbits to even lower semi-major axes, and $a \gtrsim 1\,\mathrm{AU}$ orbits to larger semi-major axes. Effects from motional resonances, not included here, are likely large. \githubicon{https://github.com/kenvantilburg/solar-basin-dynamics/blob/main/code/Stochastic_description/diffusion.nb}
    }
    \label{fig:fokkerplanck}
\end{figure}

We integrated the Fokker-Planck equation~\ref{eq:fokker_planck} with the above smoothed diffusion function $\widetilde{D}(a)$ and ejection rate $\Gamma_\mathrm{ej}$ from $t = 0$ to $t=t_\odot$. The relevant initial conditions are $p(a,0) = 0$, supplemented with the boundary conditions $p(a_\mathrm{min},t) = p(a_\mathrm{max},t) = 0$. The minimum and maximum semi-major axes were taken to be $a_\mathrm{min} = 0.3 \,\mathrm{AU}$ and $a_\mathrm{min} = 10\,\mathrm{AU}$, and the $S_2$ clamping function $S_2$ on those boundaries was given a width $\sigma_a = \mathrm{AU}/10$. The results are plotted in figure~\ref{fig:fokkerplanck}, where the effects of ejection and diffusion are apparent.

For a fully equilibrated phase space, the basin density at radius $R$ is proportional to:
\begin{align}
    \rho_\mathrm{b}(R) \propto \frac{1}{4\pi R^2} \int_0^1 \dd e \, 2e \int_{\frac{R}{1+e}}^{\frac{R}{1-e}} \dd a \, p(a,t).
\end{align}
If the results from integrating the Fokker-Planck equation are taken at face value, they would yield an effective basin time~\cite{VanTilburg:2020jvl,Lasenby:2020goo} of:
\begin{align}
\tau_\mathrm{eff} \simeq \frac{\int_0^1 \dd e \, 2e \int_{\frac{R}{1+e}}^{\frac{R}{1-e}} \dd a \, \Gamma_\mathrm{prod}(a)}{\int_0^1 \dd e \, f_\mathrm{prod}(e) \int_{\frac{R}{1+e}}^{\frac{R}{1-e}} \dd a \, p(a,t)} \approx \begin{cases}
    8.6 \times 10^8 \, \mathrm{yr} &  f_\mathrm{prod} = 2 e, \\
    6.0 \times 10^8 \, \mathrm{yr} &  f_\mathrm{prod} = \delta(e-1); \\
\end{cases}
\end{align}
by comparing the resulting density against the density that would have resulted without diffusion and ejections, either with ($f_\mathrm{prod} = 2 e$) or without ($f_\mathrm{prod} = \delta(e-1)$) phase space mixing. In either case, the resulting effective basin time is lower, but within a factor of two compared to the result obtained in section~\ref{sec:numerical}.

We caution that the results in this section only give a \emph{qualitative} idea of the dynamics at play, and serve at best as a quick, heuristic check on some of the relevant processes. Certain assumptions, such as the instantaneous, full equilibration of phase space, and ad hoc choices, most notably the  smoothing kernel of the diffusion function eq.~\ref{eq:D_tilde}, affect the results substantially: narrower smoothing kernels lead to larger $\tau_\mathrm{eff}$. Furthermore, the analysis here shows that there is preferential diffusion to $\emph{smaller}$ semi-major axes for particles produced on orbits with $a \lesssim 1\,\mathrm{AU}$. We can expect those particles to hit motional resonances with Venus and Earth, which are not included here but would drastically affect their subsequent evolution. Nevertheless, it is encouraging the diffusive analysis in this section gives roughly the correct ballpark answer for $\tau_\mathrm{eff}$. Since most of the phase space is \emph{not} initially produced in motional resonance, it would be hard to imagine an effective basin time significantly shorter than the timescale of phase space diffusion $\widetilde{\Gamma}_\mathrm{diff}$ from eq.~\ref{eq:Gamma_diff} and figure~\ref{fig:Gamma}.


\section{Analytical dynamics}
\label{sec:analytical}

In section~\ref{sec:numerical}, we presented our numerical simulations and their implications for the particle distribution at Earth in different BSM scenarios. In some sense, this is all we need to determine the effective lifetime of the basin. However, in order to better understand the behavior of our results and to gain confidence in the validity of our assumptions, it is worthwhile to study analytically as many aspects as possible of the dynamics of solar basin orbits.

In this section, we discuss salient dynamics of our simulations that can be understood in simple analytical terms, including Sun-crossing orbits affected by the finite size of the Sun, its oblateness, and relativistic effects (section~\ref{sec:solarphi}), as well as the dominant secular perturbations from Jupiter's (eccentric) orbit~(section~\ref{sec:kozai}). One should regard section~\ref{sec:solarphi} as a qualitative, analytical description of some effects included in (finite size of the Sun) or omitted from (oblateness, GR effects) our simulations. Similarly, section~\ref{sec:kozai} is superseded by the full secular perturbation theory of section~\ref{sec:secular}, but isolates the dominant effects from Jupiter, which are by far the most important. None of the treatment in this section is explicitly used in our headline results of sections~\ref{sec:summary}, \ref{sec:numerical}, and~\ref{sec:secular}, and instead is used to gain a broader qualitative understanding of the dynamics of the solar basin to verify our results.

We start by considering test particle orbits crossing the Sun, \textit{i.e.}~the initial conditions corresponding to solar basin orbits, and study certain aspects of their evolution analytically.  Since the Sun contains the vast majority of the Solar System's mass, test particle orbits are normally well-approximated by elliptical Kepler orbits around the Sun. Significant, sudden deviations occur when particles undergo sufficiently close encounters with planets (section~\ref{sec:close_encounters}), or cross the interior of the Sun, as we discuss below. Longer-range gravitational perturbations due to the planets slowly change the parameters of this Kepler orbit. This ``secular evolution'' has been studied extensively over the centuries; modern studies of secular evolution describe effects on the orbits of planets~\cite{Laskar_1988_secular, Morbidelli_2009,Brasser_2009}, asteroids~\cite{Scherer_1996,Gronchi_2011, Novakovic_2015, Novakovic_2016}, and satellites~\cite{Correia_2009, Lei_2020}.
However, scenarios relevant to solar basin dynamics do not appear to have been studied explicitly. In particular, since conventional objects which hit the Sun are destroyed, long-term perturbations of orbits which are or become Sun-crossing are usually not of interest.

\begin{figure}[t]
\begin{center}
\includegraphics[width=0.49\textwidth]{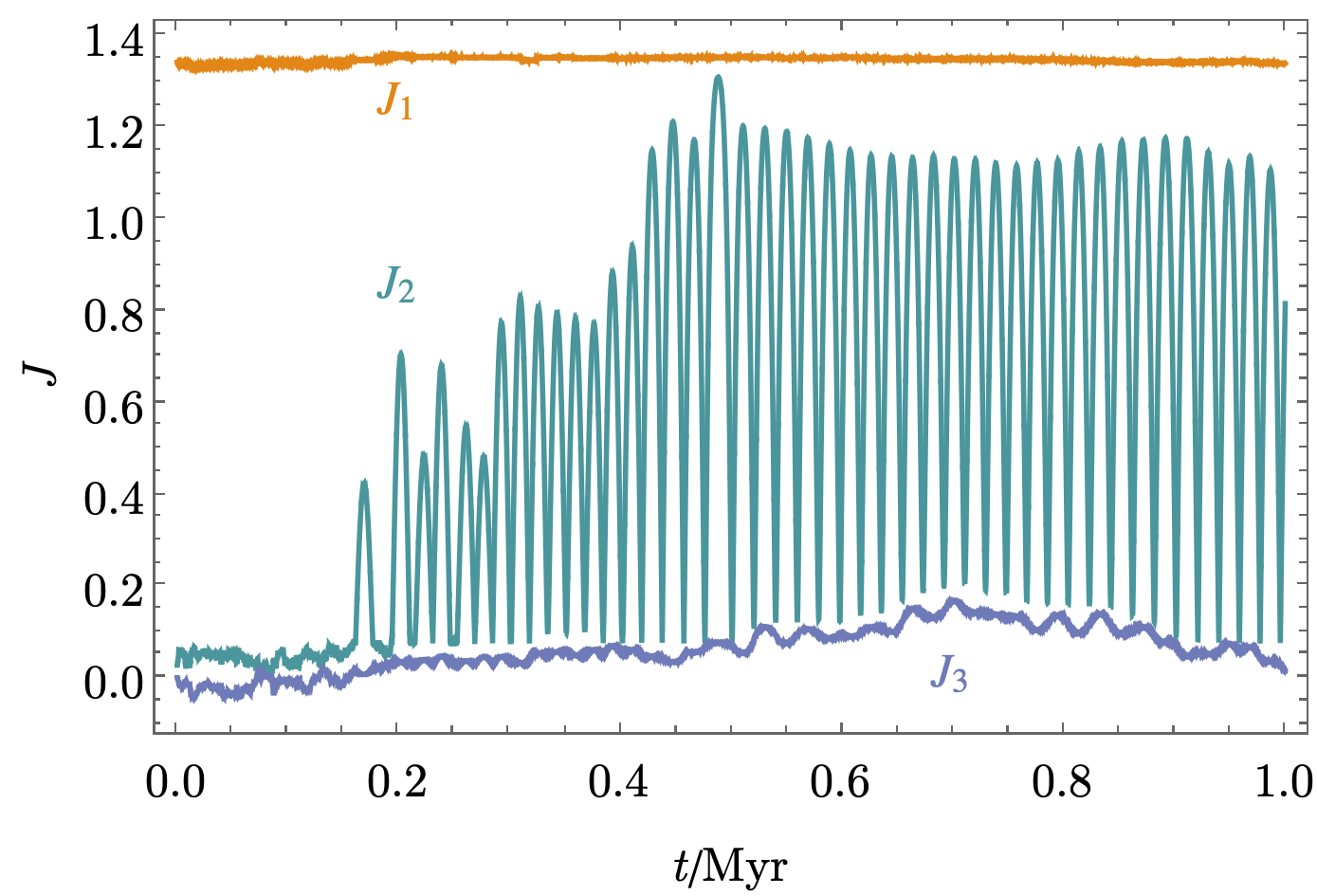}
\includegraphics[width=0.49\textwidth]{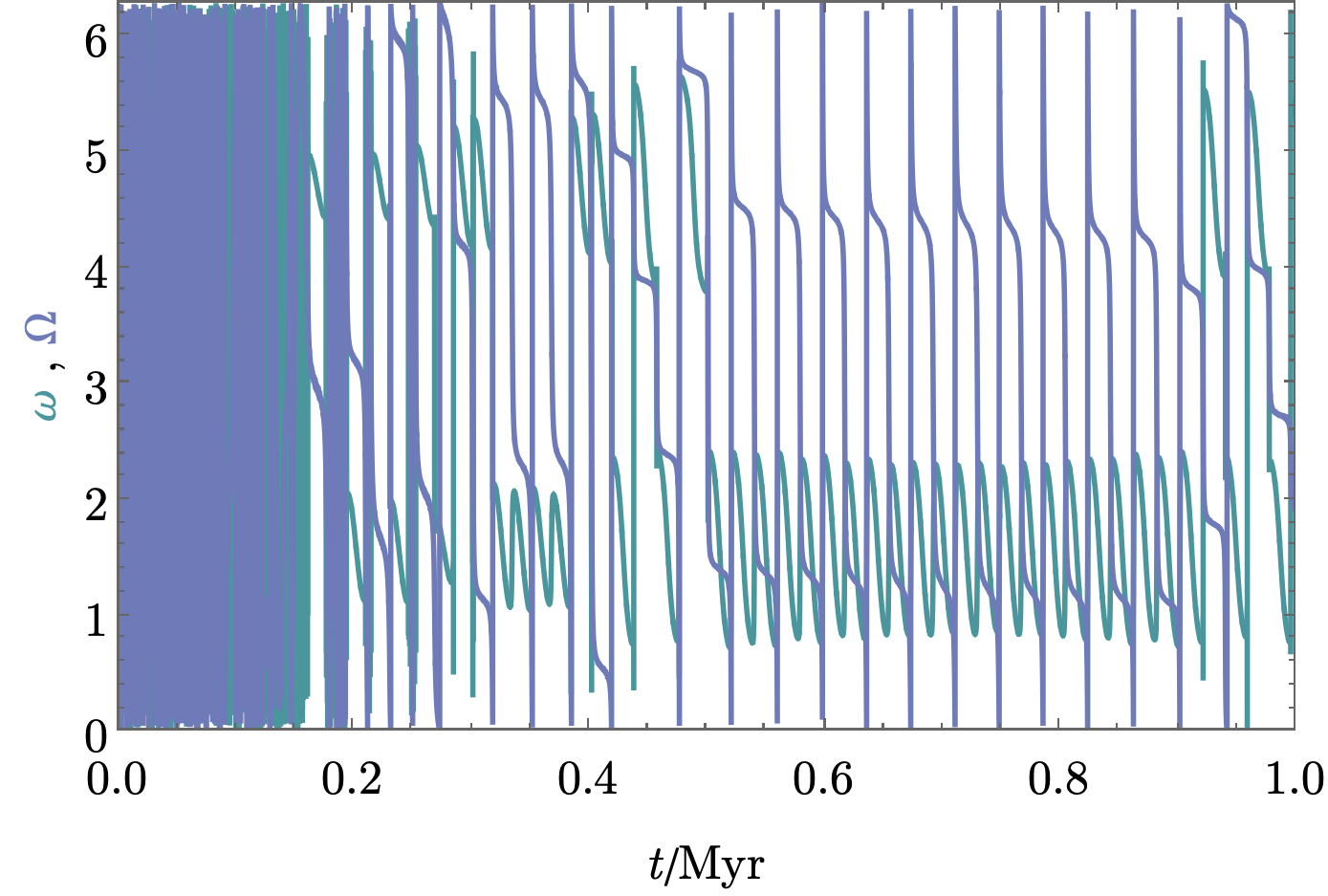}
\caption{Evolution of action (left) and angle (right) variables, in dimensionless units of eq.~\eqref{eq:action-angle-tilde} described in appendix~\ref{app_elts}, for a particular orbit starting with $a \approx 1.8 \, \mathrm{AU}$ and a perihelion inside the Sun, \textit{i.e.}~$a (1-e) < R_\odot$ and thus small $J_2$. The orbital energy---and thus $J_1\propto \sqrt{a}$---evolves stochastically, primarily due to weak close-encounter scattering with the planets (section~\ref{sec:close_encounters}), but $J_1$ changes very little overall over Myr timescales.  The orbital angular momentum, proportional to $J_2$, changes slowly for $t \lesssim 0.2\,\mathrm{Myr}$, when the particle's perihelion is still inside the Sun, due to the rapid perihelion precession (variation of $\omega$) from the Sun's non-$1/r$ potential in its interior, but undergoes Kozai oscillations thereafter. The variation of angular momentum perpendicular to the ecliptic, proportional to $J_3$, changes more slowly than the total angular momentum at all times, since it is suppressed by the planetary orbits' eccentricities. \githubicon{https://github.com/kenvantilburg/solar-basin-dynamics/blob/main/code/PaperPlots.nb}}
\label{fig_aa1}
\end{center}
\end{figure}

In figure~\ref{fig_aa1}, we show the evolution of the orbital elements for an orbit for the first Myr since production inside the Sun. We use the action-angle elements for the orbit (defined
in appendix~\ref{app_elts}), which are convenient for perturbation theory and phase space analysis. This plot illustrates a number of features:
\begin{itemize}
\item Over this timescale, the energy of the orbit ($J_1 \propto \sqrt{a}$) is almost constant.
\item For about $0.2\,\mathrm{Myr}$, the particle's perihelion is inside the Sun, and the non-$1/r$ potential inside the Sun causes fast precession of the angle of periapsis $\omega$ (see section~\ref{sec:solarphi} and figure~\ref{fig_rosette1}). This precession averages out the effects of planetary perturbations, slowing the evolution of the action variables.
\item Once the particle becomes non-Sun-crossing, the dominant form of secular evolution is an oscillation of its angular momentum $L$ over a timescale $\sim {\rm few} \times 10^4 \yr$, with corresponding evolution of $\omega$ and $\Omega$. These oscillations are mostly driven by perturbations from Jupiter, and are known as ``Kozai oscillations'' (section~\ref{sec:kozai}).
\item If Jupiter's orbit were circular, then the above Kozai oscillations would approximately preserve the angular momentum component perpendicular to the plane of Jupiter's orbit (and since Jupiter's orbit is almost in the ecliptic plane, would approximately conserve $L_z$).
However, due to the eccentricity of Jupiter's orbit, $e_\mathrm{J} \simeq 0.05$, $L_z$ does evolve, though on longer timescales than the basic Kozai cycle time (section~\ref{sec:kozai}).
\end{itemize}

\begin{figure}[t]
	\begin{center}
\includegraphics[width=\textwidth]{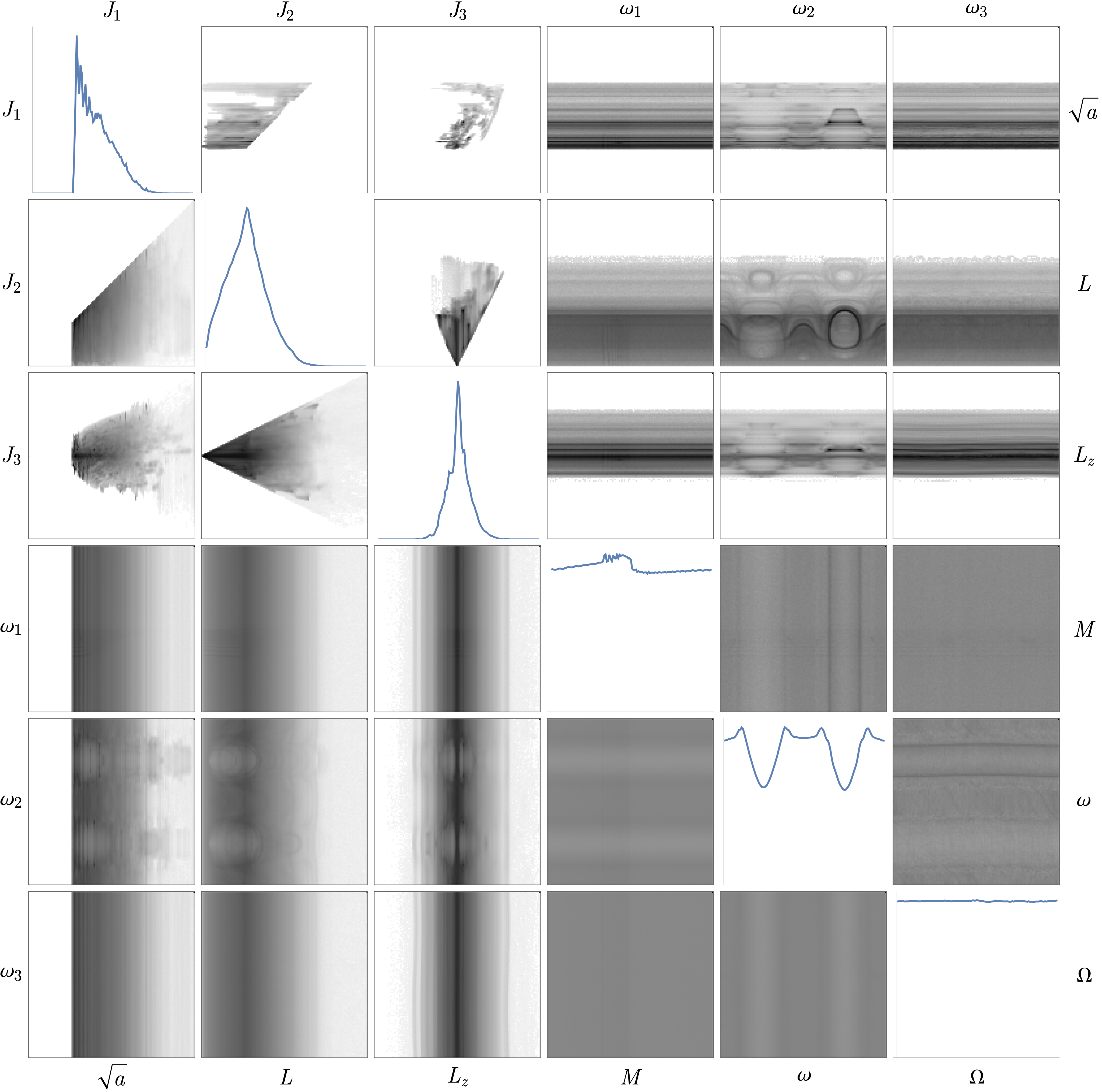}
		\caption{\emph{Upper-right panels:} Accumulated time spent
		in different phase-space bins for a specific particle
		from the forward simulations. Each panel corresponds to
		a phase-space binning in two of action-angle coordinates,
		with darker shading corresponding to larger accumulated times.
		The angle variables are plotted from $0$ to $2\pi$, while
		the action variables are plotted from
		$0$ to $2.3 \sqrt{G_N M_\odot \AU}$
		(between $\pm 2.3 \sqrt{G_N M_\odot \AU}$ for $J_3$).
		\emph{Lower-left panels:} Accumulated time spent by all 256
		forward-simulated particles in different phase space bins. 
		\emph{Diagonal panels:} Accumulated time spent 
		by all 256 particles in
		phase space bins for a single action-angle variable.  The horizontal axes correspond to the ranges of action-angle variables described above.  The units of the vertical axis are
		arbitrary.  Structure is evident for some phase space variables (\textit{i.e.}, Kozai cycles are visible in the $\omega_2$ panels), while in other phase space variables it is apparent that particles are distributed uniformly throughout phase space (\textit{i.e.}, $M$ panels). \githubicon{https://github.com/kenvantilburg/solar-basin-dynamics/blob/main/code/PaperPlots.nb}}
    \label{fig_pp1}
	\end{center}
\end{figure}

Over long timescales, the secular evolution of test particles is more complicated, with all of the orbital elements able to evolve significantly over the age of the Solar System, as illustrated for the orbital energy (the slowest-evolving orbital element) in figures~\ref{fig_asun} and~\ref{fig_aearth} and described by the formalisms of section~\ref{sec:secular} and~\ref{sec:stochastic}. The upper-right panels of figure~\ref{fig_pp1} show the accumulated time spent in different phase space regions, corresponding to pairwise combinations of the action-angle coordinates, for a single particle from the forward simulations.
If that one particle had remained on its initial Kepler orbit, then all of the elements apart from $M = \omega_1$ would have remained constant. Taking into account the precession from the non-$1/r$ potential of the Sun but ignoring planetary perturbations would imply changes only in $M$ and $\omega$.
Instead, we see that, due to planetary perturbations, all of the orbital elements change significantly over the particle's lifetime---in this case, from $4.4\,\mathrm{Gyr}$ in the past until its survival to the present---and exhibit interesting behavior through phase space.

The lower-left panels of figure~\ref{fig_pp1} show the total phase space density from all of the 256 forward runs (not just one test particle). While the initial trajectories are all at small $L$, the perturbed evolution explores the full range of orbital angular momenta. This further corroborates our finding that most of the phase space is eventually populated at similar occupation numbers, which increases the saturation density (section~\ref{sec:saturation}) and justifies the usage of the fully-equilibrated phase space (in orbital elements other than semi-major axis) in section~\ref{sec:secular}.

\subsection{Solar potential}\label{sec:solarphi}
In this subsection, we study the phenomenology of the deviations from a simple $1/r$ gravitational potential of the Sun due to its finite size, oblateness, and relativistic corrections. We will argue that finite-size corrections are important to implement for the long-term evolution of a solar basin---which is why we included the gravitational potential in the interior of the Sun in section~\ref{sec:numerical}---but that the other two effects are subdominant and can be safely neglected for our required precision. 

\paragraph*{Solar interior}
The Newtonian $1/r$ potential has the special feature that orbits are closed due to the conservation of the Laplace-Runge-Lenz vector~\cite{Goldstein_classicalmechanics}. 
A Sun-crossing particle will encounter a non-$1/r$ potential due to the finite size of the Sun, which will cause precession of the particle's perihelion. (Solar oblateness and relativistic effects cause similar effects even for non-Sun-crossing orbits, as we will show below.)

\begin{figure}[t]
	\begin{center}
\includegraphics[width=0.49\textwidth]{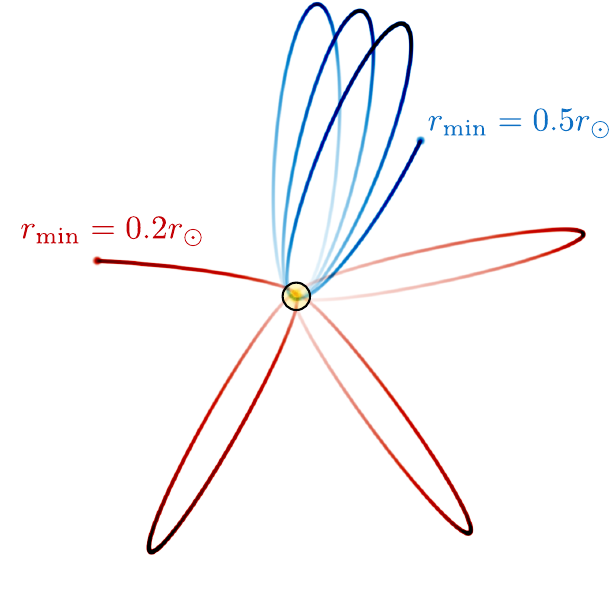}
\includegraphics[width=0.49\textwidth]{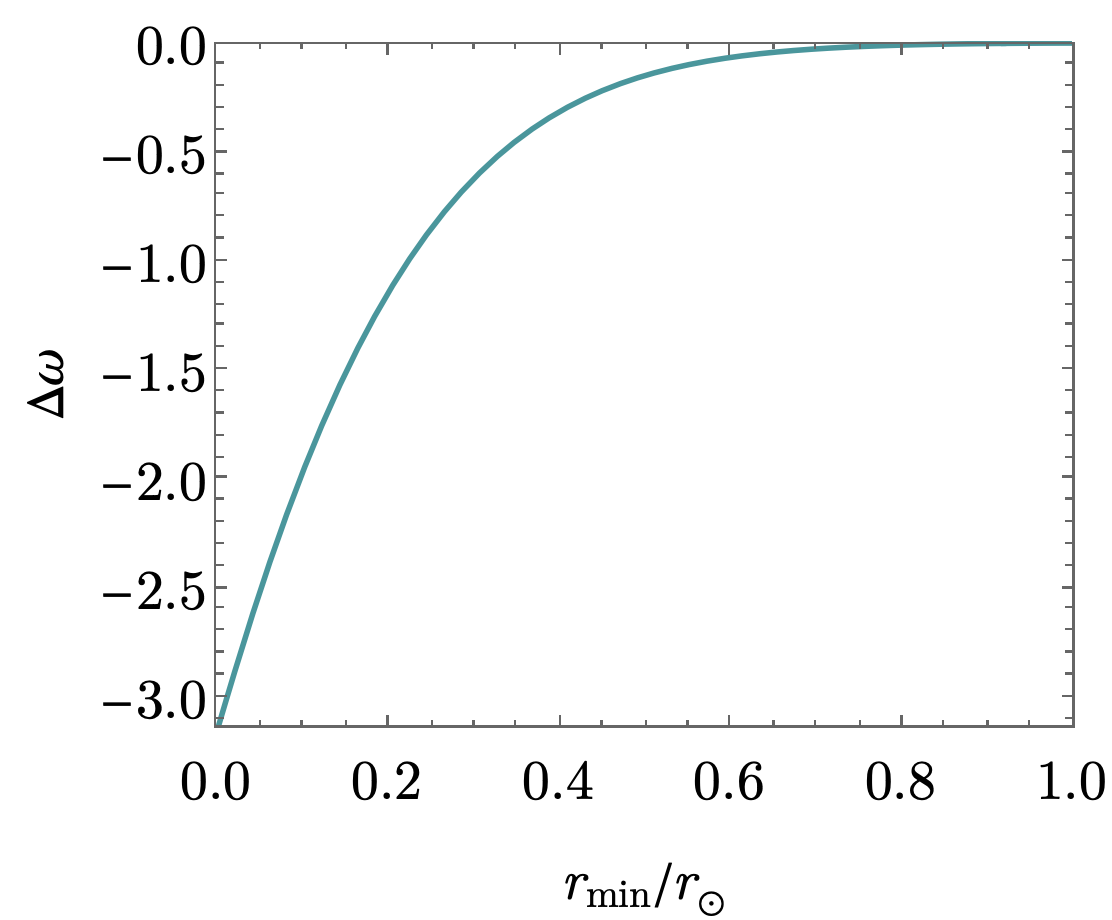}
		\caption{\emph{Left:} illustration of perihelion
		precession for Sun-crossing orbits, due to the 
		non-$1/r$ potential inside the Sun. The blue orbit 
		corresponds to $0.1$ years of evolution for a particle with semi-major axis
		$a = 0.1 \AU$ and eccentricity $e = 0.977$, giving 
		a perihelion distance of $r_{\rm min} = 0.5 \, R_\odot$ from the center of the Sun.
		The red orbit corresponds to a particle with semi-major
		axis $a = 0.1 \AU$ and eccentricity $e = 0.993$,
		giving a perihelion distance of $r_{\rm min} = 0.2 \, R_\odot$ from the center of the Sun.
		The relatively small value of $a$ is chosen to enable
		easier visualisation of the orbits.
		\emph{Right:} Perihelion precession per orbit as a function
		of perihelion distance from the center of the Sun,
		for particles with $v \simeq v_{\rm esc}$ inside
		Sun (\textit{i.e.}\ those with semi-major axis $a \gg R_\odot$).}
    \label{fig_rosette1}
	\end{center}
\end{figure}

The left-hand panel of figure~\ref{fig_rosette1} illustrates the ``rosette''-like orbital tracks of Sun-crossing orbits with small semi-major axes (very elliptical orbits with large $a$ are difficult to visualize clearly). For Sun-crossing orbits with $a \gg R_\odot$, the particle's speed inside the Sun is very close to escape velocity of the Solar System, and the amount of precession per orbit is determined by the minimum radius attained. This relationship is plotted in the right-hand panel of figure~\ref{fig_rosette1}. As expected, orbits which only just graze the Sun experience very little precession, while orbits reaching deep into the Sun can experience a precession change per orbit approaching $\pi$.

The rapid evolution of the orbit's $\omega$ parameter can have important consequences. For the secular perturbation theory calculations (reviewed below in section~\ref{sec:kozai}), averaging the secular Hamiltonian over $\omega$ renders it a function of the action variables \emph{only}, so the latter would be integrals of motion. Physically, this is because the effects of planetary perturbations are averaged out by the rapidly changing $\omega$, resulting in slower evolution of the orbit's other parameters~\cite{Damour_1999,Peter:2009mi}.

\begin{figure}[t]
\begin{center}
\includegraphics[width=0.6\textwidth]{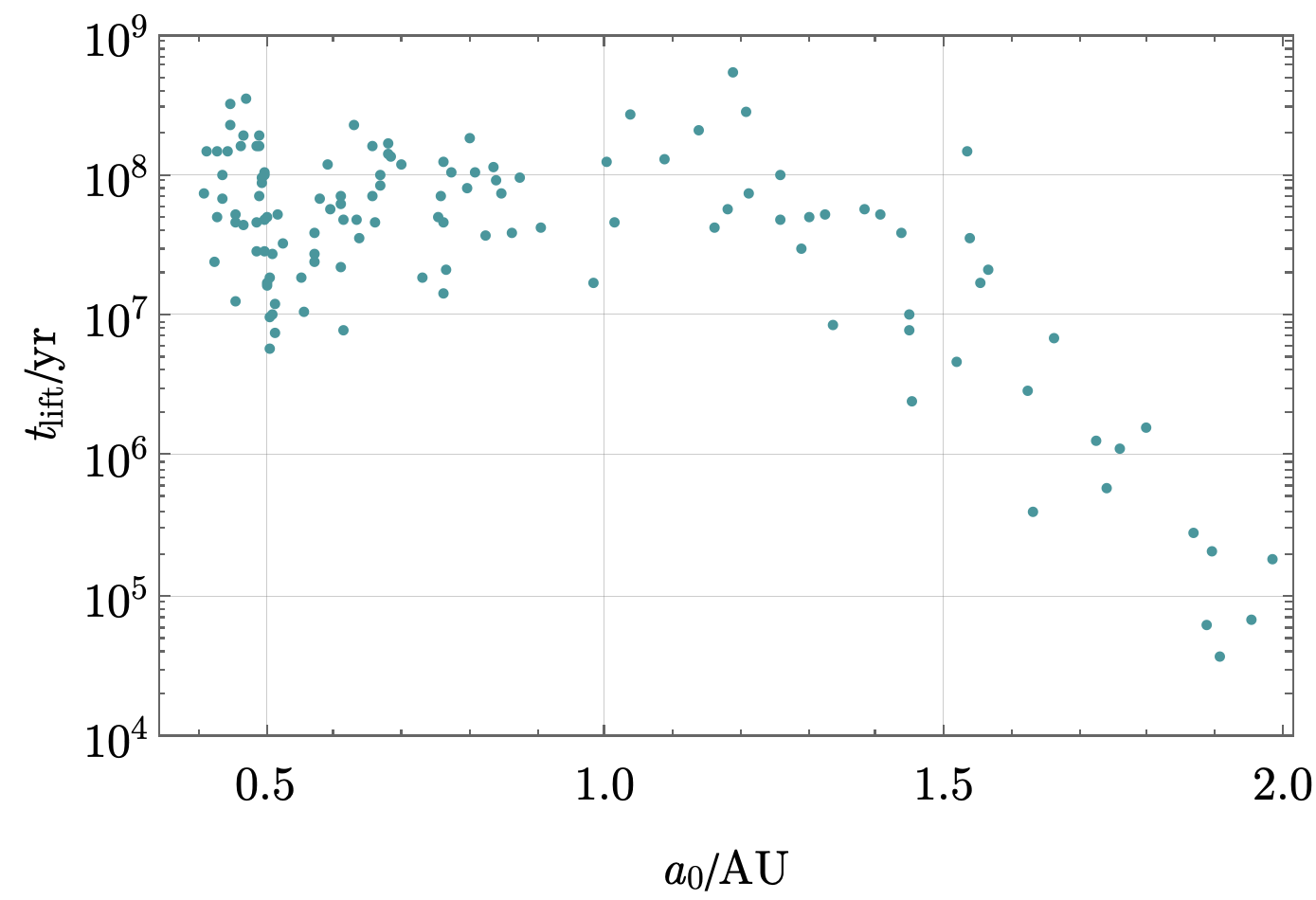}
\caption{Plot of initial semi-major axis $a_0$ against time $t_{\rm lift}$ taken for particle to become non-Sun-crossing, for 125 particles emitted from a shell of radius $0.2 R_{\odot}$ inside the Sun. The lifting process is quite efficient for $a \gtrsim 1.5\,\mathrm{AU}$ due to strong secular perturbations by Jupiter; these are much less efficient at lower semi-major axes. In the inner Solar System, the typical perihelion lifting time is $\OO(10^8\,\mathrm{yr})$, with the exception of motional resonances, where it can be much shorter (visible for orbits near $a = 0.5\,\mathrm{AU}$). \githubicon{https://github.com/kenvantilburg/solar-basin-dynamics/blob/main/code/PaperPlots.nb}}
\label{fig_lift1}
\end{center}
\end{figure}

A pertinent quantitative question for the solar basin is how long it takes planetary perturbations to ``lift'' Sun-crossing orbits into non-Sun-crossing ones, since this would be the time after which secular averaging over $\omega$ ceases and secular perturbations can alter $J_2$ and $J_3$ significantly. To investigate this, we performed a set of simulations in which we emitted particles from shells well within the Sun, and simulated them until their orbits attained perihelia sufficiently far away from the Sun. An example is shown in figure~\ref{fig_lift1}, which depicts the initial semi-major axes of these particles against the time taken for their perihelia to be lifted outside the Sun, for 125 particles emitted from a shell of radius $0.2 R_\odot$.  For initial semi-major axes $a_0 \gtrsim 1.5 \AU$, we can see that the lifting time decreases rapidly with increasing $a_0$. 

The secular evolution of the particle in figure~\ref{fig_aa1} provides an instructive example. The particle begins with $a_0 \approx 1.8 \AU$, and becomes non-Sun-crossing after $\sim 2 \times 10^5 \yr$. This lifting process is complicated, involving the joint influence of both Jupiter and the inner planets: simulations involving only Jupiter, or only the inner planets, show that the lifting process is much slower. Roughly speaking, perturbations from Jupiter ``amplify'' the secular random walk caused by perturbations from the inner planets---while Kozai oscillations (cfr.~section~\ref{sec:kozai}) are suppressed by the precession of $\omega$, these still give larger effects than inner-planet perturbations would alone. Conversely, in the absence of inner-planet perturbations, perturbations from Jupiter would average out over multiple cycles.

As we will see below in section~\ref{sec:kozai}, the characteristic rate of change of orbital elements from Jovian perturbations are suppressed by $\sim (a/a_\mathrm{J})^3$, so they are much less important at smaller $a_0$---a trend clearly visible as the larger lifting times in this regime in figure~\ref{fig_lift1}. For $a_0 \lesssim 1.5 \AU$, the lifting process is dominated by the effects of the inner planets. All of our particles become non-Sun-crossing within ${\rm few} \times 10^8 \yr$. As we showed in section~\ref{sec:sim_res}, the typical ejection time from the Solar System is $\sim 10^9 \yr$, so figure~\ref{fig_lift1} indicates that most particles will become non-Sun-crossing before they are ejected. Therefore, we expect that the non-$1/r$ potential in the solar interior does not have a significant effect on the present state of the solar basin population---in particular the temporal modulation signatures identified in section~\ref{sec:secular}---even though it substantially alters the \emph{initial} evolution of a solar basin particle produced in the Sun.

\paragraph*{Solar oblateness}
The Sun is close to a perfect sphere, with its polar radius differing from its equatorial radius by a fractional amount of $\delta \sim 10^{-5}$~\cite{Rozelot_2011}. However, since the planets have small masses compared to the Sun ($m_{\rm{Jupiter}} \approx 10^{-3} \, M_{\odot}$, and $m_{\rm{Venus}},m_{\rm{Earth}} \approx 3 \times 10^{-6} \,M_{\odot}$), the relative strengths of solar oblateness and planetary perturbations are not immediately obvious.

Outside the Sun, the dominant effect of the oblate solar mass distribution is the introduction of a quadrupole term into the gravitational potential, $\delta \Phi \sim \delta \frac{G_N M_\odot R_\odot^2}{r^3} Y_{20}$, with $\delta \sim 10^{-5}$~\cite{Rozelot_2011}, and $Y_{nm}$ are the spherical harmonics. The effects of oblateness will thus be most pronounced for orbits that pass near or cross the Sun. Since particles emitted from the Sun all start out on such orbits, this behavior is of interest.

Orbits in an almost-spherical potential have \emph{almost} conserved $J_2 = L$. More precisely, even though the solar oblatenesss breaks spherical symmetry and angular momentum is thus not conserved, the magnitude of the angular momentum $|L|$ does not undergo secular evolution in the absence of planetary perturbations. Instead, the magnitude oscillates close to its initial value, by an amount set by the deviation $\delta$ from sphericity at the location of the orbit. In contrast, the direction of the total angular momentum can drift; $\omega_2 = \omega$ and $\omega_3 = \Omega$ evolve quasi-linearly at a rate proportional to $\delta$.

To test these predictions, we evolved test particles in the potential of an oblate Sun, without planetary perturbations. Specifically, we took the potential outside the Sun\footnote{This potential could be extended to the interior of the Sun by modifying the enclosed mass function of eq.~\eqref{eq:sun_mass_function}, though since lift times are short (c.f.~figure~\ref{fig_lift1}), and because the non-$1/r$ potential terms dominate the orbital precession inside the Sun, oblateness effect are expected to be negligible.} to be
\begin{equation}
\Phi = - G_N M_\odot \left(\frac{1}{r}- \delta \frac{1}{10} \frac{R_\odot^2}{r^{3}}(3 \cos^2 \theta - 1)\right).
\end{equation}
This is the perturbation that would arise from a constant-density Sun with equatorial radius $(1 + \delta)$ times larger than its polar radius. For the real Sun, the fractional difference between these radii is $\delta \sim 10^{-5}$---the radial dependence of the density and oblateness will change the correction to $\Phi$ by an order-unity factor. For these parameters, a test particle with e.g.~a semi-major axis of $a = 1\,\mathrm{AU}$ and a small initial perihelion of $r_\mathrm{min} = 1.6\,R_\odot$ will only undergo a full cycle in $\omega$ and $\Omega$ every $2 \times 10^6\,\mathrm{yr}$, significantly slower than the Kozai cycles of section~\ref{sec:kozai} (cfr.~eq.~\eqref{eq:H_0}) over the vast majority of relevant phase space. The magnitude of $L^2$ only shows bounded peak-to-through variations of $3 \times 10^{-6}$ for this reference particle---for evolution of the action parameters, planetary perturbations will always dominate.

For low-perihelion particles with smaller $a$, the $\Omega$ evolution caused by Solar oblateness may be larger than that from Jupiter perturbations. However, since we expect the perihelion of such particles to be lifted on timescales small compared to the lifetime of the Solar System (as discussed above) and there is much stronger precession due to the solar interior potential, we do not expect solar oblateness to be significant for the overall basin population statistics. Accordingly, we did not include the effects of Solar oblateness in our main simulations of section~\ref{sec:numerical}.

\paragraph*{GR effects}
Our $N$-body simulations employ a purely Newtonian gravitational potential for the finite-size Sun, and neglect corrections from general relativity (GR). We justify the validity of this approach here.

Orbits that come close to or cross the Sun can reach fairly high speeds: $v_{\rm esc} \simeq 2 \times 10^{-3}$ at the surface of the Sun, and up to $v_{\rm esc} \simeq 4.6 \times 10^{-3}$ at its center. Since $v^2$ is comparable to the ratio between minor planet masses and the Solar mass, one might worry whether relativistic effects could be comparably important for orbital dynamics.

In a Schwarzschild metric, the general-relativistic effective potential for radial motion is 
\begin{equation}
	V(r) = - \frac{G_N M_\odot}{r} + \frac{h^2}{2 r^2}
	- \frac{G_N M_\odot h^2}{r^3},\label{eq:Schwarzschild}
\end{equation}
where $h = L / m$ is the angular momentum per unit mass,
and the final term represents the non-Newtonian contribution.
For a test particle in a bound orbit, this leads to precession of the perihelion,
by an amount
\begin{equation}
	\Delta \omega \simeq \frac{6 \pi G^2 M_\odot^2}{h^2} \label{eq:omega_GR}
\end{equation}
per orbit~\cite{Hobson_2006}.  For a mean eccentricity of $\langle e\rangle=\frac{2}{3}$, this produces a perihelion precession of roughly $10^{-6}$ radians per orbit for test particles with $a\sim\si{AU}$, with high-eccentricity orbits affected even more. The GR effect is thus somewhat larger than the effects from Solar oblateness, and furthermore affects all orbits (not just small-perihelion ones).

However, the perihelion precession both from secular perturbations and from the finite size of the Solar interior swamp the effect in eq.~\eqref{eq:omega_GR} by a considerable margin. For example, the precession of Earth's perihelion from gravitational interactions with bodies in the Solar System is roughly $12\,\mathrm{arcsec/yr}$~\cite{Standish_2006}, or roughly $6\times 10^{-5}$ radians per orbit. 

The average rate of precession from Sun-crossing orbits can be estimated by considering the fraction of particles with eccentricity $e$ such that $a(1-e)<R_{\odot}$.  The distribution function of eccentricities is expected to be $f(e)=2e$ for a fully equilibrated phase space (a reasonable approximation, cfr.~figures~\ref{fig_veldist1} and~\ref{fig_loeb1}), so the fraction of Sun-crossing particles at any one time is $2 R_\odot / a$ or about $1\%$ at $a = 1\,\mathrm{AU}$. For this small fraction of particles (at any one time in the simulation), the perihelion precession per orbit can be substantial, as shown in the right panel of figure~\ref{fig_rosette1}, and will be much larger than the GR effect of eq.~\eqref{eq:omega_GR}. 

The GR effects are therefore subdominant to two other leading causes of perihelion precession for low- and high-eccentricity orbits, and would not alter our results significantly. The incorporation of GR effects directly into our numerical simulations would have required the inclusion of velocity-dependent forces in our numerical routine, which we were unable to achieve without drastic worsening of numerical error accumulation or unacceptable speed of integration.

\subsection{Jovian secular perturbations and Kozai oscillations}
\label{sec:kozai}

Here, we will consider how the orbit of a test particle with $a \ll a_\mathrm{J} \simeq 5.2 \AU$ evolves due to the perturbing influence of Jupiter to extract the dominant secular dynamics from section~\ref{sec:secular}, which can be understood more simply in the limit that the other planets are neglected. The analysis below thus serves a useful check on the validity of our method in section~\ref{sec:secular}, and provides a partial, qualitative analytical description of the observed filling of phase space in our simulations described in section~\ref{sec:numerical} (e.g.~figure~\ref{fig_veldist1}). 

If the Jovian perturbations are small enough, then the timescales over which they change the particle's orbit will be much larger than both the particle's and Jupiter's orbital periods. Except in the case of resonances, the leading effects will be those which do not cancel upon averaging over the anomalies of both the test particle and Jupiter. Orbits can thus be treated as ellipses with appropriate mass densities and mutual gravitational interaction energies. 

\paragraph{Jupiter's quadrupole perturbation}
Expanding these interactions as a power series in $a/a_\mathrm{J}$, the lowest-order term of the secular Hamiltonian is the quadrupole~\cite{Lithwick_2011}:
\begin{align}
H_\mathrm{quad} &= H_0 F_\mathrm{quad} \\
H_0 &= J_1 \frac{3}{8} \left(\frac{G_N M_\odot}{a^3}\right)^{1/2} \frac{m_J}{M_\odot} 
\left(\frac{a}{a_\mathrm{J}}\right)^3 (1 - e_\mathrm{J})^{-3/2} \\
F_\mathrm{quad} &= \frac{1}{2}(L^2 - 1)
	+ \frac{L_z^2}{L^2} + \frac{3(1 - L^2) L_z^2}{2 L^2}
	+ \frac{5}{2} (1 - L^2)(1 - L_z^2/L^2) \cos(2\omega) ,
\end{align}
where $L \equiv \sqrt{1-e^2} = J_2/J_1$ and $L_z \equiv L \cos i = J_3/J_1$. The perturbation Hamiltonian $H_{\rm quad}$ is independent of the mean anomaly $M$ by construction, so it does not lead to evolution of the semi-major axis $a$. It is also independent of $\Omega$, so cannot cause $L_z$ evolution either. Since we have two conserved quantities, $a$ and $L_z$, and two quantities which evolve, $L$ and $\omega$, the system is integrable, and secular evolution leads to closed trajectories in $L,\omega$ space.

Since $J_1$ is conserved, we have:
\begin{align}
\dot \omega = \frac{\partial H}{\partial J_2}
= \frac{\partial(H/J_1)}{\partial L}
= \frac{H_0}{J_1} \frac{\partial F_\mathrm{quad}}{\partial L} = \frac{H_0}{J_1}  \left[L + \dots\right].
\end{align}
The timescale of $L,\omega$ evolution is thus set by the prefactor $H_0/J_1$, with a numerical value of 
\begin{equation}
\frac{H_0}{J_1} \approx 1.6 \times 10^{-5} {\rm \, yr^{-1}}
	\left(\frac{a}{\rm AU}\right)^{3/2} \label{eq:H_0}
\end{equation}
resulting in ``Kozai cycles'' with a period of order $0.1\,\mathrm{Myr}$.

\begin{figure}[t]
	\begin{center}
		\begin{tabular*}{\textwidth}{c @{\extracolsep{\fill}} cc}
			\quad\qquad Secular Quadrupole &
			Secular Octupole &
			Simulation \qquad\qquad
		\end{tabular*}\\
\includegraphics[width=0.32\textwidth]{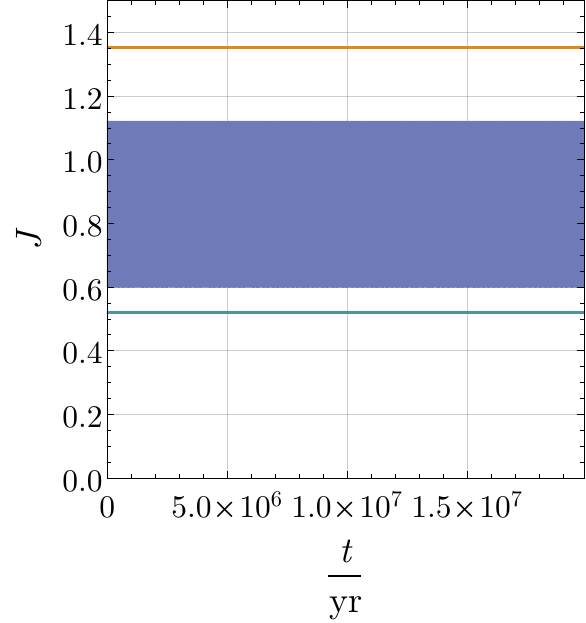}
\includegraphics[width=0.32\textwidth]{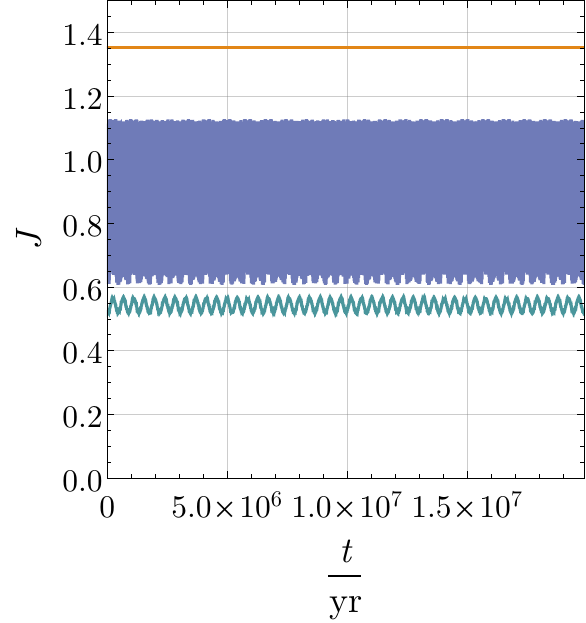}
\includegraphics[width=0.32\textwidth]{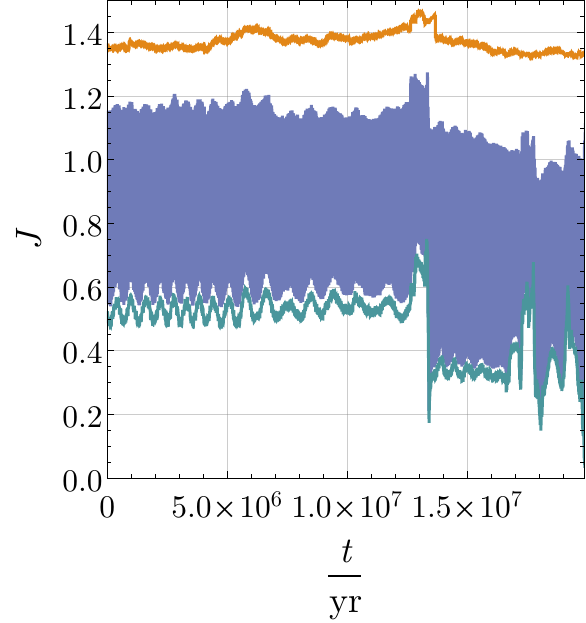}
\includegraphics[width=0.32\textwidth]{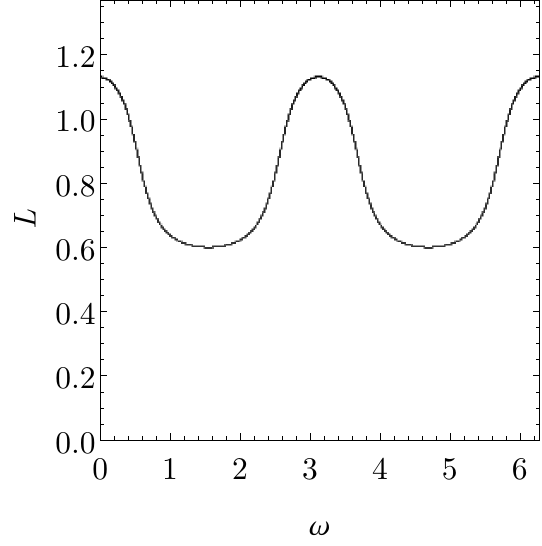}
\includegraphics[width=0.32\textwidth]{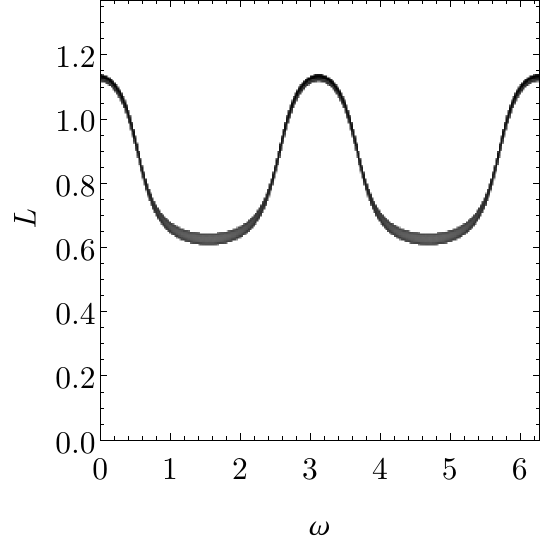}
\includegraphics[width=0.32\textwidth]{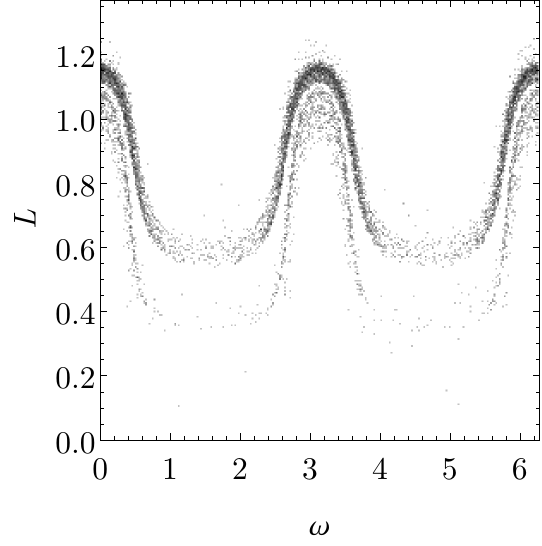}
\includegraphics[width=0.32\textwidth]{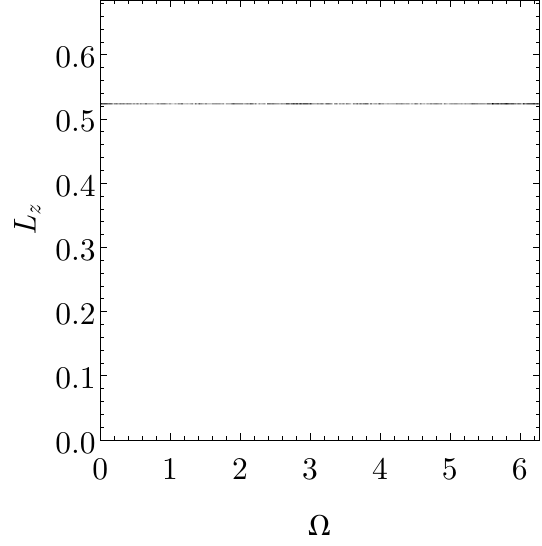}
\includegraphics[width=0.32\textwidth]{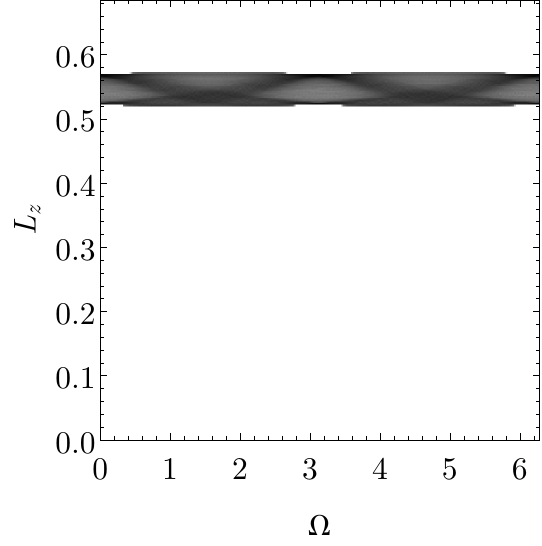}
\includegraphics[width=0.32\textwidth]{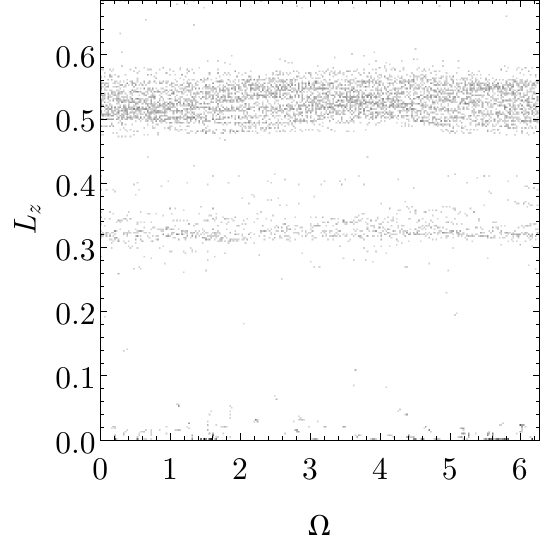}
\caption{Evolution of orbital parameters for a test particle under different approximations. The first column corresponds to secular perturbation theory evolution at quadrupole order, the second column to octupole order (both described in section~\ref{sec:kozai}), and the third column to our 4-planet numerical simulations.
The first row shows shows the evolution of the orbit's action variables $J_1, J_2, J_3$ (orange, purple, and teal curves, respectively) over a $2\times 10^7\yr$ period. The rapid oscillation of $J_2$ with a period of ${\rm few} \times 10^4 \yr$ is not resolved and appears as a solid purple range in this figure (but has a quasi-harmonic behavior as in figure~\ref{fig_aa1}. The second row shows the accumulated phase space density in different parts of $(J_2,\omega_2) = (L,\omega)$ phase space, and the third row to $(J_3,\omega_3) = (L_z,\Omega)$ phase space.
The quadrupole evolution is integrable, the octupole evolution is mildly chaotic (but bounded for these initial conditions), while the numerical integration includes situations where the orbit can undergo sudden changes due to close encounters. \githubicon{https://github.com/kenvantilburg/solar-basin-dynamics/blob/main/code/PaperPlots.nb}}
\label{fig_loeb1}
\end{center}
\end{figure}

Since $L_z = \sqrt{1-e^2} \cos i$ is conserved, these cycles involving trading off eccentricity against inclination. An eccentric orbit near the ecliptic can reduce its eccentricity by increasing its inclination, and vice versa. This provides a mechanism for the  initially highly-eccentric orbits on which particles start out, to circularize to a certain extent. Qualitatively, the $L,\omega$ evolution can take on two different forms: ``circulating'' cycles, in which $\omega$ wraps around from $0$ to $2 \pi$, and ``librating'' cycles, in which $\omega$ oscillates back and forth. The right-hand panel of figure~\ref{fig_aa1} shows, for $t \simeq 0.5$--$0.9 {\rm \, Myr}$, an example of a librating cycle, while the evolution in figure~\ref{fig_loeb1} corresponds to a circulating cycle.

\paragraph{Jupiter's octupole perturbation}
At next order in $a/a_\mathrm{J}$, there is an octupole contribution to the secular Hamiltonian
\begin{equation}
	H = H_0 \left( F_{\rm quad} + \frac{a}{a_\mathrm{J}} \frac{e_\mathrm{J}}{1 - e_\mathrm{J}^2} F_{\rm oct}\right),
\end{equation}
where $F_\mathrm{oct}(J,J_z,\omega,\Omega)$ is an order-unity function with dependence on $\Omega$~\cite[eq.~11]{Lithwick_2011}. 
Since $H$ depends on $\Omega$, $L_z$ is no longer constant at the octupole order, so the system is no longer integrable and chaotic behavior can occur.
The strength of this non-integrability is controlled by $\epsilon \equiv \frac{a}{a_\mathrm{J}} \frac{e_\mathrm{J}}{1 - e_\mathrm{J}^2} \approx 0.01 \frac{a}{\rm AU}$ (since $e_\mathrm{J} \approx 0.05$~\cite{williams2021jupiter}). 
Ref.~\cite{Li_2014} showed that the threshold for chaotic behavior is $\epsilon \simeq 0.01$, so we can expect a large fraction of the Earth-crossing basin energy density to exhibit this chaotic phase space evolution. We illustrate this for a specific orbit in  figure~\ref{fig_loeb1}, which compares the evolution under the quadrupole Hamiltonian to that including the octupole term. While the quadrupole evolution is restricted to a definite surface in phase space, the octupole evolution explores a restricted but non-negligible volume---in particular, including a finite $L_z$ range.

\medskip

In summary, over most of the relevant (Earth-crossing) phase space of the solar basin, the dominant effect over ``short'' time scales ($\ll \mathrm{Gyr}$) is the quadrupole perturbation $H_\mathrm{quad}$ by Jupiter due to its large mass and because the semi-major axis ratio $a/a_\mathrm{J}$ is not tiny. The octupole perturbation introduces chaos and smears out the secular evolution in the angular momentum components. 
Jupiter's dominance is illustrated in figure~\ref{fig_loeb1}, which compares the evolution of a particular test particle under secular evolution from Jupiter only, to the evolution of the same particle in our full 4-planet simulation, over a $2 \times 10^7 \yr$ interval. Over short timescales, the full simulation is well-approximated by the secular evolution of this section. Over longer timescales, extra perturbations accumulate and move the orbit away from the secular solution---most dramatically, close encounters (section~\ref{sec:diffusion}) can suddenly change the orbital parameters, moving the orbit to a new part of orbital phase space.

The secular perturbation theory investigated here neglects numerous effects. Even for interactions  with Jupiter only, it does not capture motional resonances, and the perturbative expansion is not under control beyond the regime of $a / a_\mathrm{J} \ll 1$. Multiplanetary secular perturbations will introduce further diffusion through phase space at the characteristic frequencies $f_l$ and $g_l$ introduced in the treatment of eqs.~\ref{eq:sec:1}--\ref{eq:sec:8}. 

For our purposes, the main conclusion regarding these secular perturbations is that test particles explore an $\OO(1)$ range in $L$, $\omega$, and $\Omega$ over timescales parametrically shorter than those of gravitational ejection or diffusion.

\section{Conclusions}
\label{sec:conclusions}

In this work, we have conducted simulations of test particle orbits within a model four-planet Solar System over its entire history of $4.5\,\mathrm{Gyr}$. While the physical world is significantly more complicated, we have argued in sections~\ref{sec:close_encounters} and~\ref{sec:analytical} that the simulation setup described in section~\ref{sec:setup} should account for the most important features affecting the statistical properties of the solar basin's orbital evolution. We found that the orbital dynamics are governed by a complicated interplay of initially non-Keplerian orbits through the solar interior (section~\ref{sec:solarphi}), relatively rapid secular perturbations by Jupiter's orbit made chaotic on longer timescales by Jupiter's eccentricity (section~\ref{sec:kozai}) and secular perturbations by the other planets (section~\ref{sec:secular}). On even longer timescales (but shorter than the age of the Sun), energy-changing processes also occur, primarily driven by many quasi-random gravitational scattering encounters, leading to a diffusive evolution of the test particles' semi-major axis~(section~\ref{sec:diffusion}) over most of the solar basin's phase space (away from motional resonances). While that treatment could not encapsulate the full long-term evolution, primarily due to the absence of motional resonances and the (weakly-violated) phase space equidistribution assumption, we found it to be a good qualitative descriptor for the rates of change in semi-major axes.\footnote{We surmise that the methods of section~\ref{sec:stochastic} may be used for rough determinations of the evolution and survival timescales of stellar basins around other stars.} While we can grasp the essence of all of the above effects in isolation, their full combination on the long-term evolution can only be studied via direct numerical integration, whose results we show in section~\ref{sec:sim_res}.

Our simulations also shed light on other physical scenarios, such as capture of halo DM particles, via gravitational scattering with the planets or by nongravitational scattering in the Sun. Our results show that previous analyses~\cite{Lundberg:2004dn, Damour:1998rh, Peter:2009mi, Peter:2009mm}, which assumed that large volumes of phase space would remain unmixed over the lifetime of the Solar System (the ``hole'' proposed in~\cite{1991ApJ...368..610G}), ignored important physical effects. Mixing is efficient enough to connect the unbound halo with almost all of the Earth-crossing velocity space, as figures~\ref{fig_veldist1} and~\ref{fig_ejv} indicate. These results may have some implications for DM-related phenomena that are enhanced at low velocities, such as those of refs.~\cite{Peter:2009mi,Essig:2022dfa,Berlin:2019uco,VanTilburg:2024xib,Iles_2024}, but we leave a detailed analysis to future work.

The phenomenological implications of our work are presented in the executive summary of section~\ref{sec:summary}, but we will recount the headline results here. The effective solar basin accumulation time is $\tau_{\rm eff} = 1.20\pm 0.09 \,\mathrm{Gyr}$.  We estimate that effects due to \emph{not} including solar evolution, solar oblateness, GR effects, and exclusion of planets (Mercury, Mars, Neptune, and Uranus) lead to a lower systematic error than the statistical error reported in our determination of $\tau_{\rm eff}$. When saturation effects (section~\ref{sec:saturation}) are negligible, i.e.~at sufficiently low coupling, the computation of the present-day solar basin density is then $\rho_\mathrm{b} \simeq \dot{\rho}_\mathrm{b} \tau_\mathrm{eff}$ with $\dot{\rho}_\mathrm{b}$ the present-day solar basin density production rate of the BSM particle under consideration, as calculated in e.g.~refs.~\cite{VanTilburg:2020jvl, Lasenby:2020goo, Berlin:2021kcm, DeRocco:2022jyq}. 

Using this new result for the effective solar basin lifetime, we can now robustly exclude a wide range of parameter space of kinetically-mixed dark photons (figure~\ref{fig_epsDP}) and axion-like particles coupled to electrons (figure~\ref{fig_gaee}), independent of cosmology. For dark photons in particular, our recasted bounds from DM experiments are the most stringent irreducible bounds on the parameter space around the mass range $\SI{10}{eV}-\SI{e3}{eV}$. These constraints are especially important given the dearth of phenomenologically viable early-universe production mechanisms for dark photons~\cite{East:2022rsi} (although see ref.~\cite{Cyncynates:2023zwj} for loopholes). We also found characteristic annual and semi-annual fractional modulation of the solar basin density with amplitudes of 6.5\% and 2.2\%, respectively, and known phases (figure~\ref{fig:modulation_real} and eq.~\ref{eq:modulation_real_2}). This temporal variation calls for dedicated analysis strategies---especially in experiments that have a significant level of background events---and may be leveraged for a discovery of a particle beyond the Standard Model through its solar basin.


\acknowledgments{We thank Asher Berlin, Andrea Caputo, Alexander Dittmann, Andrei Gruzinov, David Hogg, Yuri Levin, and Katelin Schutz for helpful conversations. We thank Andrea Caputo and Katelin Schutz for comments on our manuscript.
Some of the computing for this project was performed on the Sherlock and Farmshare clusters. We would like to thank Stanford University and the Stanford Research Computing Center for providing computational resources and support that contributed to these research results.
RL's research is supported in part by the National Science Foundation under Grant No.~PHYS-2014215, and the Gordon and Betty Moore Foundation Grant GBMF7946.  This material is based upon work supported by the NSF Graduate Research Fellowship under Grant No.~DGE1839302.  
This material is based upon work supported by the National Science Foundation under Grant No.~PHY-2210551. 
This research was supported in part by Perimeter Institute for Theoretical Physics. Research at Perimeter Institute is supported by the Government of Canada through the Department of Innovation, Science and Economic Development and by the Province of Ontario through the Ministry of Research, Innovation and Science.  This research was supported in part by grant NSF PHY-2309135 to the Kavli Institute for Theoretical Physics (KITP).}

\appendix

\section{Orbital elements and action-angle variables}
\label{app_elts}
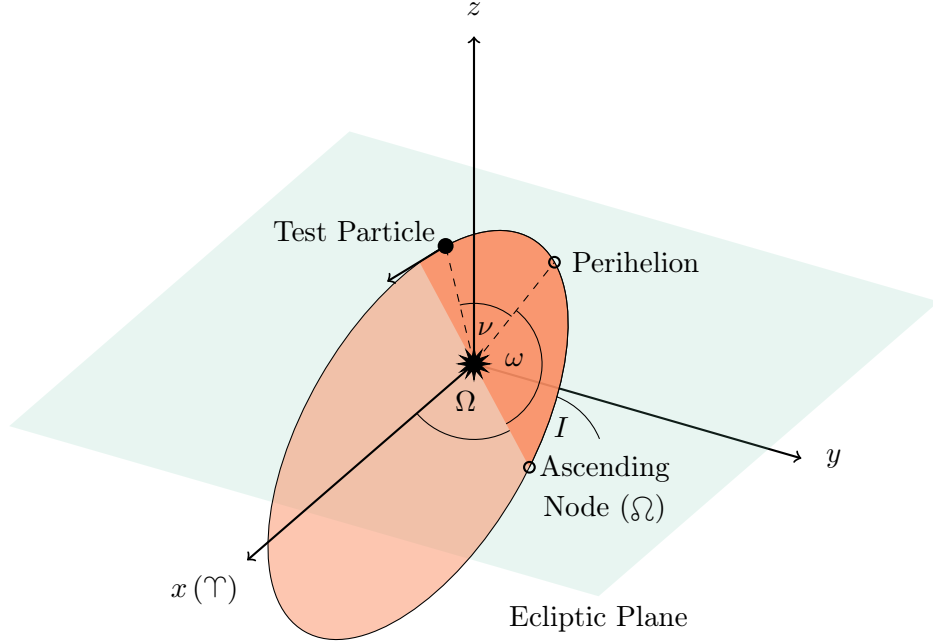
\begin{figure}[t]
    \centering
    \tdplotsetmaincoords{60}{120}

    \begin{tikzpicture}[tdplot_main_coords]

        \coordinate (O) at (0,0,0);
        \draw[thick,->] (0,0,0) -- (0,5,0) node[right=.5]{$y$};

        \coordinate (A) at (-4.5,-4.5,0);
        \coordinate (B) at (4.5,-4.5,0);
        \coordinate (C) at (4.5,4.5,0);
        \coordinate (D) at (-4.5,4.5,0);
        \coordinate (Dplus) at (-4.5,4.5,6);
    
        \fill[eclipticcolor, opacity=0.15] (A) -- (B) -- (C) -- (D) -- cycle;

        \node[yshift=-.3cm] at (C.south) {Ecliptic Plane};

        \coordinate (center) at (2,.3,0); 
        \begin{scope}
            \clip (A) -- (C) -- (Dplus) ;
            \filldraw[fill=orbitcolor, draw=black, fill opacity=0.8,rotate=60] (center) ellipse [x radius=3cm, y radius=1.5cm];
        \end{scope}
        \filldraw[fill=orbitcolor, fill opacity=0.5,draw=black, rotate=60] (center) ellipse [x radius=3cm, y radius=1.5cm];

        \draw[thick,->] (0,0,0) -- (6,0,0) node[below left=-0.1]{$x \,(\vernal)$};
        \draw[thick,->] (0,0,0) -- (0,0,5) node[above=-.2]{$z$};

        \path[name path=ellipsepath] (center) ellipse [x radius=3cm, y radius=1.5cm, rotate around={60:(center)}];
        \path[name path=planepath] (A) -- (C);
        \path [name intersections={of=ellipsepath and planepath, by={I1,I2}}];
        \draw[rotate around={28:(center)},draw=none] (I2) -- ++(-4,0) coordinate (inclinationline);
        \pic [draw=black,angle radius=1cm, "$I$"  {yshift=.1cm}] {angle = D--I2--inclinationline};

        \draw[thick] (I2) circle [radius=2pt];
        \node[align=center][anchor=west,yshift=-.3cm] at (I2) {Ascending\\Node ($\ascnode$)};

        \path[name path=particletrajectory] (0,0) -- (-4,-3);
        \path[name intersections={of=ellipsepath and particletrajectory, by={particlepoint}}];
        \fill[black] (particlepoint) circle [radius=3pt];
        \draw[->,thick,rotate around={0:(center)}] (particlepoint) -- ++(1.2,-.2);
        \draw[dashed](O) -- (particlepoint);
        \node[left,yshift=0.2cm] at (particlepoint.north) {Test Particle};

        \coordinate (xpoint) at (1,0,0);
        \pic [draw=black,angle radius=1cm, "$\Omega$"  {yshift=.1cm}] {angle = xpoint--O--I2};

        \path[name path=majoraxis](center) -- ++(-5,-.1,1);
        \path [name intersections={of=ellipsepath and majoraxis, by={perihelion}}];
        \draw[dashed](O) -- (perihelion);
        \draw[thick] (perihelion) circle [radius=2pt];
        \node[right][xshift=.1cm] at (perihelion.east) {Perihelion};
    
        \pic [draw=black,angle radius=.9cm, "$\omega$"  {yshift=.1cm}] {angle = I2--O--perihelion};

        \pic [draw=black,angle radius=.8cm, "$\nu$"  {xshift=.05cm}] {angle = perihelion--O--particlepoint};

        \node[star, star points=12, star point ratio=2.25, minimum size=12pt, inner sep=0pt, fill=black, draw=black] at (O) {};
        
    \end{tikzpicture}
    \caption{Diagram of angular orbital elements for a Keplerian orbit.  The reference direction, $\vernal$, is defined along the $x$-axis.  The test particle orbit is shown as an ellipse.  The longitude of the ascending node, $\Omega$, is the angle between $\vernal$ and the ascending node, the latter denoted $\ascnode$ and defined as the point where the particle passes upwards through the ecliptic plane.  The inclination of the orbit is $I$.  The perihelion (the point of the orbit closest to the sun) is indicated on the diagram---$\omega$ is the argument of periapsis, or the angle between the ascending node and perihelion, and the mean anomaly, $\nu$, is the angle between the test particle's current position and the perihelion.}
    \label{fig_elts}
\end{figure}

An orbit in a Keplerian ($\propto 1/r$) potential can be
parameterised by 5 orbital elements which determine
the shape of the orbit, along with a 6th element,
the ``anomaly'', which determines the particle's position along the orbit.
The standard set of orbital elements is
\begin{itemize}
	\item $a$, the semi-major axis,
	\item $e$, the eccentricity,
	\item $I$, the inclination, defined
		as the angle between the ecliptic plane
		and the particle velocity at the ascending 
		node,
	\item $\Omega$, the longitude of the ascending node,
		defined as
		the angle between a reference direction in the ecliptic
		plane and the ascending node,
	\item $\omega$, the argument of periapsis,
		defined as the angle between the ascending node and the
		perihelion
		(in the sense defined by the particle's angular velocity),
	\item $\nu$, the true anomaly, defined
		as the angle between the perihelion 
		and the particle's position (in the sense defined
		by the particle's angular velocity).
\end{itemize}
The angular elements, $I,\,\Omega, \,\omega$ and $\nu$, are illustrated in figure~\ref{fig_elts}.  

In a pure $1/r$ potential, all of the orbital elements are constant
apart from $\nu$. Instead of $\nu$, the `mean anomaly' $M$
is often used; this is a fictitious angle which increases at
a constant rate with time, increasing by $2\pi$ every orbital period.

For dynamical purposes, it is often more convenient
to use action-angle variables. For a gravitational potential
$\Phi = - G_N M_\odot / R$, these are given by
\begin{alignat}{3}
        J_1 &= m \sqrt{G_N M_\odot a}, \qquad & \omega_1 &= M, \\
	J_2 &= L, \qquad  & \omega_2 &= \omega, \\
        J_3 &= L_z, \qquad & \omega_3 &= \Omega;
\end{alignat}
where $L$ is the total angular momentum of the particle (\emph{not}
per unit mass), $L_z$ is the angular momentum component
in the $z$ direction, $m$ is its mass, and the orbital elements
are as above. Since we are considering test particles,
it will be helpful to divide the action variables by the particle mass
and define $\tilde J_i \equiv J_i / m$.
In terms of the orbital elements, this gives
\begin{equation}
	\tilde J_1 = \sqrt{G_N M_{\odot} a}; 
	\qquad
	\tilde J_2 = \tilde L = \tilde J_1 \sqrt{1 - e^2};
	\qquad
	\tilde J_3 = \tilde L_z = \tilde J_2 \cos I. \label{eq:action-angle-tilde}
\end{equation}
If we define $\tilde H \equiv H / m$,
then 
$\dot \omega_i = \partial H / \partial J_i = \partial \tilde H / \partial
\tilde J_i$,
so $\tilde J_i, \omega_i$ are action-angle variables for the
(dimensionless) Hamiltonian $\tilde H$.
In the text, we will usually elide the tildes on the $\tilde J_i$.

\section{Estimates of basin density from simulations}\label{app:basin_density}
In this appendix, we describe how we use our simulation data to estimate the solar basin density at Earth.

\subsection{Statistical estimators of effective basin lifetime} \label{app:density_estimators}
In the small-coupling limit where re-absorption can be neglected, we use eqs.~\ref{eq_tball} and~\ref{eq_tau_eff_forward} to estimate $\tau_\mathrm{eff}$ from the forward simulations, which in turn allows for solar basin density estimates at Earth via eq.~\ref{eq:tau_eff}. Similarly, eq.~\ref{eq_tau_eff_backward} is the statistic used to estimate $\tau_\mathrm{eff}$ from the backward simulations. We provide derivations of eqs.~\ref{eq_tball},~\ref{eq_tau_eff_forward}, and~\ref{eq_tau_eff_backward} below.

\paragraph*{Forward runs}
Here we derive eq.~\ref{eq_tball} for the average time $\bar{t}_{\rm{ball}}(R)$ spent per particle in a ball of radius $r_{\rm{ball}}$ located a distance $R$ from the Sun. First, we consider the one-dimensional (radial) probability density that a particle on an orbit with semi-major axis $a$ is at a distance $R$ from the Sun:
\begin{equation}
    p_1(R,a) = \frac{C_1}{v_R}.
\end{equation}
Above, $v_R=\sqrt{2G_N M\left(\frac{1}{R}-\frac{1}{2a}\right)}$ is the particle's radial velocity magnitude, assuming orbits are very nearly radial.  Normalizing the integral of $p_1(R,a)$ from $R = 0$ to $R = 2 a$ to unity fixes $C_1$ and yields
\begin{equation}
    p_1(R,a) = \frac{1}{\pi a}\frac{1}{\sqrt{\frac{2a}{R} - 1}}.
\end{equation}
The 3-dimensional probability density that a particle on an orbit with semi-major axis $a$ is at a distance $R$ from the Sun is $p_1(R,a)$ divided by the area of the spherical surface at that radius:
\begin{equation}
    p_3(R,a) = \frac{1}{4\pi R^2} p_1(R,a).
\end{equation}
This can be integrated against the distribution of semi-major axes to find the 3-dimensional probability density that a particle is at a radius $R$ from the Sun, $p_3(R)$. With our assumptions in section~\ref{sec:numerical}, that distribution is 
\begin{equation}
    f(a) = \frac{C_f}{a^2}\Theta(a-a_{\rm{min}})\Theta(a_{\rm{max}} - a),
\end{equation}
where $a_{\rm{min}}$ and $a_{\rm{max}}$ describe the range of semi-major axes of the particle population, and $C_f = \frac{1}{1/a_{\rm{min}} - 1/a_{\rm{max}}}$ so that the integral of $f(a)$ over all $a$ is unity.  Assuming $a_{\rm{min}} < \frac{R}{2}$, the population-averaged 3-dimensional probability density is
\begin{align}
    p_3(R) &= \int_0^{\infty} \dd a \,f(a) p_3(R,a) 
    = \int_{\frac{R}{2}}^{a_{\rm{max}}} \dd a \,\frac{C_f}{a^2} \frac{1}{4\pi R^2} \frac{1}{a\pi}\frac{1}{\sqrt{\frac{2a}{R} - 1}} 
    \simeq \frac{3 C_f}{8\pi R^4},
\end{align}
where in the last equality we also assumed $a_{\rm{max}} \gg R$.

The probability that a particle at radius $R$ from the Sun is located inside a ball of radius $r_{\rm{ball}} \ll R$ is then
\begin{align}
    p_{\rm{ball}} &\simeq p_3(R) \frac43 \pi r_{\rm{ball}}^3
    = \frac12 \frac{r_{\rm{ball}}^3}{R^4}\frac{1}{1/a_{\rm{min}} - 1/a_{\rm{max}}}.
\end{align}
Finally, the average time spent in in a ball of radius $r_{\rm{ball}}$ located a distance $R$ from the Sun is given by the probability of a particle being located in this region, multiplied by the total simulation time $t_f$:
\begin{equation}
    \bar{t}_{\rm{ball}} = t_f p_{\rm{ball}} = \frac12 \frac{r_{\rm{ball}}^3}{R^4}\frac{t_f}{1/a_{\rm{min}} - 1/a_{\rm{max}}},
\end{equation}
which is exactly eq.~\ref{eq_tball} from the main text. 

The \emph{realized} average time spent in $r_\mathrm{ball}$ in the forward simulations is $\hat{t}_\mathrm{ball} = \sum_p \hat{t}_{\mathrm{ball},p} / N_p$. The ratio $\hat{t}_\mathrm{ball} / \bar{t}_\mathrm{ball}$ is thus a direct measure of the terrestrial solar basin density ratio with and without perturbations, and thus of $\tau_\mathrm{eff} / t_\odot$. This yields our statistical estimator $\hat{\tau}_\mathrm{eff}$ from the forward simulations in eq.~\ref{eq_tau_eff_forward}.

\paragraph*{Backward runs}
To estimate the effective basin lifetime in the backward simulations, we first compute the average occupation number $\bar{f}_0$ expected in absence of planetary perturbations, and then give a statistical estimator for the average occupation number $\hat{f}_0$ across all particles in the simulations. In what follows, we assume production occurs only in a single shell at $R_\mathrm{prod}$ (cfr.~eq.~\ref{eq:delta_f}) and disregard re-absorption; the latter assumption will be relaxed in appendix~\ref{app:sat_density}. The ratio of the realized average occupation number with perturbations and expected average occupation number without perturbations is $\hat{\tau}_\mathrm{eff} / t_\odot = \hat{f}_0 / \bar{f}_0$, thus establishing eq.~\ref{eq_tau_eff_backward}.

The average expected occupation number over velocity phase space is
\begin{align}
    \bar{f}_0 = \frac{1}{\mathcal{V}_R} \int_{\mathcal{V}_R} \dd^3 v \, f_0(\vect{v}), \label{eq:avg_f0_unperturbed}
\end{align}
with $\mathcal{V}_R = (4\pi/3) v_{\mathrm{esc},R}^3$ the total velocity phase space volume and $v_{\mathrm{esc},R}$ is the escape velocity at a distance $R$ from the Sun. Spherical symmetry (in absence of the planets) dictates that the phase space density function only depends on the magnitudes of the radial and transverse velocities, $v_R$ and $v_\perp$, respectively:
\begin{align}
    f_0(\vect{v}) = f_0(v_R,v_\perp) \simeq N_\mathrm{cross}(v_R) \delta f(v_\perp).
\end{align}
Here $N_{\rm{cross}}(v_R)$ is the number of times a particle with radial velocity $v_R$ crosses a shell with radius $R$, and $\delta f(v_\perp)$ is the change in occupation number per crossing.

Because particle trajectories would be constant in the static potential of the Sun, the phase space occupation is proportional to $N_\mathrm{cross}(v_R)$ and $\delta f(v_\perp)$.  Given that a particle penetrates the shell twice per orbit, with an orbital period $P(a)$, the number of crossings is
\begin{align}
    N_\mathrm{cross}(v_R) \simeq 2 \frac{t_\odot}{P(a)} = t_\odot \frac{\sqrt{G_N M_\odot}}{\pi a^{3/2}},
\end{align} 
while $\delta f(v_\perp)$ is obtained from eq.~\ref{eq:delta_f}: 
\begin{align}
    \delta f(v_\perp) = \frac{\Gamma_\mathrm{prod} \Delta R_\mathrm{prod}}{v_{R_\mathrm{prod}}(v_\perp)} = \frac{\Gamma_\mathrm{prod} \Delta R_\mathrm{prod}}{ \sqrt{v_{\mathrm{esc},R}^2 - v_\perp^2 R^2 / R_\mathrm{prod}^2}}.
\end{align}
The maximum transverse velocity magnitude for the perihelion to cross the shell is, by angular momentum conservation, $v_\perp^\mathrm{max} \equiv v_{\mathrm{esc},R} R_\mathrm{prod}/R$.

Assembling all of these pieces, we can evaluate the integral in eq.~\ref{eq:avg_f0_unperturbed} as:
\begin{align}
    \bar{f}_0 
    &= \frac{1}{\mathcal{V}_R} \left[2 \int_{0}^{v_{\mathrm{esc},R}} \dd v_R \, N_\mathrm{cross}(v_R) \right] \left[ 2 \pi \int_0^{v_\perp^\mathrm{max}} \dd v_\perp \, v_\perp \delta f(v_\perp) \right] \\
    &=\frac{3}{4\pi v_{\mathrm{esc},R}^3} \left[ \frac{3}{2} \frac{G_N M_\odot t_\odot}{R^2}\right] \left[ 2 \pi v_{\mathrm{esc},R_\mathrm{prod}} \frac{R_\mathrm{prod}^2}{R^2} \Gamma_\mathrm{prod} \Delta R_\mathrm{prod}\right] \\
    &= \Gamma_\mathrm{prod} \Delta R_\mathrm{prod} t_\odot \frac{9}{8} \frac{v_{\mathrm{esc},R_\mathrm{prod}}}{v_{\mathrm{esc},R}} \frac{R_\mathrm{prod}^2}{R^3}, \label{eq:avg_f0_unperturbed_result}
\end{align}
where to go to the second line, we used 
\begin{align}
    \frac{\dd v_R}{\dd a} = \frac{\sqrt{G_N M_\odot}}{2\sqrt{2} a^2} \frac{1}{\sqrt{1/R - 1/(2a)}}
\end{align}
to convert the radial-velocity integral to one over $a \in [R/2, \infty]$.

The estimate for the average occupation number over all of phase space in the backward simulations is:
\begin{align}
    \hat{f}_0 = \frac{1}{N_p} \sum_p^{N_p} \sum_i^{(p)} \delta f = \Gamma_\mathrm{prod} \Delta R_\mathrm{prod} \frac{1}{N_p} \sum_p^{N_p} \sum_i^{(p)} \frac{1}{v_{R_\mathrm{prod}}}.\label{eq:hat_f0_backward}
\end{align}
Therefore, our statistical estimator for $\tau_\mathrm{eff}$ from the backward runs is
\begin{align}
    \hat{\tau}_\mathrm{eff} = t_\odot \frac{\hat{f}_0}{\bar{f}_0}, 
\end{align}
which yields eq.~\ref{eq_tau_eff_backward} after insertion of eqs.~\ref{eq:avg_f0_unperturbed_result} and~\ref{eq:hat_f0_backward}.

\subsection{Full estimate of basin density from backward simulations}
\label{app:sat_density}
Here we detail how we use the time particles spend at different production radii within the sun to account for saturation effects in our estimates of the basin lifetime for the forward simulations.  

For the 2048 backward-run particles, we tabulate the sum $t_{\rm tot}(R_{\rm{prod}}) = C \sum \frac{1}{v_{R_\mathrm{prod}}}$, where $C$ is a normalization constant, $v_{R_\mathrm{prod}}$ is the radial  velocity magnitude and the sum is over the number of times the particle hit the shell of radius $R_{\rm{prod}}$ during its backward evolution. This quantity is tabulated at six values of $R_{\rm{prod}}$: $[0.1, 0.3, 0.5, 0.7, 0.9, 1.0]\,R_\odot$.  The total time spent by the particle within a shell of some small radius around $R_{\rm{prod}}$ is thus equal to $t_\mathrm{tot}(R_\mathrm{prod})$ with appropriate normalization constant $C$.

Given the coupling $g$ and mass $m$ of particle, we can use this data to compute the density of particles at Earth. While a particle is inside a particular shell (indexed by $i$), the rate of change of phase space density for that trajectory is
\begin{equation}
	\dot f = \Gamma_i (1 - f/f_i),
\end{equation}
where $\Gamma_i$ is the production rate inside the shell, and $f_i = 1/(e^{m/T_i} - 1)$ is the thermal occupation number for the temperature at that shell.
If $g$ is sufficiently small so that all of the $\Gamma_i \propto g^2$ are very small, then the $f/f_i$ term will always be $\ll 1$, and we have
\begin{equation}
	f_{\rm final} = f_0 \equiv \sum_i \Gamma_i t_{\textrm{tot}}(R_{\textrm{prod},i}),
\end{equation}
where the sum can be made into an integral by considering infinitesimally thin shells.

If instead $g$ is large enough that the $f/f_i$ terms cannot be neglected, then the number of times we strike each shell and the order in which we hit the different shells over the whole simulation will matter for $f_{\rm final}$. However, we can make the approximation that we hit the different shells sufficiently often, relative to the total trajectory time, that we can average their contributions. In that case, we can solve the equation
\begin{equation}
	\dot f = \frac{1}{t_\mathrm{tot}} \left[\sum_i \Gamma_i t_{\textrm{tot}}(R_{\textrm{prod},i})
	- f \sum_i \frac{\Gamma_i}{f_i} t_{\textrm{tot}}(R_{\textrm{prod},i}) \right]
\end{equation}
where $t_{\rm tot}$ is the total integration time for the particle.
Setting $f(t=0) = 0$, we obtain
\begin{equation}
	f_{\rm final} = f(t = t_{\rm tot}) = f_{\rm av} (1 - e^{-f_0/f_{\rm av}}),
\end{equation}
where
\begin{equation}
	f_{\rm av} \equiv \frac{f_0}{\sum_i \Gamma_i t_{\textrm{tot}}(R_{\textrm{prod},i})/f_i}
	= \frac{\sum_i \Gamma_i t_{\textrm{tot}}(R_{\textrm{prod},i})}{\sum_i \Gamma_i t_{\textrm{tot}}(R_{\textrm{prod},i})/f_i}.
\end{equation}
This gives us an estimate for $f_{\rm final}$ for each particle. 

We finally estimate the energy density $\rho_\mathrm{b}$ of basin particles at Earth by averaging this over the 2048 different particles, sampled from the sphere of bound velocities at Earth:
\begin{equation}
	\rho_\mathrm{b} = m \int \frac{\dd^3 k}{(2\pi)^3} f(k)
	= m^4 \int \frac{\dd^3 v}{(2 \pi)^3} f(v)
	\simeq \frac{m^4 \mathcal{V}_R}{(2\pi)^3} \frac{1}{N_p}\sum_p f_{\textrm{final},p}, \label{eq:rho_b}
\end{equation}
where $\mathcal{V}_R$ is the volume in velocity space sampled by the backward-running particles as below eq.~\ref{eq:avg_f0_unperturbed}, and $N_p = 2048$ is the number of particles. 

The results of this procedure, culminating in eq.~\ref{eq:rho_b}, are used to calculate the solar basin energy density as a function of mass and effective coupling for both the dark photon and axion benchmark models in figure~\ref{fig_rhoDP}. For couplings of $|\epsilon| \lesssim 10^{-12}$ and $|g_{aee}| \lesssim 10^{-12}$ respectively, re-absorption can be ignored to a good approximation, so the solar basin density computation amounts to calculating the energy density production rate multiplied by the effective basin time, now determined to be approximately $\tau_\mathrm{eff} = 1.2 \pm 0.09\,\mathrm{Gyr}$ based on the first 2048 backward-run particles. The terrestrial absorption rates of dark photons and axions are directly proportional to their solar basin energy density, and are thus also an input for the parameter space constraints (blue regions) of figures~\ref{fig_epsDP} and~\ref{fig_gaee}. 

\section{Gravitational scattering}\label{sec:scattering}
\begin{figure}
    \centering
    \begin{tikzpicture}
        \draw (0,0) circle [radius=3cm];
        \draw (0,0) circle [radius=5cm];
        
        \coordinate (particle) at (4,0);
        \coordinate (planet) at (0,0);

        \fill[particlecolor] (particle) circle [radius=2pt];
        \fill[planetcolor] (planet) circle [radius=5pt];

        \draw[gray] (particle) circle [radius=2.5cm];

        \draw[->,thick,planetcolor] (planet) -- ($(particle.west)+(-.1,0)$) node[midway, below, sloped, align=center] {$\vec{v}_{\textrm{P}}$};

        \coordinate (vin_outer) at (-5,5);
        \path[draw=none, name path=vinarrow] (planet) -- (vin_outer);
        \path[name path=innercircle] (0,0) circle [radius=3cm];
        \path [name intersections={of=vinarrow and innercircle, by=vin}];
        \draw[<->,thick, gray] ($(planet)+(-.1,.1)$) -- (vin) node[midway, above, sloped,align=center] {$v_{\textrm{in}}$};

        \coordinate (vout_outer) at (-6,3);
        \path[draw=none, name path=voutarrow] (planet) -- (vout_outer);
        \path[name path=outercircle] (0,0) circle [radius=5cm];
        \path [name intersections={of=voutarrow and outercircle, by=vout}];
        \draw[<->,thick, gray] ($(planet)+(-.1,.1)$) -- (vout) node[midway, above, sloped,align=center] {$v_{\textrm{out}}$};

        \path[name path=particlecircle] (particle) circle [radius=2.5 cm];
        \path [name intersections={of=particlecircle and innercircle, by={i,j}}];
        \path [name intersections={of=particlecircle and outercircle, by={k,l}}];

        \draw[->,thick,particlecolor] (particle) -- (i) node[midway, above, sloped,align=center,xshift=-0.2cm] {$\vec{w}_{\textrm{in}}$};
        \draw[->,thick,particlecolor] (particle) -- (k) node[midway, below, sloped,align=center] {$\vec{w}_{\textrm{out}}$};

        \draw[<->,thick,gray] ($(particle.south)+(0,-.1)$) -- ++(0,-2.4) node[midway, below,align=center,xshift=0.2cm,yshift=0.1cm] {$w$};

        \draw[->,thick,particlecolor] (planet) -- (i) node[midway, above, sloped,align=center] {$\vec{v}_{\textrm{in}}$};
        
        \coordinate (C) at (5,0);
        \pic [draw=black,angle radius=1cm, "$\alpha_{\rm{in}}$"  {xshift=0.1cm}] {angle = C--planet--i};
        \pic [draw=black,angle radius=.6cm, "$\phi_{\rm{out}}$"{xshift=0.2cm,yshift=-0.4cm}] {angle = C--particle--k};
        \pic [draw=black,angle radius=.8cm, "$\phi_{\rm{in}}$"{xshift=-0.4cm,yshift=0.6cm}] {angle = C--particle--i};

        \draw[-,thick,gray,dashed] (particle) -- ++(1,0);

    \end{tikzpicture}
    \caption{The relevant scalars, vectors, and angles for the setup of the gravitational scattering problem in velocity space.  We will find the differential rate for a particle to scatter from semi-major axis $a_{\rm{in}}$, with velocity $\vec{w}_{\rm{in}}$, to semi-major axis $a_{\rm{out}}$, with velocity $\vec{w}_{\rm{out}}$.  Vectors associated with the motion of the test particle are shaded blue-green, while vectors associated with the motion of the planet are shaded orange.}
    \label{fig:scattering_diagram}
\end{figure}
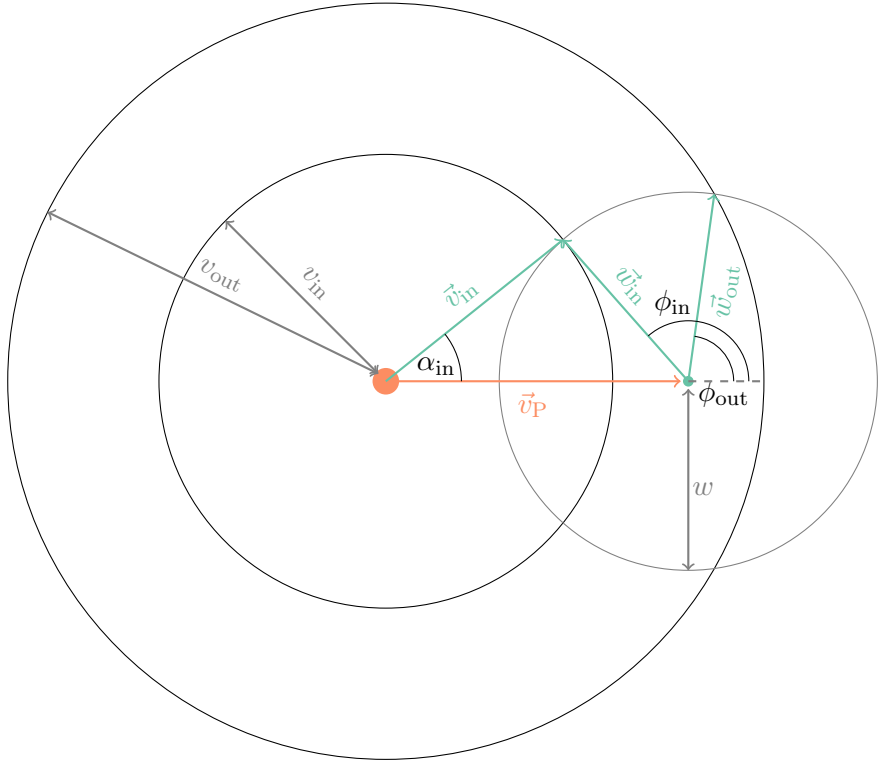

We derive the necessary components to find the differential rate for a particle to scatter from a semi-major axis $a_{\rm{in}}$ to $a_{\rm{out}}$ in appendix~\ref{sec:scattering_derivations} below.  This calculation is adapted from ref.~\cite{Levin_2006}, wherein the differential rate for a particle to be ejected from a semi-major axis $a_{\rm{in}}$ is computed instead (a result we have verified). The problem setup and variables in velocity space are compiled in figure~\ref{fig:scattering_diagram}. Closed-form analytic expressions for the numerator function $F(a)$ (and thus the diffusion function $D(a)$) and the ejection rate $\Gamma_\mathrm{ej}$ are collected in appendix~\ref{sec:scattering_analytic}.

\subsection{Derivations} \label{sec:scattering_derivations}

We begin with a particle of velocity $\vect{v}_{\text {in }}$, which will be boosted to a velocity $\vect{v}_{\text {out }}$ after scattering. Following \cite{Levin_2006}, we define new vectors in the planet frame:
\begin{align}
\vect{w}_{\text {in }} & =\vect{v}_{\text {in }}-\vect{v}_{\mathrm{P}}, \\
\vect{w}_{\text {out }} & =\vect{v}_{\text {out }}-\vect{v}_{\mathrm{P}}, \\
w & =\left|\vect{w}_{\text {in }}\right|=\left|\vect{w}_{\text {out }}\right| .
\end{align}
We explicitly write the components of these vectors to define angles, with the velocity of the planet defining the $\hat{x}$ direction, and the plane of scattering defining the $x-y$ plane:
\begin{equation}
\vect{w}_\mathrm{in} = w (\cos \phi_\mathrm{in}, \sin \phi_\mathrm{in}, 0).
\end{equation}
Then the incoming velocity can be related to the outgoing velocity after scattering $\vect{w}_\mathrm{out}$ via
\begin{equation}
\vect{w}_\mathrm{out} = w (\cos \phi_\mathrm{out}, \sin \phi_\mathrm{out} \cos \phi_2, \sin \phi_\mathrm{out} \sin \phi_2) = \vect{w}_\mathrm{in} \cos \theta - \hat{\vect{b}} w \sin \theta,\label{eq:wout}
\end{equation}
where $\hat{\vect{b}}$ is the impact parameter for the scattering and is given by
\begin{align}
b &= \frac{(G_N M_{\rm{P}})^2}{2w^4} \tan^{-2}\left(\frac{\theta}{2}\right)\\
\hat{\vect{b}} &= (-\sin \phi_\mathrm{in} \cos \phi_b, \cos \phi_\mathrm{in} \cos \phi_b, \sin \phi_b).
\end{align}
$M_{\textrm{P}}$ denotes the planet mass.  Note eq.~\eqref{eq:wout} defines the angle $\theta$.

To complete our geometrical setup of the problem, the angles can be expressed in terms of velocities, which are either known or can be related to $a_{\rm{in}}$ and $a_{\rm{out}}$:
\begin{align}
\cos \phi_\mathrm{out} &= \frac{v_\mathrm{out}^2 - v_\mathrm{P}^2 - w^2}{2 v_\mathrm{P} w} =  \cos \phi_\mathrm{in} \cos \theta +  \sin \phi_\mathrm{in} \sin \theta \cos \phi_b \label{eq:defCosPhiOut}\\
\cos \phi_\mathrm{in} &= \frac{v_\mathrm{in}^2 - v_\mathrm{P}^2 - w^2}{2 v_\mathrm{P} w}\\
\cos \phi_2 &= \frac{\sin \phi_\mathrm{in} \cos \theta - \cos \phi_\mathrm{in} \sin \theta \cos \phi_b}{\sin \phi_\mathrm{out}}.
\end{align}

We can write $\dd\sigma = b\,\dd b \,\dd\phi_b$, since $\dd\sigma$ is the differential area element in the plane of $\vect b$. So we can 
rewrite
\begin{align}
\frac{\dd\sigma}{\dd\cos\phi_{\textrm{out}}\dd\phi_b} &= b\frac{\dd b}{\dd \cos \phi_{\textrm{out}}}\nonumber\\
& = \frac{1}{2}\frac{\dd b^2}{\dd \cos{\phi_{\textrm{out}}}}\nonumber\\
&=\frac{(G_N M_{\textrm{P}})^2}{2w^4}\frac{\dd\tan^{-2}\left(\frac{\theta}{2}\right)}{\dd \cos\phi_{\textrm{out}}}\label{eq:crossSectionAsArea}.
\end{align}
Writing
\begin{equation}
\tan^2\left(\frac{\theta}{2}\right) = \frac{1-\cos\theta}{1+\cos\theta}
\end{equation}
and using the above definitions to solve for $\cos\theta$:
\begin{equation}
\cos\theta_{\pm} = \frac{\cos\phi_{\textrm{in}} \cos\phi_{\textrm{out}} \pm \cos\phi_b\sin\phi_{\textrm{in}} \sqrt{\cos^2\phi_{\textrm{in}} - \cos^2\phi_{\textrm{out}} + \cos^2\phi_b\sin^2\phi_{\textrm{in}}}}{\cos^2\phi_\textrm{in} + \cos^2\phi_b\sin^2\phi_{\textrm{in}}},
\end{equation}
we can rewrite the differential cross section in 
terms of $\phi_{\textrm{in}}$ and $\phi_{\textrm{out}}$ 
only.  In order to do so, we will count contributions from 
both of the two branches of $\cos\theta$, so
\begin{equation}
\frac{\dd\tan^{-2}\left(\frac{\theta}{2}\right)}{\dd \cos\phi_{\textrm{out}}} = 4\frac{\cos^2\phi_{\textrm{in}} - \cos\phi_{\textrm{in}}\cos\phi_{\textrm{out}} + 2\cos^2\phi_b\sin^2\phi_{\textrm{in}}}{\left(\cos\phi_{\textrm{in}} - \cos\phi_{\textrm{out}}\right)^3}.
\end{equation}
Comparing with eq.~\eqref{eq:crossSectionAsArea}, we find
\begin{equation}
\frac{\dd\sigma}{\dd \cos\phi_{\textrm{out}}} = 2\pi \frac{(G_N M_{\textrm{P}})^2}{w^4}\frac{1-\cos\phi_{\textrm{in}}\cos\phi_{\textrm{out}}}{\left(\cos\phi_{\textrm{in}} - \cos\phi_{\textrm{out}}\right)^3},\label{eq:cross_section_Rutherford}
\end{equation}
where we have integrated $\phi_b$ only over half
of the $b$-plane to avoid double-counting after
summing both branches of $\cos\theta$.

Finally, we are ready to write down the differential rate for scattering from
$a_{\rm{in}}$ to $a_{\rm{out}}$.  The integrated rate is given by
\begin{equation}
\Gamma = n\sigma v;
\end{equation}
with
\begin{equation}
n(R) = \int\dd^3 \mathbf{v}_{\textrm{in}} f(\mathbf{v}_{\textrm{in}},R).
\end{equation}
Then we can set up the differential rate by using the differential cross section above:
\begin{equation}
\frac{\dd \Gamma}{\dd^3\mathbf{v}_{\textrm{in}}\dd\cos\phi_{\textrm{out}}} = f(\mathbf{v}_{\textrm{in}},R) \frac{\dd\sigma}{\dd \cos\phi_{\textrm{out}}}  w.
\end{equation}
We express $\dd^3\mathbf{v}_{\textrm{in}}$ 
as $v_{\textrm{in}}^2\dd v_{\textrm{in}}\dd\Omega_{in}$, 
so that
\begin{equation}
\frac{\dd \Gamma}{\dd v_{\textrm{in}}\dd\cos\phi_{\textrm{out}}} = \int \dd\Omega_{\textrm{in}} \,  f(\vect{v}_{\textrm{in}},R) \frac{\dd\sigma}{\dd \cos\phi_{\textrm{out}}}  w v_{\textrm{in}}^2.
\end{equation}
To convert $\dd v_{\textrm{in}}$ to $\dd a_{in}$, we 
use the fact that energy conservation implies
\begin{equation}
-\frac{G_N M_{\odot}}{2a} = \frac{v^2(R)}{2} - \frac{G_N M_{\odot}}{R}.
\end{equation}
We evaluate this expression at the solar distance
at which scattering occurs (in the present case, 
$\sim\SI{1}{AU}$) to express $v_\mathrm{in}$ in terms of $a$.  
Differentiating, we find
\begin{equation}
\dd v_{\textrm{in}} = \frac{1}{2}\frac{G_N M_{\odot}}{\sqrt{2G_N M_{\odot}\left(\frac{1}{R} - \frac{1}{2 a_{\textrm{in}}}\right)}}\frac{\dd a_{\textrm{in}}}{a_{\textrm{in}}^2}\label{eq:dvInTermsOfda},
\end{equation}
and so
\begin{equation}
\frac{\dd \Gamma}{\dd a_{\textrm{in}}\dd\cos\phi_{\textrm{out}}} = \int \dd\Omega_{\textrm{in}}\,  f(\vect{v}_{\textrm{in}},R) \frac{\dd\sigma}{\dd \cos\phi_{\textrm{out}}} w v_{\textrm{in}} \frac{G_N M_{\odot}}{2 a_\mathrm{in}^2} .
\end{equation}
Next we express $\dd \cos\phi_{\textrm{out}}$ as 
$\dd a_{\textrm{out}}$.  We use eq.~\eqref{eq:defCosPhiOut}
to express $\cos\phi_{\textrm{out}}$ in terms of
$v_{\textrm{out}}$, and then use eq.~\eqref{eq:dvInTermsOfda}
once again to express this velocity differential 
in terms of semi-major axis $a$, giving:
\begin{align}
\left.\frac{\dd \cos \phi_\mathrm{out}}{\dd a_\mathrm{out}}\right|_w = \frac{G_N M_\odot}{2 a_\mathrm{out}^2} \frac{1}{v_\mathrm{P} w}.
\end{align}
With this final manipulation, one finds the desired result for the scattering rate in eq.~\ref{eq:dGammadaout}.

\subsection{Analytic expressions}\label{sec:scattering_analytic}

The numerator function $F(a)$ in eq.~\ref{eq:dPscatter} evaluates to:
\begin{align}
    F(a) 
    &\equiv 
    \frac{4}{3\pi} \frac{M_\mathrm{P}^2}{M_\odot^2} \frac{1}{R} \sqrt{\frac{G_N M_\odot}{a (2 a-R)}} \label{eq:F}\\
    &\phantom{\equiv}\times \left\lbrace \frac{\left[4 a^2-2 a R- (3 a-R) \sqrt{a (2 a-R)}\right] \left[5 a^2-4 a R +R^2 -(3 a-R)\sqrt{a (2 a-R)}\right]}{\left(3 a-R -2 \sqrt{a (2 a-R)}\right)^{3/2}} \right. \nonumber \\
    &\phantom{\equiv \times }\left. +\frac{\left[4 a^2-2 a R+(3 a-R)\sqrt{a (2 a-R)} \right] \left[5
   a^2-4 a R+R^2 +(3 a-R)\sqrt{a (2 a-R)} \right]}{\left[3a - R + 2\sqrt{a (2 a-R)}\right]^{3/2}}
   \right\rbrace \nonumber.
\end{align}
The ejection rate in eq.~\ref{eq:Gamma_ej} can also be written in closed form:
\begin{alignat}{2}
\Gamma_\mathrm{ej} (a_\mathrm{in})
&= \frac{1}{12 \pi} \frac{M_\mathrm{P}^2}{M_\odot^2}\frac{\sqrt{G_N M_\odot}}{a_\mathrm{in}^2 R} \Theta(2a_\mathrm{in}-R) \left[T_1(a_\mathrm{in})-T_2(a_\mathrm{in})\right] \label{eq:Gamma_ej_1}\\
T_1(a) &\equiv \frac{40 a^2 - R^2 - 8 a R + 4 (6a+R) \sqrt{a(2a-R)}}{\left[3 a - R + 2 \sqrt{a(2a-R)} \right]^{1/2}} \label{eq:Gamma_ej_2} \\
T_2(a) &\equiv 
\begin{cases}
\frac{40 a^2 - R^2 - 8 a R - 4 (6a+R) \sqrt{a(2a-R)}}{\left[3 a - R - 2 \sqrt{a(2a-R)} \right]^{1/2}} & a < \frac{1+\sqrt{2}}{4} R \\
8 \left(2 \sqrt{2}-1\right) a^{3/2}  & a \geq \frac{1+\sqrt{2}}{4} R.
\end{cases}
\label{eq:Gamma_ej_3}
\end{alignat}

\bibliography{solar_basin_dynamics}

\providecommand{\href}[2]{#2}\begingroup\raggedright\begin{thebibliography}{100}

\bibitem{Sikivie_1983}
P.~Sikivie, \emph{Experimental tests of the "invisible" axion},
  \href{http://dx.doi.org/10.1103/PhysRevLett.51.1415}{\emph{Phys. Rev. Lett.}
  {\bf 51} (Oct, 1983) 1415--1417}.

\bibitem{Bibber_1989}
K.~van Bibber, P.~M. McIntyre, D.~E. Morris and G.~G. Raffelt, \emph{Design for
  a practical laboratory detector for solar axions},
  \href{http://dx.doi.org/10.1103/PhysRevD.39.2089}{\emph{Phys. Rev. D} {\bf
  39} (Apr, 1989) 2089--2099}.

\bibitem{PASCHOS_1994}
E.~Paschos and K.~Zioutas, \emph{A proposal for solar axion detection via bragg
  scattering}, {\emph{Physics Letters B} {\bf 323} (1994) 367--372}.

\bibitem{Moriyama_1995}
S.~Moriyama, \emph{Proposal to search for a monochromatic component of solar
  axions using ${}^{57}$fe},
  \href{http://dx.doi.org/10.1103/PhysRevLett.75.3222}{\emph{Phys. Rev. Lett.}
  {\bf 75} (Oct, 1995) 3222--3225}.

\bibitem{Arik_2011}
M.~Arik, S.~Aune, K.~Barth, A.~Belov, S.~Borghi et~al., \emph{Search for
  {Sub-eV} mass solar axions by the {CERN Axion Solar Telescope} with {$^3$He}
  buffer gas},
  \href{http://dx.doi.org/10.1103/physrevlett.107.261302}{\emph{Physical Review
  Letters} {\bf 107} (Dec., 2011) }.

\bibitem{Redondo_2013a}
J.~Redondo, \emph{Solar axion flux from the axion-electron coupling},
  \href{http://dx.doi.org/10.1088/1475-7516/2013/12/008}{\emph{Journal of
  Cosmology and Astroparticle Physics} {\bf 2013} (Dec., 2013) 008–008}.

\bibitem{Armengaud_2014}
E.~Armengaud, F.~T. Avignone, M.~Betz, P.~Brax, P.~Brun, G.~Cantatore et~al.,
  \emph{Conceptual design of the {International Axion Observatory (IAXO)}},
  \href{http://dx.doi.org/10.1088/1748-0221/9/05/t05002}{\emph{Journal of
  Instrumentation} {\bf 9} (May, 2014) T05002–T05002}.

\bibitem{Giannotti_2016}
M.~Giannotti, I.~Irastorza, J.~Redondo and A.~Ringwald, \emph{Cool wisps for
  stellar cooling excesses},
  \href{http://dx.doi.org/10.1088/1475-7516/2016/05/057}{\emph{Journal of
  Cosmology and Astroparticle Physics} {\bf 2016} (May, 2016) 057–057}.

\bibitem{Giannotti_2017}
M.~Giannotti, I.~G. Irastorza, J.~Redondo, A.~Ringwald and K.~Saikawa,
  \emph{Stellar recipes for axion hunters},
  \href{http://dx.doi.org/10.1088/1475-7516/2017/10/010}{\emph{Journal of
  Cosmology and Astroparticle Physics} {\bf 2017} (Oct., 2017) 010–010}.

\bibitem{Mastrototaro_2020}
L.~Mastrototaro, A.~Mirizzi, P.~D. Serpico and A.~Esmaili, \emph{Heavy sterile
  neutrino emission in core-collapse supernovae: constraints and signatures},
  \href{http://dx.doi.org/10.1088/1475-7516/2020/01/010}{\emph{Journal of
  Cosmology and Astroparticle Physics} {\bf 2020} (Jan., 2020) 010–010}.

\bibitem{Di_Luzio_2022}
L.~Di~Luzio, M.~Fedele, M.~Giannotti, F.~Mescia and E.~Nardi, \emph{Stellar
  evolution confronts axion models},
  \href{http://dx.doi.org/10.1088/1475-7516/2022/02/035}{\emph{Journal of
  Cosmology and Astroparticle Physics} {\bf 2022} (Feb., 2022) 035}.

\bibitem{Chang_2022}
J.~H. Chang, D.~E. Kaplan, S.~Rajendran, H.~Ramani and E.~H. Tanin, \emph{Dark
  solar wind},
  \href{http://dx.doi.org/10.1103/PhysRevLett.129.211101}{\emph{Phys. Rev.
  Lett.} {\bf 129} (Nov, 2022) 211101}.

\bibitem{Engel_1990}
J.~Engel, D.~Seckel and A.~C. Hayes, \emph{Emission and detectability of
  hadronic axions from {SN 1987A}},
  \href{http://dx.doi.org/10.1103/PhysRevLett.65.960}{\emph{Phys. Rev. Lett.}
  {\bf 65} (Aug, 1990) 960--963}.

\bibitem{lella2023}
A.~Lella, P.~Carenza, G.~Co', G.~Lucente, M.~Giannotti, A.~Mirizzi et~al.,
  \emph{Getting the most on supernova axions},  2023.

\bibitem{carenza2023}
P.~Carenza, G.~Co', M.~Giannotti, A.~Lella, G.~Lucente, A.~Mirizzi et~al.,
  \emph{Cross section for supernova axion observation in neutrino water
  {Cherenkov} detectors},  2023.

\bibitem{Pospelov_2021}
M.~Pospelov and H.~Ramani, \emph{Earth-bound millicharge relics},
  \href{http://dx.doi.org/10.1103/physrevd.103.115031}{\emph{Physical Review D}
  {\bf 103} (June, 2021) }.

\bibitem{berlin_2023}
A.~Berlin, H.~Liu, M.~Pospelov and H.~Ramani, \emph{The terrestrial density of
  strongly-coupled relics},  2023.

\bibitem{VanTilburg:2020jvl}
K.~Van~Tilburg, \emph{{Stellar basins of gravitationally bound particles}},
  \href{http://dx.doi.org/10.1103/PhysRevD.104.023019}{\emph{Phys. Rev. D} {\bf
  104} (2021) 023019}, [\href{http://arxiv.org/abs/2006.12431}{{\tt
  2006.12431}}].

\bibitem{Lasenby:2020goo}
R.~Lasenby and K.~Van~Tilburg, \emph{{Dark photons in the solar basin}},
  \href{http://dx.doi.org/10.1103/PhysRevD.104.023020}{\emph{Phys. Rev. D} {\bf
  104} (2021) 023020}, [\href{http://arxiv.org/abs/2008.08594}{{\tt
  2008.08594}}].

\bibitem{Berlin:2021kcm}
A.~Berlin and K.~Schutz, \emph{{Helioscope for gravitationally bound
  millicharged particles}},
  \href{http://dx.doi.org/10.1103/PhysRevD.105.095012}{\emph{Phys. Rev. D} {\bf
  105} (2022) 095012}, [\href{http://arxiv.org/abs/2111.01796}{{\tt
  2111.01796}}].

\bibitem{DeRocco:2022jyq}
W.~DeRocco, S.~Wegsman, B.~Grefenstette, J.~Huang and K.~Van~Tilburg,
  \emph{{First Indirect Detection Constraints on Axions in the Solar Basin}},
  \href{http://dx.doi.org/10.1103/PhysRevLett.129.101101}{\emph{Phys. Rev.
  Lett.} {\bf 129} (2022) 101101}, [\href{http://arxiv.org/abs/2205.05700}{{\tt
  2205.05700}}].

\bibitem{Hannestad_2002}
S.~Hannestad and G.~G. Raffelt, \emph{Stringent neutron-star limits on large
  extra dimensions},
  \href{http://dx.doi.org/10.1103/physrevlett.88.071301}{\emph{Physical Review
  Letters} {\bf 88} (feb, 2002) }.

\bibitem{DiLella_2003}
L.~DiLella and K.~Zioutas, \emph{Observational evidence for gravitationally
  trapped massive axion(-like) particles},
  \href{http://dx.doi.org/10.1016/s0927-6505(02)00186-x}{\emph{Astroparticle
  Physics} {\bf 19} (apr, 2003) 145--170}.

\bibitem{Morgan_2005}
B.~Morgan, N.~Spooner, M.~Armel-Funkhouser, D.~Hoffmann, J.~Jacoby,
  D.~Snowden-Ifft et~al., \emph{Searches for solar {K}aluza{\textendash}{K}lein
  axions with gas {TPCs}},
  \href{http://dx.doi.org/10.1016/j.astropartphys.2005.01.002}{\emph{Astroparticle
  Physics} {\bf 23} (apr, 2005) 287--302}.

\bibitem{XMASS:2017sij}
{\scshape XMASS} collaboration, N.~Oka, K.~Abe, K.~Hiraide, K.~Ichimura,
  Y.~Kishimoto et~al., \emph{{Search for solar Kaluza\textendash{}Klein axions
  by annual modulation with the {XMASS-I} detector}},
  \href{http://dx.doi.org/10.1093/ptep/ptx137}{\emph{PTEP} {\bf 2017} (2017)
  103C01}, [\href{http://arxiv.org/abs/1707.08995}{{\tt 1707.08995}}].

\bibitem{Laskar_1989}
J.~Laskar, \emph{A numerical experiment on the chaotic behaviour of the {Solar
  System}}, {\emph{Nature} {\bf 338} (1989) 237--238}.

\bibitem{LASKAR_1990}
J.~Laskar, \emph{The chaotic motion of the {Solar System}: A numerical estimate
  of the size of the chaotic zones},
  \href{http://dx.doi.org/https://doi.org/10.1016/0019-1035(90)90084-M}{\emph{Icarus}
  {\bf 88} (1990) 266--291}.

\bibitem{Sussman_1992}
G.~J. Sussman and J.~Wisdom, \emph{Chaotic evolution of the {Solar System}},
  \href{http://dx.doi.org/10.1126/science.257.5066.56}{\emph{Science} {\bf 257}
  (1992) 56--62},
  [\href{http://arxiv.org/abs/https://www.science.org/doi/pdf/10.1126/science.257.5066.56}{{\tt
  https://www.science.org/doi/pdf/10.1126/science.257.5066.56}}].

\bibitem{Mogavero_2021}
F.~{Mogavero} and J.~{Laskar}, \emph{{Long-term dynamics of the inner planets
  in the Solar System}},
  \href{http://dx.doi.org/10.1051/0004-6361/202141007}{\emph{\aap} {\bf 655}
  (Nov., 2021) A1}, [\href{http://arxiv.org/abs/2105.14976}{{\tt 2105.14976}}].

\bibitem{mogavero2023timescales}
F.~Mogavero, N.~H. Hoang and J.~Laskar, \emph{Timescales of chaos in the inner
  {Solar System}: {L}yapunov spectrum and quasi-integrals of motion},
  {\emph{Physical Review X} {\bf 13} (2023) 021018}.

\bibitem{1991ApJ...368..610G}
A.~{Gould}, \emph{{Gravitational Diffusion of Solar System WIMPs}},
  \href{http://dx.doi.org/10.1086/169726}{\emph{\apj} {\bf 368} (Feb., 1991)
  610}.

\bibitem{Anderson:2020rdk}
N.~B. Anderson, A.~Partenheimer and T.~D. Wiser, \emph{{Direct detection
  signatures of a primordial Solar dark matter halo}},
  \href{http://arxiv.org/abs/2007.11016}{{\tt 2007.11016}}.

\bibitem{Essig:2022dfa}
R.~Essig, G.~K. Giovanetti, N.~Kurinsky, D.~McKinsey, K.~Ramanathan et~al.,
  \emph{{Snowmass2021 Cosmic Frontier: The landscape of low-threshold dark
  matter direct detection in the next decade}},  in \emph{{Snowmass 2021}}, 3,
  2022.
\newblock \href{http://arxiv.org/abs/2203.08297}{{\tt 2203.08297}}.

\bibitem{Bahcall:1995bt}
J.~N. Bahcall and M.~H. Pinsonneault, \emph{{Solar models with helium and heavy
  element diffusion}},
  \href{http://dx.doi.org/10.1103/RevModPhys.67.781}{\emph{Rev. Mod. Phys.}
  {\bf 67} (1995) 781--808}, [\href{http://arxiv.org/abs/hep-ph/9505425}{{\tt
  hep-ph/9505425}}].

\bibitem{2012Sci...338..651C}
J.~N. {Connelly}, M.~{Bizzarro}, A.~N. {Krot}, {\r{A}}.~{Nordlund},
  D.~{Wielandt} and M.~A. {Ivanova}, \emph{{The Absolute Chronology and Thermal
  Processing of Solids in the Solar Protoplanetary Disk}},
  \href{http://dx.doi.org/10.1126/science.1226919}{\emph{Science} {\bf 338}
  (Nov., 2012) 651}.

\bibitem{Bloch_2017}
I.~M. Bloch, R.~Essig, K.~Tobioka, T.~Volansky and T.-T. Yu, \emph{Searching
  for dark absorption with direct detection experiments},
  \href{http://dx.doi.org/10.1007/jhep06(2017)087}{\emph{Journal of High Energy
  Physics} {\bf 2017} (jun, 2017) }.

\bibitem{aprile2019light}
E.~Aprile, J.~Aalbers, F.~Agostini, M.~Alfonsi, L.~Althueser, F.~Amaro et~al.,
  \emph{Light dark matter search with ionization signals in {XENON1T}},
  {\emph{Physical Review Letters} {\bf 123} (2019) 251801}.

\bibitem{barak2020sensei}
L.~Barak, I.~M. Bloch, M.~Cababie, G.~Cancelo, L.~Chaplinsky, F.~Chierchie
  et~al., \emph{{SENSEI}: Direct-detection results on sub-{GeV} dark matter
  from a new skipper {CCD}}, {\emph{Physical Review Letters} {\bf 125} (2020)
  171802}.

\bibitem{Aprile_2020}
E.~Aprile, J.~Aalbers, F.~Agostini, M.~Alfonsi, L.~Althueser et~al.,
  \emph{Excess electronic recoil events in {XENON1T}},
  \href{http://dx.doi.org/10.1103/physrevd.102.072004}{\emph{Physical Review D}
  {\bf 102} (Oct., 2020) }.

\bibitem{Aprile_2022}
E.~Aprile, K.~Abe, F.~Agostini, S.~Ahmed~Maouloud, L.~Althueser et~al.,
  \emph{Search for new physics in electronic recoil data from {XENONnT}},
  \href{http://dx.doi.org/10.1103/physrevlett.129.161805}{\emph{Physical Review
  Letters} {\bf 129} (Oct., 2022) }.

\bibitem{Darkside50}
{\scshape 50} collaboration, P.~Agnes, I.~Albuquerque, T.~Alexander, A.~Alton,
  M.~Ave et~al., \emph{Search for dark matter particle interactions with
  electron final states with {DarkSide-50}},  2023.

\bibitem{Knapen_2018}
S.~Knapen, T.~Lin, M.~Pyle and K.~M. Zurek, \emph{Detection of light dark
  matter with optical phonons in polar materials},
  \href{http://dx.doi.org/10.1016/j.physletb.2018.08.064}{\emph{Physics Letters
  B} {\bf 785} (oct, 2018) 386--390}.

\bibitem{Baryakhtar_2018}
M.~Baryakhtar, J.~Huang and R.~Lasenby, \emph{Axion and hidden photon dark
  matter detection with multilayer optical haloscopes},
  \href{http://dx.doi.org/10.1103/physrevd.98.035006}{\emph{Physical Review D}
  {\bf 98} (aug, 2018) }.

\bibitem{supercdmscollaboration_2023}
{\scshape SuperCDMS} collaboration, M.~F. Albakry, I.~Alkhatib, D.~W.~P.
  Amaral, T.~Aralis, T.~Aramaki et~al., \emph{A strategy for low-mass dark
  matter searches with cryogenic detectors in the {SuperCDMS SNOLAB} facility},
   2023.

\bibitem{an2020new}
H.~{An}, M.~{Pospelov} and J.~{Pradler}, \emph{{New stellar constraints on dark
  photons}},
  \href{http://dx.doi.org/10.1016/j.physletb.2013.07.008}{\emph{Physics Letters
  B} {\bf 725} (Oct., 2013) 190--195},
  [\href{http://arxiv.org/abs/1302.3884}{{\tt 1302.3884}}].

\bibitem{li2023}
S.-P. Li and X.-J. Xu, \emph{Production rates of dark photons and {$Z'$} in the
  {S}un and stellar cooling bounds},  2023.

\bibitem{Dolan:2023cjs}
M.~J. Dolan, F.~J. Hiskens and R.~R. Volkas, \emph{{Constraining dark photons
  with self-consistent simulations of globular cluster stars}},
  \href{http://dx.doi.org/10.1088/1475-7516/2024/05/099}{\emph{JCAP} {\bf 05}
  (2024) 099}, [\href{http://arxiv.org/abs/2306.13335}{{\tt 2306.13335}}].

\bibitem{Redondo:2013wwa}
J.~Redondo, \emph{{Solar axion flux from the axion-electron coupling}},
  \href{http://dx.doi.org/10.1088/1475-7516/2013/12/008}{\emph{JCAP} {\bf 12}
  (2013) 008}, [\href{http://arxiv.org/abs/1310.0823}{{\tt 1310.0823}}].

\bibitem{2020PhRvD.102h3007C}
F.~{Capozzi} and G.~{Raffelt}, \emph{{Axion and neutrino bounds improved with
  new calibrations of the tip of the red-giant branch using geometric distance
  determinations}},
  \href{http://dx.doi.org/10.1103/PhysRevD.102.083007}{\emph{\prd} {\bf 102}
  (Oct., 2020) 083007}, [\href{http://arxiv.org/abs/2007.03694}{{\tt
  2007.03694}}].

\bibitem{MillerBertolami:2014rka}
M.~M. Miller~Bertolami, B.~E. Melendez, L.~G. Althaus and J.~Isern,
  \emph{{Revisiting the axion bounds from the Galactic white dwarf luminosity
  function}},
  \href{http://dx.doi.org/10.1088/1475-7516/2014/10/069}{\emph{JCAP} {\bf 10}
  (2014) 069}, [\href{http://arxiv.org/abs/1406.7712}{{\tt 1406.7712}}].

\bibitem{Langhoff:2022bij}
K.~Langhoff, N.~J. Outmezguine and N.~L. Rodd, \emph{{Irreducible Axion
  Background}},
  \href{http://dx.doi.org/10.1103/PhysRevLett.129.241101}{\emph{Phys. Rev.
  Lett.} {\bf 129} (2022) 241101}, [\href{http://arxiv.org/abs/2209.06216}{{\tt
  2209.06216}}].

\bibitem{Aalbers_2023}
J.~Aalbers, D.~S. Akerib, A.~K. Al~Musalhi, F.~Alder, C.~S. Amarasinghe et~al.,
  \emph{Search for new physics in low-energy electron recoils from the first
  {LZ} exposure},
  \href{http://dx.doi.org/10.1103/physrevd.108.072006}{\emph{Physical Review D}
  {\bf 108} (Oct., 2023) }.

\bibitem{Takahashi:2020bpq}
F.~Takahashi, M.~Yamada and W.~Yin, \emph{{XENON1T Excess from Anomaly-Free
  Axionlike Dark Matter and Its Implications for Stellar Cooling Anomaly}},
  \href{http://dx.doi.org/10.1103/PhysRevLett.125.161801}{\emph{Phys. Rev.
  Lett.} {\bf 125} (2020) 161801}, [\href{http://arxiv.org/abs/2006.10035}{{\tt
  2006.10035}}].

\bibitem{Feulner_2012}
G.~Feulner, \emph{The faint young {Sun} problem},
  \href{http://dx.doi.org/10.1029/2011rg000375}{\emph{Reviews of Geophysics}
  {\bf 50} (may, 2012) }.

\bibitem{2005Natur.435..466G}
R.~{Gomes}, H.~F. {Levison}, K.~{Tsiganis} and A.~{Morbidelli}, \emph{{Origin
  of the cataclysmic Late Heavy Bombardment period of the terrestrial
  planets}}, \href{http://dx.doi.org/10.1038/nature03676}{\emph{\nat} {\bf 435}
  (May, 2005) 466--469}.

\bibitem{Tsiganis_2005}
K.~Tsiganis, R.~Gomes, A.~Morbidelli and H.~F. Levison, \emph{Origin of the
  orbital architecture of the giant planets of the {Solar System}},
  {\emph{Nature} {\bf 435} (2005) 459--461}.

\bibitem{Morbidelli_2005}
A.~Morbidelli, H.~F. Levison, K.~Tsiganis and R.~Gomes, \emph{{Chaotic capture
  of Jupiter's Trojan asteroids in the early Solar System}}, {\emph{Nature}
  {\bf 435} (2005) 462--465}.

\bibitem{2001Icar..152..205C}
J.~E. {Chambers}, \emph{{Making More Terrestrial Planets}},
  \href{http://dx.doi.org/10.1006/icar.2001.6639}{\emph{\icarus} {\bf 152}
  (Aug., 2001) 205--224}.

\bibitem{2009Icar..203..644R}
S.~N. {Raymond}, D.~P. {O'Brien}, A.~{Morbidelli} and N.~A. {Kaib},
  \emph{{Building the terrestrial planets: Constrained accretion in the inner
  Solar System}},
  \href{http://dx.doi.org/10.1016/j.icarus.2009.05.016}{\emph{\icarus} {\bf
  203} (Oct., 2009) 644--662}, [\href{http://arxiv.org/abs/0905.3750}{{\tt
  0905.3750}}].

\bibitem{2005Sci...309.1847S}
R.~G. {Strom}, R.~{Malhotra}, T.~{Ito}, F.~{Yoshida} and D.~A. {Kring},
  \emph{{The Origin of Planetary Impactors in the Inner Solar System}},
  \href{http://dx.doi.org/10.1126/science.1113544}{\emph{Science} {\bf 309}
  (Sept., 2005) 1847--1850}, [\href{http://arxiv.org/abs/astro-ph/0510200}{{\tt
  astro-ph/0510200}}].

\bibitem{2017OLEB...47..261Z}
N.~E.~B. {Zellner}, \emph{{Cataclysm No More: New Views on the Timing and
  Delivery of Lunar Impactors}},
  \href{http://dx.doi.org/10.1007/s11084-017-9536-3}{\emph{Origins of Life and
  Evolution of the Biosphere} {\bf 47} (Sept., 2017) 261--280},
  [\href{http://arxiv.org/abs/1704.06694}{{\tt 1704.06694}}].

\bibitem{vinyoles2017new}
N.~Vinyoles, A.~M. Serenelli, F.~L. Villante, S.~Basu, J.~Bergstr{\"o}m and
  Gothers, \emph{A new generation of standard solar models}, {\emph{The
  Astrophysical Journal} {\bf 835} (2017) 202}.

\bibitem{Preto_1999}
M.~{Preto} and S.~{Tremaine}, \emph{{A Class of Symplectic Integrators with
  Adaptive Time Step for Separable Hamiltonian Systems}},
  \href{http://dx.doi.org/10.1086/301102}{\emph{\aj} {\bf 118} (Nov., 1999)
  2532--2541}, [\href{http://arxiv.org/abs/astro-ph/9906322}{{\tt
  astro-ph/9906322}}].

\bibitem{Hairer_2005}
E.~Hairer and G.~S\"{o}derlind, \emph{Explicit, time reversible, adaptive step
  size control}, \href{http://dx.doi.org/10.1137/040606995}{\emph{SIAM Journal
  on Scientific Computing} {\bf 26} (2005) 1838--1851},
  [\href{http://arxiv.org/abs/https://doi.org/10.1137/040606995}{{\tt
  https://doi.org/10.1137/040606995}}].

\bibitem{Richardson_2011}
A.~S. Richardson and J.~M. Finn, \emph{Symplectic integrators with adaptive
  time steps},
  \href{http://dx.doi.org/10.1088/0741-3335/54/1/014004}{\emph{Plasma Physics
  and Controlled Fusion} {\bf 54} (Dec., 2011) 014004}.

\bibitem{duruisseaux_2021}
V.~Duruisseaux, J.~Schmitt and M.~Leok, \emph{Adaptive {H}amiltonian
  variational integrators and symplectic accelerated optimization},  2021.

\bibitem{2015MNRAS.446.1424R}
H.~{Rein} and D.~S. {Spiegel}, \emph{{IAS15: a fast, adaptive, high-order
  integrator for gravitational dynamics, accurate to machine precision over a
  billion orbits}},
  \href{http://dx.doi.org/10.1093/mnras/stu2164}{\emph{\mnras} {\bf 446} (Jan.,
  2015) 1424--1437}, [\href{http://arxiv.org/abs/1409.4779}{{\tt 1409.4779}}].

\bibitem{dprkn}
J.~Dormand, M.~El-Mikkawy and P.~Prince, \emph{High-order embedded
  {Runge-Kutta-Nystrom} formulae}, {\emph{IMA Journal of Numerical Analysis}
  {\bf 7} (1987) 423–430}.

\bibitem{diffeqjl}
C.~Rackauckas and Q.~Nie, \emph{Differentialequations.jl--a performant and
  feature-rich ecosystem for solving differential equations in julia},
  {\emph{Journal of Open Research Software} {\bf 5} (2017) }.

\bibitem{Efron_1979}
B.~Efron, \emph{{Bootstrap Methods: Another Look at the Jackknife}},
  \href{http://dx.doi.org/10.1214/aos/1176344552}{\emph{The Annals of
  Statistics} {\bf 7} (1979) 1 -- 26}.

\bibitem{Peter:2009mi}
A.~H.~G. Peter, \emph{{Dark matter in the solar system I: The distribution
  function of WIMPs at the Earth from solar capture}},
  \href{http://dx.doi.org/10.1103/PhysRevD.79.103531}{\emph{Phys. Rev. D} {\bf
  79} (2009) 103531}, [\href{http://arxiv.org/abs/0902.1344}{{\tt 0902.1344}}].

\bibitem{gould1988direct}
A.~Gould, \emph{Direct and indirect capture of weakly interacting massive
  particles by the earth}, {\emph{The Astrophysical Journal} {\bf 328} (1988)
  919--939}.

\bibitem{Berlin:2019uco}
A.~Berlin, R.~T. D'Agnolo, S.~A.~R. Ellis, P.~Schuster and N.~Toro,
  \emph{{Directly Deflecting Particle Dark Matter}},
  \href{http://dx.doi.org/10.1103/PhysRevLett.124.011801}{\emph{Phys. Rev.
  Lett.} {\bf 124} (2020) 011801}, [\href{http://arxiv.org/abs/1908.06982}{{\tt
  1908.06982}}].

\bibitem{VanTilburg:2024xib}
K.~Van~Tilburg, \emph{{Wake forces in a background of quadratically coupled
  mediators}}, \href{http://dx.doi.org/10.1103/PhysRevD.109.096036}{\emph{Phys.
  Rev. D} {\bf 109} (2024) 096036}.

\bibitem{Iles_2024}
E.~Iles, S.~Heeba and K.~Schutz, \emph{Direct detection of the millicharged
  background},  2024.

\bibitem{1985ApJ...296..679P}
W.~H. {Press} and D.~N. {Spergel}, \emph{{Capture by the sun of a galactic
  population of weakly interacting, massive particles}},
  \href{http://dx.doi.org/10.1086/163485}{\emph{\apj} {\bf 296} (Sept., 1985)
  679--684}.

\bibitem{2009PhRvD..79j3532P}
A.~H.~G. {Peter}, \emph{{Dark matter in the Solar System. II. WIMP annihilation
  rates in the Sun}},
  \href{http://dx.doi.org/10.1103/PhysRevD.79.103532}{\emph{\prd} {\bf 79}
  (May, 2009) 103532}, [\href{http://arxiv.org/abs/0902.1347}{{\tt
  0902.1347}}].

\bibitem{1987NuPhB.279..804S}
M.~{Srednicki}, K.~A. {Olive} and J.~{Silk}, \emph{{High-energy neutrinos from
  the sun and cold dark matter}},
  \href{http://dx.doi.org/10.1016/0550-3213(87)90020-4}{\emph{Nuclear Physics
  B} {\bf 279} (Jan., 1987) 804--823}.

\bibitem{Lundberg:2004dn}
J.~Lundberg and J.~Edsjo, \emph{{WIMP diffusion in the solar system including
  solar depletion and its effect on earth capture rates}},
  \href{http://dx.doi.org/10.1103/PhysRevD.69.123505}{\emph{Phys. Rev. D} {\bf
  69} (2004) 123505}, [\href{http://arxiv.org/abs/astro-ph/0401113}{{\tt
  astro-ph/0401113}}].

\bibitem{2012PhRvD..85l3514S}
S.~{Sivertsson} and J.~{Edsj{\"o}}, \emph{{WIMP diffusion in the Solar System
  including solar WIMP-nucleon scattering}},
  \href{http://dx.doi.org/10.1103/PhysRevD.85.123514}{\emph{\prd} {\bf 85}
  (June, 2012) 123514}.

\bibitem{fitzpatrick2012introduction}
R.~Fitzpatrick, \emph{An introduction to celestial mechanics}.
\newblock Cambridge University Press, 2012.

\bibitem{lee2014effect}
S.~K. Lee, M.~Lisanti, A.~H. Peter and B.~R. Safdi, \emph{Effect of
  gravitational focusing on annual modulation in dark-matter direct-detection
  experiments}, {\emph{Physical review letters} {\bf 112} (2014) 011301}.

\bibitem{Levin_2006}
Y.~Levin, \emph{Ejection of high‐velocity stars from the galactic center by
  an inspiraling intermediate‐mass black hole},
  \href{http://dx.doi.org/10.1086/507830}{\emph{The Astrophysical Journal} {\bf
  653} (Dec., 2006) 1203–1209}.

\bibitem{Laskar_1988_secular}
J.~{Laskar}, \emph{{Secular evolution of the solar system over 10 million
  years}}, {\emph{\aap} {\bf 198} (June, 1988) 341--362}.

\bibitem{Morbidelli_2009}
A.~{Morbidelli}, R.~{Brasser}, K.~{Tsiganis}, R.~{Gomes} and H.~F. {Levison},
  \emph{{Constructing the secular architecture of the solar system I: The giant
  planets}}, \href{http://dx.doi.org/10.1051/0004-6361/200912876}{\emph{\aap}
  {\bf 507} (Nov., 2009) 1041--1052}.

\bibitem{Brasser_2009}
R.~{Brasser}, A.~{Morbidelli}, R.~{Gomes}, K.~{Tsiganis} and H.~F. {Levison},
  \emph{{Constructing the secular architecture of the solar system II: The
  terrestrial planets}},
  \href{http://dx.doi.org/10.1051/0004-6361/200912878}{\emph{\aap} {\bf 507}
  (Nov., 2009) 1053--1065}, [\href{http://arxiv.org/abs/0909.1891}{{\tt
  0909.1891}}].

\bibitem{Scherer_1996}
K.~{Scherer} and W.~{Neutsch}, \emph{{On the Secular Evolution of Asteroids}},
  in \emph{Completing the Inventory of the Solar System} (T.~{Rettig} and J.~M.
  {Hahn}, eds.), vol.~107 of \emph{Astronomical Society of the Pacific
  Conference Series}, pp.~69--74, Jan., 1996.

\bibitem{Gronchi_2011}
G.~F. Gronchi and C.~Tardioli, \emph{Secular evolution of the orbit distance
  and asteroid hazard}, .

\bibitem{Novakovic_2015}
B.~{Novakovi{\'c}}, C.~{Maurel}, G.~{Tsirvoulis} and Z.~{Kne{\v{z}}evi{\'c}},
  \emph{{Asteroid Secular Dynamics: {C}eres{\textquoteright} Fingerprint
  Identified}},
  \href{http://dx.doi.org/10.1088/2041-8205/807/1/L5}{\emph{\apjl} {\bf 807}
  (July, 2015) L5}, [\href{http://arxiv.org/abs/1506.01586}{{\tt 1506.01586}}].

\bibitem{Novakovic_2016}
B.~Novaković, G.~Tsirvoulis, S.~Marò, V.~Đošović and C.~Maurel,
  \emph{Secular evolution of asteroid families: the role of {C}eres},
  \href{http://dx.doi.org/10.1017/s1743921315008595}{\emph{Proceedings of the
  International Astronomical Union} {\bf 10} (Aug., 2015) 46–54}.

\bibitem{Correia_2009}
A.~C.~M. {Correia}, \emph{{Secular Evolution of a Satellite by Tidal Effect:
  Application to Triton}},
  \href{http://dx.doi.org/10.1088/0004-637X/704/1/L1}{\emph{\apjl} {\bf 704}
  (Oct., 2009) L1--L4}, [\href{http://arxiv.org/abs/0909.4210}{{\tt
  0909.4210}}].

\bibitem{Lei_2020}
H.~Lei, \emph{Dynamical models for secular evolution of navigation satellites},
  {\emph{Astrodynamics} {\bf 4} (2020) 57--73}.

\bibitem{Goldstein_classicalmechanics}
H.~Goldstein, C.~Poole, J.~Safko and S.~R. Addison, \emph{{Classical Mechanics,
  3rd ed.}}, \href{http://dx.doi.org/10.1119/1.1484149}{\emph{American Journal
  of Physics} {\bf 70} (07, 2002) 782--783},
  [\href{http://arxiv.org/abs/https://pubs.aip.org/aapt/ajp/article-pdf/70/7/782/7530963/782\_1\_online.pdf}{{\tt
  https://pubs.aip.org/aapt/ajp/article-pdf/70/7/782/7530963/782\_1\_online.pdf}}].

\bibitem{Damour_1999}
T.~Damour and L.~M. Krauss, \emph{New {WIMP} population in the solar system and
  new signals for dark-matter detectors},
  \href{http://dx.doi.org/10.1103/physrevd.59.063509}{\emph{Physical Review D}
  {\bf 59} (feb, 1999) }.

\bibitem{Rozelot_2011}
{Rozelot, J.-P.} and {Damiani, C.}, \emph{History of solar oblateness
  measurements and interpretation},
  \href{http://dx.doi.org/10.1140/epjh/e2011-20017-4}{\emph{Eur. Phys. J. H}
  {\bf 36} (2011) 407--436}.

\bibitem{Hobson_2006}
M.~P. Hobson, G.~P. Efstathiou and A.~N. Lasenby, \emph{General Relativity: An
  Introduction for Physicists}.
\newblock Cambridge University Press, 2006,
  \href{http://dx.doi.org/10.1017/CBO9780511790904}{10.1017/CBO9780511790904}.

\bibitem{Standish_2006}
M.~Standish and J.~Williams, \emph{Orbital Ephemerides of the Sun, Moon, and
  Planets}.
\newblock 01, 2006.

\bibitem{Lithwick_2011}
Y.~{Lithwick} and S.~{Naoz}, \emph{{The Eccentric Kozai Mechanism for a Test
  Particle}}, \href{http://dx.doi.org/10.1088/0004-637X/742/2/94}{\emph{\apj}
  {\bf 742} (Dec., 2011) 94}, [\href{http://arxiv.org/abs/1106.3329}{{\tt
  1106.3329}}].

\bibitem{williams2021jupiter}
D.~R. Williams, ``Jupiter fact sheet.''
  \url{https://nssdc.gsfc.nasa.gov/planetary/factsheet/jupiterfact.html},
  December 23, 2021.

\bibitem{Li_2014}
G.~Li, S.~Naoz, M.~Holman and A.~Loeb, \emph{Chaos in the test particle
  eccentric {Kozai-Lidov} mechanism},
  \href{http://dx.doi.org/10.1088/0004-637X/791/2/86}{\emph{The Astrophysical
  Journal} {\bf 791} (jul, 2014) 86}.

\bibitem{Damour:1998rh}
T.~Damour and L.~M. Krauss, \emph{{A new Solar System population of WIMP dark
  matter}}, \href{http://dx.doi.org/10.1103/PhysRevLett.81.5726}{\emph{Phys.
  Rev. Lett.} {\bf 81} (1998) 5726--5729},
  [\href{http://arxiv.org/abs/astro-ph/9806165}{{\tt astro-ph/9806165}}].

\bibitem{Peter:2009mm}
A.~H.~G. Peter, \emph{{Dark matter in the Solar System III: The distribution
  function of WIMPs at the Earth from gravitational capture}},
  \href{http://dx.doi.org/10.1103/PhysRevD.79.103533}{\emph{Phys. Rev. D} {\bf
  79} (2009) 103533}, [\href{http://arxiv.org/abs/0902.1348}{{\tt 0902.1348}}].

\bibitem{East:2022rsi}
W.~E. East and J.~Huang, \emph{{Dark photon vortex formation and dynamics}},
  \href{http://dx.doi.org/10.1007/JHEP12(2022)089}{\emph{JHEP} {\bf 12} (2022)
  089}, [\href{http://arxiv.org/abs/2206.12432}{{\tt 2206.12432}}].

\bibitem{Cyncynates:2023zwj}
D.~Cyncynates and Z.~J. Weiner, \emph{{Detectable, defect-free dark photon dark
  matter}},  \href{http://arxiv.org/abs/2310.18397}{{\tt 2310.18397}}.

\end{thebibliography}\endgroup
\bibliographystyle{JHEP}

\end{document}